\documentclass[twocolumn,showpacs,superscriptaddress,preprintnumbers,prd]{revtex4-1}
\usepackage{graphicx,amsmath,amssymb,dcolumn,bm}
\usepackage{color}
\usepackage{braket}

\usepackage{slashed}
\newcommand{\pslash}{\slashed{p}}

\begin{document}
\preprint{KEK-TH-1959, J-PARC-TH-0086}
\title{Hadron tomography by generalized distribution amplitudes \\ 
in pion-pair production process \boldmath$\gamma^* \gamma \rightarrow \pi^0 \pi^0 $ \\
and gravitational form factors for pion}
\author{S. Kumano}
\affiliation{KEK Theory Center,
             Institute of Particle and Nuclear Studies, \\
             High Energy Accelerator Research Organization (KEK), \\
             1-1, Ooho, Tsukuba, Ibaraki, 305-0801, Japan}
\affiliation{J-PARC Branch, KEK Theory Center,
             Institute of Particle and Nuclear Studies, KEK, \\
           and Theory Group, Particle and Nuclear Physics Division, 
           J-PARC Center, \\
           203-1, Shirakata, Tokai, Ibaraki, 319-1106, Japan}
\affiliation{Department of Particle and Nuclear Physics, \\
             Graduate University for Advanced Studies (SOKENDAI), \\
             1-1, Ooho, Tsukuba, Ibaraki, 305-0801, Japan} 
\author{Qin-Tao Song}
\affiliation{KEK Theory Center,
             Institute of Particle and Nuclear Studies, \\
             High Energy Accelerator Research Organization (KEK), \\
             1-1, Ooho, Tsukuba, Ibaraki, 305-0801, Japan}
\affiliation{Department of Particle and Nuclear Physics, \\
             Graduate University for Advanced Studies (SOKENDAI), \\
             1-1, Ooho, Tsukuba, Ibaraki, 305-0801, Japan}
\author{O. V. Teryaev}
\affiliation{KEK Theory Center,
             Institute of Particle and Nuclear Studies, \\
             High Energy Accelerator Research Organization (KEK), \\
             1-1, Ooho, Tsukuba, Ibaraki, 305-0801, Japan}
\affiliation{Bogoliubov Laboratory of Theoretical Physics, \\
            Joint Institute for Nuclear Research, 141980 Dubna, Russia}
\date{January 10, 2019}

\begin{abstract}
Hadron tomography can be investigated by three-dimensional structure functions
such as generalized parton distributions (GPDs),
transverse-momentum-dependent parton distributions,
and generalized distribution amplitudes (GDAs). 
Here, we extract the GDAs, which are $s$-$t$ crossed quantities of
the GPDs, from cross-section measurements of hadron-pair production process 
$\gamma^* \gamma \rightarrow \pi^0 \pi^0$ at KEKB. 
This work is the first attempt to obtain the GDAs from the actual
experimental data.
The GDAs are expressed by a number of parameters and they are determined 
from the data of $\gamma^* \gamma \rightarrow \pi^0 \pi^0$ 
by including intermediate scalar- and tensor-meson contributions 
to the cross section.
Our results indicate that the dependence of parton-momentum fraction $z$ 
in the GDAs is close to the asymptotic one.
The timelike gravitational form factors $\Theta_1$ and $\Theta_2$ 
are obtained from our GDAs, and they are converted to the spacelike 
ones by the dispersion relation.
From the spacelike $\Theta_1$ and $\Theta_2$, 
gravitational densities of the pion are calculated.
Then, we obtained the mass (energy) radius 
and the mechanical (pressure and shear force) radius
from $\Theta_2$ and $\Theta_1$, respectively.
They are calculated as
$\sqrt {\langle r^2 \rangle _{\text{mass}}} =0.32 \sim 0.39$ fm,
whereas the mechanical radius is larger 
$\sqrt {\langle r^2 \rangle _{\text{mech}}} =0.82 \sim 0.88$ fm.
This is the first report on the gravitational radius of a hadron
from actual experimental measurements.
It is interesting to find the possibility 
that the gravitational mass and mechanical radii 
could be different from the experimental charge radius 
$\sqrt {\langle r^2 \rangle _{\text{charge}}} =0.672 \pm 0.008$ fm
for the charged pion.
For drawing a clear conclusion on the GDAs 
of hadrons, accurate experimental data are needed, and it should 
be possible, for example, by future measurements of super-KEKB
and international linear collider.
Accurate measurements will not only provide important information 
on hadron tomography but also possibly shed light on 
gravitational physics in the quark and gluon level.
\end{abstract}
\maketitle

\section{Introduction}
\label{intro}

Internal structure of hadrons has been investigated 
in terms of form factors and parton distribution functions (PDFs). 
Now, the field of hadron tomography, namely hadron-structure studies
by three-dimensional (3D) structure functions 
\cite{gpds-gdas,gpds,tmds,kk2014}, 
is one of fast developing areas in particle and nuclear physics. 
The 3D structure functions contain information on both the form factors 
and the PDFs, and they are ultimate quantities for understanding 
the nature of hadrons from low to high energies. Furthermore, 
it is essential to investigate the 3D structure of the nucleon 
for understanding the origin of nucleon spin because orbital 
angular momenta of partons could play an important role.
The 3D structure functions could be also useful for clarifying internal 
quark-gluon configurations of exotic-hadron candidates \cite{kk2014}.

Among the 3D structure functions, generalized parton distributions (GPDs)
\cite{gpds-gdas,gpds} and transverse-momentum-dependent parton distributions (TMDs) 
\cite{tmds} have been investigated extensively in recent years. 
We now have crude idea on these distributions. 
There are also generalized distribution amplitudes (GDAs) \cite{gpds-gdas,kk2014}
as one of the 3D structure functions, and it is rather an unexplored field 
in comparison with the GPD and TMD studies.
The GDAs can be obtained theoretically by the $s$-$t$ crossing of the GPDs.
Here, $s$ and $t$ are Mandelstam variables. Therefore, the GDA studies
should also be valuable for the GPD understanding.
In particular, both GPDs and GDAs can be expressed by common 
double distributions (DDs) with different Radon transforms
as discussed later in Sec.\,\ref{radon}. Therefore, the GDA studies are 
valuable also for understanding the GPDs through the DDs
and simply by the $s$-$t$ crossing.

The GDAs are key quantities for probing 3D structure of hadrons
by timelike processes. In addition, one of the other important advantages 
of the GDAs is that 3D tomography is possible in principle for exotic-hadron 
candidates \cite{kk2014} because they can be produced in a pair in the final state, 
whereas no stable exotic hadron exists as a fixed target for measuring 
their GPDs and TMDs. The constituent counting rule can be used for
identifying the number of elementary constituents in exotic hadron
candidates at high energies. We should be able to distinguish
exotic multiquark states from the ordinary $q\bar q$ and $qqq$ ones
by the counting rule \cite{counting,kk2014}. Furthermore, form factors 
contained in the GDAs should provide information whether exotic 
hadron candidates are diffuse molecular states or compact multiquark 
ones \cite{kk2014}.

Another advantage is that the GDAs and GPDs contain 
information on form factors of the energy-momentum tensor so that
the gravitational-interaction radius can be investigated.
Although the root-mean-square charge radii are well known
for the nucleons, the gravitational radius has never been 
measured experimentally. 
We try to extract the gravitational-interaction sizes,
namely mass and mechanical radii,
from existing experimental data in this work.
Of course, the gravitational interactions are too weak to be measured 
directly for hadrons and elementary particles, such as quarks and gluons,
``usually" in accelerator experiments,
and there is no reliable quantum theory for the gravitational interactions
at this stage. Nonetheless, it is interesting that the hadron tomography 
studies can access the gravitational information in hadrons through
the energy-momentum tensor.

Fortunately, the Belle collaboration recently reported 
the cross sections for the pion-pair production 
in two-photon process $\gamma^* \gamma \rightarrow \pi^0 \pi^0 $
at KEKB with various kinematical conditions \cite{pion-da-exp,Masuda:2015yoh}. 
It is our research purpose of this paper to extract the pion GDAs 
from the Belle measurements. Our studies should be the first attempt
to extract any hadron GDAs from actual experimental measurements.
Now, other hadron production processes $\gamma^* \gamma \rightarrow h\bar h$
are being analyzed in the Belle collaboration, so that other
GDAs can be extracted in future.
Furthermore, the KEKB accelerator has just upgraded and 
accurate measurements are expected in future for the two-photon processes.
The two-photon processes have been used for investigating
existence and properties of new hadrons in electron-positron
annihilation reactions \cite{two-photon}.
The same two-photon processes should be possible 
at the future international linear collider \cite{ilc},
and the GDAs will be investigated in the PANDA project \cite{gsi}.
This work is merely the first step for determining the GDAs; however,
much progress is expected in the near future.

In this article, the generalized TMD (GTMD) or the Wigner distribution 
is explained first as a generating function for the 3D structure 
functions in Sec.\,\ref{3d}.
Then, the GPDs and GDAs are introduced, and the form factors
of energy-momentum tensor in the GDAs are explained in connection
with the gravitational radii.
Next, our theoretical formalism is developed for the 
$\gamma^* \gamma \rightarrow \pi^0 \pi^0 $ cross section
and the pion GDAs in Sec.\,\ref{formalism}.
The cross section of $\gamma^* \gamma \to \pi^0 \pi^0$
is expressed in terms of the GDAs.
For extracting the GDAs from the experimental data, we introduce 
a parametrization of the GDAs, which are then determined by 
the analysis of the Belle measurement.
Our analysis method is described in Sec.\,\ref{formalism},
results are shown in Sec.\,\ref{results}.
Finally, our studies are summarized in Sec.\,\ref{summary}.

\section{Three-dimensional structure functions of hadrons}
\label{3d}

The 3D structure of hadrons becomes one of hot topics in hadron physics, 
and it can be investigated by the GPDs, TMDs, and GDAs.
First, we explain the Wigner phase-space distribution 
and the GTMD in Sec.\,\ref{wigner-3d} as generating functions
for form factors, PDFs, and the 3D structure functions.
Then, we discuss the details of the GPDs and GDAs which are relevant
to our studies including their relations
in Secs.\,\ref{gpd}, \ref{gda}, and \ref{gpd-gda}.
Both GPDs and GDAs are expressed by double distributions through
Radon transforms as explained in Sec.\,\ref{radon}.
The GDAs are related to the timelike form factors of 
the energy-momentum tensor, and then the spacelike
gravitational form factors and radii are
explained in Sec.\,\ref{gravitation-radius}.

\subsection{Wigner distribution and three-dimensional structure functions}
\label{wigner-3d}

The 3D structure functions originate from the generating function,
called the Wigner distribution, which is a phase space distribution
$W (\vec r, \vec k \,)$ expressed by the space coordinate $\vec r$
and momentum $\vec k$. In the classical limit of $\hbar \to 0$,
it becomes the $\delta$ function $\delta (H(\vec r, \vec k \,)-E)$,
which is the classical trajectory in the phase space.
Therefore, its delocalization indicates quantum effects,
and the Wigner function contains full information
for describing quantum systems.

For the nucleon, the Wigner distribution was originally defined
in Ref.~\cite{Wigner} as the 6-dimensional phase-space distribution
$W (x, \vec k_T, \vec r \,)$, where $x$ is the Bjorken scaling variable
and $\vec k_T$ is the transverse momentum.
However, it was defined in a special Lorentz frame, so that 
a new definition was proposed in the infinite momentum frame 
\cite{lp-2011} to express it by 5-dimensional phase-space 
distribution $W (x, \vec k_T, \vec r_T)$.
It is equal to the $\Delta^+ = 0$ ($\xi =0$) limit of the 
generalized transverse-momentum-dependent parton distribution 
(GTMD)~\cite{lplp}.

\begin{figure}[t!]
\vspace{+0.20cm}
\begin{center}
   \includegraphics[width=8.5cm]{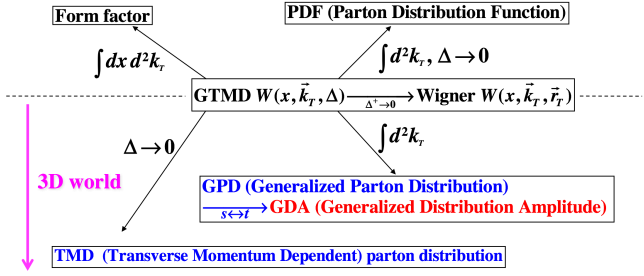}
\end{center}
\vspace{-0.6cm}
\caption{Wigner distribution, GTMD, and 3D structure functions.}
\label{fig:wigner-3d-sfs}
\vspace{-0.3cm}
\end{figure}

In Fig.~\ref{fig:wigner-3d-sfs}, relations of the GTMD
and the Wigner distribution to the form factor, PDF,
and 3D structure functions are shown ~\cite{Wigner,lp-2011,lplp}
by integrating the GTMD by various kinematical variables.
The form factors and the PDFs have been investigated until recently,
and now the nucleon-structure studies focus on 3D structure functions,
the GPDs and TMDs. However, there are few recent research activities 
on the GDAs which have a close connection to the GPDs 
by the $s$-$t$ crossing.

\subsection{Generalized parton distributions}
\label{gpd}

\begin{figure}[b!]
\vspace{-0.00cm}
\begin{center}
   \includegraphics[width=6.5cm]{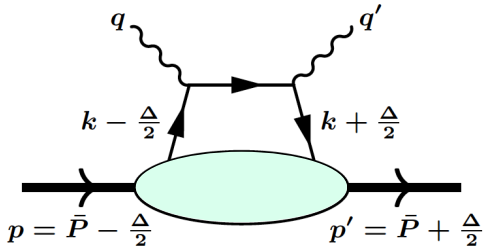}
\end{center}
\vspace{-0.5cm}
\caption{Kinematics for GPDs in deeply virtual Compton 
scattering process.}
\label{fig:gpd-fig}
\vspace{-0.0cm}
\end{figure}

The GPDs have been investigated by deeply virtual Compton scattering
(DVCS) process as shown in Fig.\,\ref{fig:gpd-fig} and also
by meson and lepton-pair production processes 
\cite{transition-gpd,pire-gpd}.
In addition, there are possibilities to study them at hadron-beam
facilities, for example, by $N + N \to N +\pi + B$ where $N$ and $B$ are
the nucleon and baryon \cite{kss-gpd}
and by an exclusive Drell-Yan process 
$\pi^- + p \to \mu^+ \mu^- + n$ 
\cite{pire-gpd,Goloskokov:2015zsa,j-parc-gpd}.
Here, we explain the definition of the GPDs by using the DVCS process
($\gamma^* + h \to \gamma + h$)
because its $s$-$t$ crossing is the two-photon process 
($\gamma^* + \gamma \to h +\bar h$)
which is analyzed in this work in terms of the GDAs.

We define kinematical variables for expressing the GPDs
of the nucleon. The initial and final momenta of the nucleon 
are $p$ and $p'$, respectively, as shown in Fig.\,\ref{fig:gpd-fig},
and they are $q$ and $q'$ for the photon. Then, their average momenta
and the momentum transfer are given as 
\cite{gpds-gdas,gpds,kk2014,freund-nlo-dvcs-2001}
\begin{align}
\bar P = \frac{p+p'}{2} , \ \ 
\bar q = \frac{q+q'}{2} , \ \ 
\Delta = p'-p = q-q' .
\end{align}
Expressing the momentum squared quantities as 
$Q^2 = -q^2$ and $\bar Q^2 = - \bar q^2$, we define
the Bjorken scaling variable $x$, momentum-transfer-squared $t$,
and the skewdness parameter $\xi$ as
\begin{align}
x = \frac{Q^2}{2p \cdot q} , \ \ \ \ 
t = \Delta^2 , \ \ \ \ 
\xi = \frac{\bar Q^2}{2 \bar P \cdot \bar q} .
\end{align}
If the kinematical condition $Q^2 \gg |t|$ is satisfied,
the skewdness parameter is expressed by the lightcone-coordinate
expression as 
\begin{align}
\! \! \!
\xi = \frac{x + x \, t / (2 Q^2)}{2 - x + x \, t / Q^2 } 
    \simeq \frac{x}{2 - x} 
    = - \frac{\Delta^+}{2\bar P^+}
    \ \  \text{for} \ Q^2 \gg |t| .
\end{align}
The lightcone notation is given by
$a=(a^+, \, a^-, \, \vec a_\perp)$
with $a^\pm = (a^0 \pm a^3)/\sqrt{2}$ and
the transverse vector $\vec a_\perp$. 
Then, the momenta are expressed as
\begin{alignat}{2}
& \! \! 
p \simeq \left ( p^+, \, 0, \, \vec 0_\perp \right ) , \ \ \, &
& p' \simeq \left ( {p'}^+, \, 0, \, \vec 0_\perp \right ) ,
\nonumber \\
& \! \! 
q \simeq \left ( -x p^+, \, \frac{Q^2}{2xp^+}, \, \vec 0_\perp \right ) , \ \ \, &
& q' \simeq \left ( 0,     \,  \frac{Q^2}{2xp^+}, \, \vec 0_\perp \right ) ,
\end{alignat}
by using the relation
$(p^+)^2 , \, Q^2 \gg M^2, \, |t|$.
The scaling variable $x$ is the lightcone momentum fraction 
carried by a quark in the nucleon, whereas 
the skewdness parameter $\xi$ or the momentum $\Delta$ 
indicates the momentum transfer from the initial nucleon
to the final one or the momentum transfer between the initial
and final quarks.
The cross section of the DVCS $\gamma^* h \rightarrow \gamma h$
can be factorized into the hard part of quark interactions 
and the soft one expressed by the GPDs as shown in Fig.\,\ref{fig:gpd-fig}
if the kinematical condition
\begin{align}
Q^2 \gg |t|, \, \Lambda^2_{\text{QCD}} ,
\label{eqn:hard-gpd}
\end{align}
is satisfied.
Here, $\Lambda_{\text{QCD}}$ is the QCD scale parameter.

The GPDs for the nucleon are defined by off-forward matrix elements
of quark and gluon operators with a lightcone separation,
and quark GPDs are defined by
\begin{align}
 & \! \! \! \! \! \! \! \! 
 \int\frac{d y^-}{4\pi}e^{i x \bar P^+ y^-}
 \left< N(p') \left| 
\overline{q}(-y/2) \gamma^+ q(y/2) 
 \right| N(p) \right> \Big |_{y^+ = \vec y_\perp =0}
\nonumber \\
 & \! \! \! \! \! \! \! \! \! 
 = \frac{1}{2  \bar P^+} \, \overline{u} (p') \!
 \left [ H_q (x,\xi,t) \gamma^+
   \! + \!  E_q (x,\xi,t)  \frac{i \sigma^{+ \alpha} \Delta_\alpha}{2 \, M}
 \right ] u (p) .
\label{eqn:gpd-n}
\end{align}
Here, $q(y/2)$ is the quark field, $M$ is the nucleon mass, and
$\sigma^{\alpha\beta}$ is given by
$\sigma^{\alpha\beta}=(i/2)[\gamma^\alpha, \gamma^\beta]$.
The functions $H_q (x,\xi,t)$ and $E_q (x,\xi,t)$ are 
the unpolarized GPDs of the nucleon, and there are also gluon
GPDs $H_g (x,\xi,t)$ and $E_g (x,\xi,t)$
defined in a similar way \cite{gpds}.
To be precise, the link operator needs to be introduced
in the left-hand side of Eq.\,(\ref{eqn:gpd-n})
to satisfy the color gauge invariance.
In this article, it is simply ignored.

The advantages of the GPDs are that they contain both longitudinal
momentum distributions for partons and transverse form factors.
In fact, the GPDs $H_q (x,\xi,t)$ become unpolarized PDFs 
for the nucleon in the forward limit ($\Delta,\, \xi,\, t \rightarrow 0$):
\begin{equation}
H_q (x, 0, 0) = \theta (x) q(x) - \theta(-x) \bar q(-x)  ,
\label{eqn:gpd-pdf}
\end{equation}
where $\theta (x)$ is the step function, 
$\theta(x)=1$ for $x>0$ and 
$\theta(x)=0$ for $x<0$.
Their first moments become Dirac and Pauli form factors
$F_1 (t)$ and $F_2 (t)$, respectively:
\begin{align}
\! \! \! \! \!
\int_{-1}^{1} dx            H_q(x,\xi,t)  = F_1 (t), \ 
\int_{-1}^{1} dx            E_q(x,\xi,t)  = F_2 (t),
\label{eqn:gpd-form}
\end{align}
Another important feature, actually the most important for high-energy
spin physicists, of the GPDs is that a second moment indicates
a quark orbital-angular-momentum contribution ($L_q$) 
to the nucleon spin:
\begin{align}
   J_q & = \frac{1}{2} \int dx \, x \, [ H_q (x,\xi,t=0) +E_q (x,\xi,t=0) ]
\nonumber \\
       & = \frac{1}{2} \Delta q^+ + L_q ,
\label{eqn:Jq}
\end{align}
because we know the quark contribution 
$\Delta q^+ =\Delta q +\Delta \bar q$
from polarized charged-lepton DIS measurements.

The GPDs have been mainly investigated for the nucleon. However, since
the pion GDAs are investigated in this work and they are related to
the pion GPDs by the $s$-$t$ crossing, we also show the definition
of the pion GPDs in the same way with Eq.\,(\ref{eqn:gpd-n})
for the nucleon \cite{teryaev-2001}:
\begin{align}
 & 
 \int\frac{d y^-}{4\pi}e^{i x \bar P^+ y^-}
 \left< \pi (p') \left| 
\overline{q}(-y/2) \gamma^+ q(y/2) 
 \right| \pi (p) \right> \Big |_{y^+ = \vec y_\perp =0}
\nonumber \\
 & 
 =  H_q^\pi (x,\xi,t) .
\label{eqn:gpd-pi}
\end{align}
The pion is a scalar particle, so that the function $E_q (x,\xi,t)$
does not exist.

In comparison with PDF parametrizations, such studies are still premature 
for the GPDs due to the lack of experimental information. 
The simplest idea is to use the factorized form into the longitudinal
PDF $q(x)$ and the transverse form factor $F_T (t, x)$
at $x$ \cite{gpd-paramet}.
For example, it is expressed as
\begin{align}
H_q (x,\xi=0,t)= q (x) \, F_T (t, x) ,
\label{eqn:gpd-paramet1}
\end{align}
at $\xi=0$ for $x>0$.
Namely, the GPDs contain information on both the PDFs and
the form factors as already shown by the sum rules in 
Eqs.\,(\ref{eqn:gpd-pdf}) and (\ref{eqn:gpd-form}).

\subsection{Generalized distribution amplitudes}
\label{gda}

\begin{figure}[t!]
\vspace{+0.00cm}
\begin{center}
\hspace{+0.20cm}
   \includegraphics[width=5.5cm]{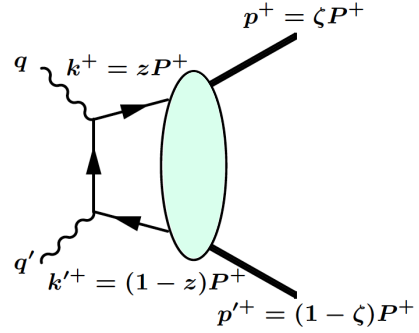}
\end{center}
\vspace{-0.6cm}
\caption{Kinematics for GDAs in two-photon process
$\gamma^* + \gamma \rightarrow h + \bar h$.
This process corresponds to the $s$-$t$ crossed one of
the Compton scattering process in Fig.\,\ref{fig:gpd-fig}.}
\label{fig:gda-fig}
\vspace{-0.0cm}
\end{figure}

If we exchange the $s$ and $t$ channels in the Compton scattering
in Fig.\,\ref{fig:gpd-fig}, it becomes the two-photon process
$\gamma^* + \gamma \rightarrow h + \bar h$ in Fig.\,\ref{fig:gda-fig}.
The GDAs describe the production of the hadron pair $h\bar h$ 
from a $q\bar q$ or gluon pair. 
We explain kinematical variables for describing the two-photon process 
and the GDAs 
\cite{gpds-gdas,muller-1994,Polyakov-1999,diehl-2000,nlo-2gamma,ee-eerhorho}
as shown in Fig.\,\ref{fig:gda-fig}.
The initial photon momenta are denoted as $q$ and $q'$,
the final hadron momenta are $p$ and $p'$, $P$ is their total momentum $P=p+p'$, 
and $k$ and $k'$ are quark and antiquark momenta.
One of the photon is taken as a real one with ${q'}^2 = 0$,
and another one should satisfy the condition 
\begin{align}
Q^2 = - q^2 \gg \Lambda^2_{\text{QCD}} , \ W^2
\label{eqn:hard-gpa}
\end{align}
so that the two-photon process is factorized into
a hard part and a soft one in terms of the GDAs
as shown in Fig.\,\ref{fig:gda-fig} \cite{factorization}.
Here, $W^2$ is one of the variables in the GDAs, 
and it is the invariant-mass squared $W^2$
of the final-hadron pair. It is also equal to the center-of-mass (c.m.) 
energy squared $s$:
\begin{align}
W^2 = (p+p')^2 = (q+q')^2 = s .
\end{align}

The second variable $\zeta$ indicates the lightcone momentum fraction
for one of the final hadrons in the total momentum $P$ as shown 
in Fig.\,\ref{fig:gda-fig}:
\begin{align}
\zeta = \frac{p \cdot q'}{P \cdot q'}
      = \frac{p^+}{P^+} = \frac{1+\beta \cos\theta}{2} .
\label{eqn:zeta-frac}
\end{align}
Here, $\theta$ is the scattering angle in the c.m. frame
of the final hadrons with the momentum assignments:
\begin{align}
q  & = \left ( q^0,      \, 0, \, 0, \,  |\vec q \,| \right ) , \ \ \ 
q' = \left ( |\vec q \,|, \, 0, \, 0, \, -|\vec q \,| \right ) ,
\nonumber \\
p  & = \left ( p^0,      \,  |\vec p \,| \sin \theta, \, 0, \,  
                             |\vec p \,| \cos \theta \right ) ,
\nonumber \\
p' & = \left ( p^0,      \, -|\vec p \,| \sin \theta, \, 0, \,  
                            -|\vec p \,| \cos \theta \right ) ,
\label{eqn:qqpp}
\end{align}
and $\beta$ is the hadron velocity defined by
\begin{align}
\beta = \frac{|\vec p \, |}{p^0} = \sqrt{1-\frac{4 m_h^2}{W^2}} ,
\label{eqn:beta}
\end{align}
with the final-hadron mass $m_h$.
The third variable $z$ is the lightcone momentum fraction for a quark
in the total hadron-pair momentum $P$, and it is defined by
\begin{align}
z = \frac{k \cdot q'}{P \cdot q'} = \frac{k^+}{P^+} .
\end{align}
The GDAs are expressed by these three variables, $z$, $\zeta$, 
and $W^2=s$. 

The quark GDAs are defined by the matrix element 
of the same operators used in defining the GPDs in 
Eq.\,(\ref{eqn:gpd-n}) between the the vacuum 
and the hadron pair:
\begin{align}
& \Phi_q^{h\bar h} (z,\zeta,W^2) 
= \int \frac{d y^-}{2\pi}\, e^{i (2z-1)\, P^+ y^- /2}
   \nonumber \\
& \ \ \ 
  \times \langle \, h(p) \, \bar h(p') \, | \, 
 \overline{q}(-y/2) \gamma^+ q(y/2) 
  \, | \, 0 \, \rangle \Big |_{y^+=\vec y_\perp =0} \, .
\label{eqn:gda-def}
\end{align}
We use the notation $\Phi_q^{h\bar h}$ for one specific quark ($q$)
without the summation over the quark flavor.
Here, the kinematical range of $z$ is $0 \le z \le 1$,
whereas the variable $z'=2z-1$ is often used with the same notation $z$
(or $x$) in the range $-1 \le z' \le 1$ for the distribution amplitude
as explained in Ref.\,\cite{j-parc-gpd}. However, because 
many articles of the GDAs use the notation $z$ in the range $0 \le z \le 1$,
we follow this convention in this work.
The expression 
$e^{i (2z-1)\, P^+ y^- /2} \langle \, h (p) \, \bar h (p') \, | \, 
 \overline{\psi}(-y/2) \gamma^+ \psi(y/2)   \, | \, 0 \, \rangle $
is sometimes written by the equivalent one as
$e^{- i z\, P^+ y^-}$ $\langle \, h (p) \, \bar h (p') \, | \, 
 \overline{\psi}(y) \gamma^+ \psi(0)   \, | \, 0 \, \rangle $.
Furthermore, the gauge link should be introduced in the nonlocal
operator to satisfy the color gauge invariance; however, it is 
simply neglected in this paper.
There are sum rules for the quark GDAs of the isospin $I=0$ 
two-meson final states 
\cite{Polyakov-1999,diehl-2000}:
\begin{align}
\int_0^1 dz \, & \Phi_q^{h\bar h (I=0)} (z,\zeta,W^2) = 0, 
\nonumber \\
\int_0^1 dz \, & (2z -1) \, \Phi_q^{h\bar h (I=0)} 
(z,\zeta,0) 
\nonumber \\
& \ \ 
= - 4 \, M_{2(q)}^h  \, \zeta (1-\zeta) ,
\label{eqn:gda-sum-I=0}
\end{align}
where $M_{2(q)}^h$ is the momentum fraction carried 
by flavor-$q$ quarks and antiquarks in the hadron $h$
(note: total quark fraction $\sum_q M_{2(q)}^h$).
As shown in Eq.\,(\ref{eqn:integral-over-z}), this integral
is expressed by the energy-momentum tensor of a quark,
so that the right-hand-side of Eq.\,(\ref{eqn:gda-sum-I=0})
should be described by the form factors of the energy-momentum tensor
at finite $W^2$ \cite{gda-form-factor-W2}.
There are recent theoretical studies on the energy-momentum tensor
for the nucleon \cite{energy-momentum} and on its lattice QCD estimate
\cite{lattice-emt-gpd}.
In general, there are two energy-momentum tensor form factors 
for the pion \cite{emt-pion,emt-pion1}, and they are explained in 
Secs.\ref{gravitation-radius} and \ref{gravitational-ffs}.

Since the GDAs contain intermediate-meson contributions as explained
in Sec.\,\ref{gda-expression}, the second sum of 
Eq.\,(\ref{eqn:gda-sum-I=0}) should be 
a complex value at finite $W^2$.
There are resonance terms and the continuum one which 
contains a quark part of the form factor $F_q^h (W^2)$ defined in
Eq.\,(\ref{eqn:form-factorw2}).
The explicit expression is shown later in 
Eqs.\,(\ref{eqn:gda-parametrization}) and (\ref{eqn:B10B12-final})
for analyzing actual experimental data.
Therefore, our studies can suggest the optimum form factor $F_q^h (W^2)$
of the energy-momentum tensor for the continuum part of the hadron $h$, 
and they are related to the size of gravitational interaction. 
The gravitational radii of a hadron 
are discussed in more details in Sec.\,\ref{gravitation-radius}.
The sum rule of Eq.\,(\ref{eqn:gda-sum-I=0}) was derived 
for the kinematical point of $W^2=0$ \cite{Polyakov-1999,diehl-2000},
and then it was considered even at finite $W^2$ 
as the form of form factor of the energy-momentum tensor
\cite{gda-form-factor-W2}.
However, since there are two gravitational form factors 
for the pion in general,
a relation between the GDAs and the form factors is newly derived
in Sec.\,\ref{gravitational-ffs} of this article.

\begin{figure}[t!]
\vspace{+0.00cm}
\begin{center}
\hspace{+0.20cm}
   \includegraphics[width=7.0cm]{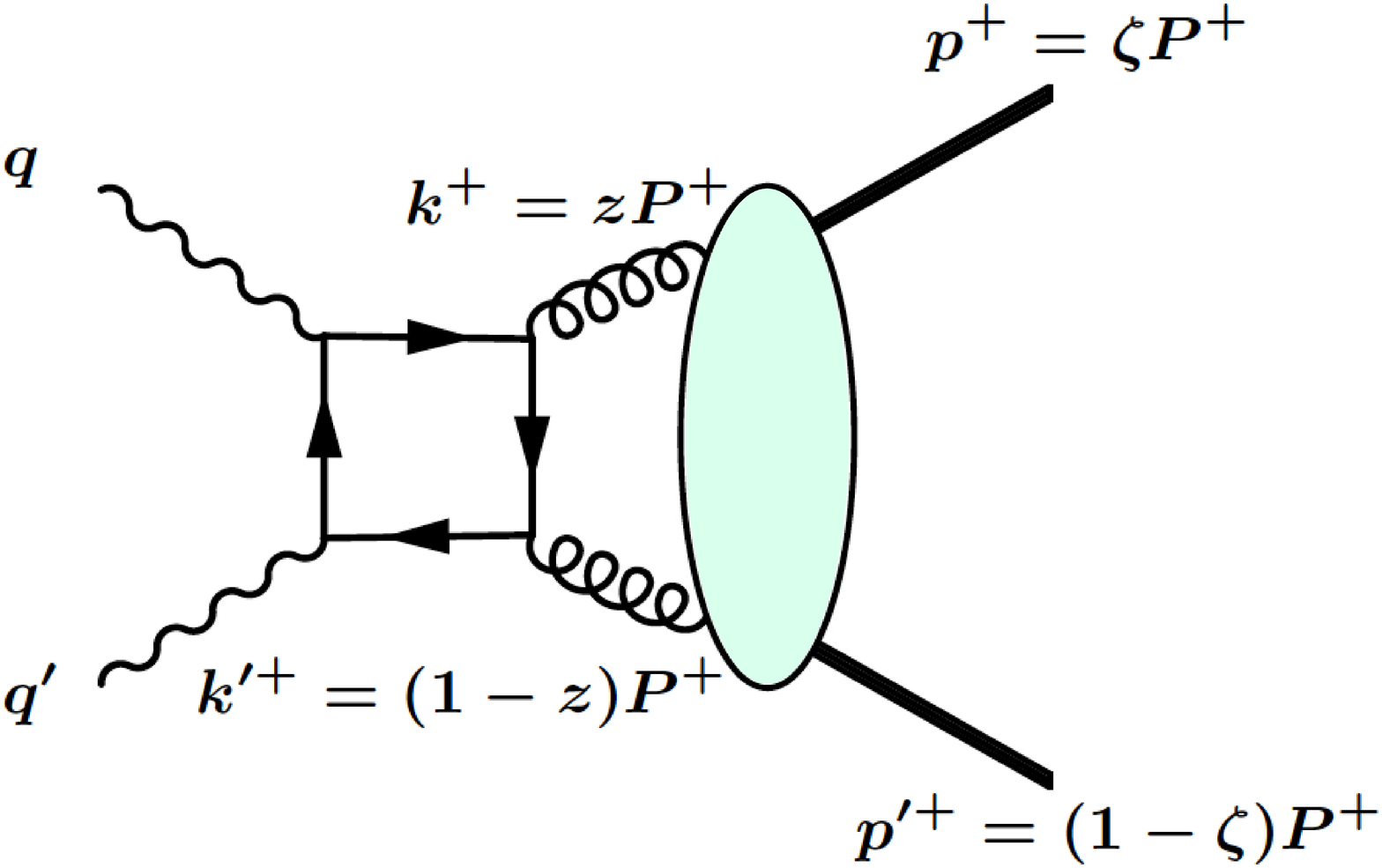}
\end{center}
\vspace{-0.6cm}
\caption{Contribution to the two-photon cross section
from the gluon GDA.}
\label{fig:gda-g}
\vspace{-0.0cm}
\end{figure}

The GDAs are defined for the hadron-antihadron system,
so that they satisfy the charge-conjugation invariance \cite{gpds}:
\begin{align}
\Phi_q^{h\bar h} (1-z,\zeta,W^2) & = - C \, \Phi_q^{h\bar h} (z,\zeta,W^2) 
\nonumber \\
& = - \Phi_q^{h\bar h} (z,1-\zeta,W^2) , 
\label{eqn:charge-conjugation}
\end{align}
where $C$ is the charge-conjugation operator.
We may note that the gluon GDA should satisfy the condition
\begin{align}
\Phi_g^{h\bar h} (z,\zeta,W^2) & = \Phi_g^{h\bar h} (1-z,\zeta,W^2) 
\nonumber \\
& =\Phi_g^{h\bar h} (z,1-\zeta,W^2) ,
\label{eqn:charge-conjugation-g}
\end{align}
due to the translational invariance in 
defining the gluon GDA and $C$ invariance. 
As shown in Fig.\,\ref{fig:gda-g}, the gluon GDA contributes
to the two-photon cross section as a next-to-leading order term 
\cite{diehl-2000,nlo-2gamma},
so that it is neglected in our current leading-order analysis.

\vspace{0.50cm}
\noindent
{\bf Generalized distribution amplitudes for pions}
\vspace{0.20cm}

The pion GDAs are investigated in this work, and there are 
two notation types for them. One is the representation based
on the $C$-parity eigenstates, and the other is 
by the isospin. In order to avoid confusion,
we explain them here in details.

First, we consider possible two-pion states.
Denoting $I$ for the isospin, we have the $I=1$ $\pi \pi$ state 
which is antisymmetric under the exchange of the pions.
On the other hand, the $I=0$ and $I=2$  $\pi \pi$ states are symmetric:
\begin{align}
\Ket{0 \, 0} & =\frac{1}{\sqrt{3}} \left( \pi^+ \pi^- + \pi^- \pi^+ - \pi^0 \pi^0 \right),  
\nonumber \\
\Ket{2 \, 0} & =\frac{1}{\sqrt{6}} \left( \pi^+ \pi^- + \pi^- \pi^+ +2 \pi^0 \pi^0 \right).
\label{eqn:i20}
\end{align}
Since the $C$ parity of $\gamma^* \gamma$ is even, 
the $\pi \pi$ state needs to satisfy $C=(-1)^{L+S}=(-1)^L=+1$
with $S=0$. Therefore, $L$ should be even. The Pauli principle indicates
\begin{align}
(-1)^L (-1)^I (-1)^S=1,
\label{eqn:pau}
\end{align}
so that the isospin states should be $I=0$ or 2. 
The GDAs are defined in Eq.\,(\ref{eqn:gda-def}) by the 
the matrix element of the vector-type nonlocal operator.
Since the isospin of $\bar{q}q$ is 0  or 1, the only possible 
choice for the $\pi \pi$ isospin is $I=0$. In this way, only
the $\pi\pi$ sates allowed in the $\gamma^*\gamma$ process should
have $I=0$ with $L=\text{even numbers (0,\,2,\,$\cdots$)}$.

For the actual pion GDAs which are investigated in this work, 
we may express them as $C$-parity eigenstates
\cite{diehl-2000}
\begin{align}
\Phi_q^{\pi\pi (\pm)} (z,\zeta,W^2) 
= \frac{1}{2} \, \bigg [ \, &  \Phi_q^{\pi^+ \pi^-} (z,\zeta,W^2) 
\nonumber \\[-0.3cm]
&   \pm \Phi_q^{\pi^+ \pi^-} (z,1-\zeta,W^2) \, \bigg ] ,
\label{eqn:pion-gpdas-c}
\end{align}
where $(\pm)$ indicates the $C$ parity.
Therefore, the $\pi^+ \pi^-$ GDAs are given by
$\Phi_q^{\pi^+ \pi^-} = \Phi_q^{\pi\pi (+)} + \Phi_q^{\pi\pi (-)}$,
and the $C$-even part satisfies 
$ \Phi_q^{\pi\pi (+)} (z,\zeta,W^2) = - \Phi_q^{\pi\pi (+)} (1-z,\zeta,W^2) $.
The $\pi^0 \pi^0$ GDAs contain only the $C$-even function:
\begin{align}
\Phi_q^{\pi^0 \pi^0} (z,\zeta,W^2) = \Phi_q^{\pi\pi (+)} (z,\zeta,W^2) .
\label{eqn:pi0pi0-gpdas-c}
\end{align}
Then, the isospin invariance leads to the relations between
the $u$- and $d$-quark GDAs as
$\Phi_u^{\pi\pi (+)} = \Phi_d^{\pi\pi (+)}$
and $\Phi_u^{\pi\pi (-)} =-\Phi_d^{\pi\pi (-)}$.

On the other hand, the isospin decomposition of the pion GDAs 
is discussed in Ref.\,\cite{Polyakov-1999} first 
by defining them as the twist-2 chiral-even amplitudes by
\begin{align} 
\! \! \! \! \! \! \! \! \! \!
& \! \! \! \! \! \!
\Phi^{\pi^a \pi^b} (z,\zeta,W^2) 
= \int \frac{d y^-}{2\pi}\, e^{i (2z-1) P^+ y^- /2}
   \nonumber \\
& \! \! \!
  \times \langle \, \pi^a(p) \, \bar \pi^b(p') \, | \, 
 \overline{\psi}(-y/2) \gamma^+ \widetilde T \psi(y/2) 
  \, | \, 0 \rangle \Big |_{y^+=\vec y_\perp =0} \, ,
\label{eqn:pion-gdas-def1}
\end{align}
where $\psi$ is the quark field with $u$ and $d$ quark components
$\psi = \left(   
 \begin{matrix}
 u \\
 d
 \end{matrix}
\right)$,
$\widetilde T$ is the flavor matrix: $\widetilde T=\hat I/2$ ($\widetilde T=\tau^3 /2$) 
for the isosinglet (isovector) GDA.
The notation $\hat I$ is the identity matrix.
They are expressed by the isoscalar and isovector GDAs as
\begin{align}
\Phi^{\pi^a \pi^b} 
& = \delta^{ab} \, \text{Tr} \big ( \widetilde T \big ) \, \Phi^{\pi\pi (I=0)} 
\nonumber \\
& \ \ 
+ \frac{1}{2} \, \text{Tr} 
   \left( \, [\tau^a,\tau^b] \, \widetilde T \, \right) \Phi^{\pi\pi (I=1)}  .
\label{eqn:pion-gdas-isospin}
\end{align}
They satisfy the symmetry relations due to the charge conjugation:
\begin{align}
\Phi^{\pi\pi (I=0)} (z,\zeta,W^2) & = - \Phi^{\pi\pi (I=0)} (1-z,\zeta,W^2) 
\nonumber \\
                           & =   \Phi^{\pi\pi (I=0)} (z,1-\zeta,W^2) ,
\nonumber \\
\Phi^{\pi\pi (I=1)} (z,\zeta,W^2) & =  \Phi^{\pi\pi (I=1)} (1-z,\zeta,W^2) 
\nonumber \\
                           & = -  \Phi^{\pi\pi (I=1)} (z,1-\zeta,W^2) .
\label{eqn:I=0-1-relations}
\end{align}
If the isospin-symmetry relations are satisfied for the pion GDAs, 
the isosinglet and isovector GDAs are related 
to the $C$ even and odd GDAs as
\begin{align}
\Phi^{\pi\pi (I=0)} & = \Phi_{u}^{\pi\pi (+)} = \Phi_{d}^{\pi\pi (+)} , 
\nonumber \\
\Phi^{\pi\pi (I=1)} & = \Phi_{u}^{\pi\pi (-)} = - \Phi_{d}^{\pi\pi (-)} .
\label{eqn:pion-gdas-relatiohns}
\end{align}

In this work of the $\pi^0 \pi^0$ production process,
only the following isoscalar or $C$-even GDAs are involved
in the cross section $\gamma^* \gamma \to \pi^0\pi^0$:
\begin{align}
\Phi_q^{\pi^0 \pi^0} (z,\zeta,W^2) & = \Phi^{\pi\pi (I=0)} (z,\zeta,W^2)  
\nonumber \\
& = \Phi_q^{\pi\pi (+)} (z,\zeta,W^2)  ,
\label{eqn:2pion0-gdas}
\end{align}
where $q$ indicates $u$ or $d$.
This function is parametrized and used for the analysis of $\pi^0\pi^0$
production data later by using Eq.\,(\ref{eqn:cross2}).

\subsection{Relation between GPDs and GDAs}
\label{gpd-gda}

As obvious from the diagrams of the DVCS and two-photon process
in Figs.\,\ref{fig:gpd-fig} and \ref{fig:gda-fig}, respectively,
the GPDs and GDAs are related with each other by the $s$-$t$ crossing
as long as the factorization conditions are satisfied.
Namely, the scale $Q^2$ should be large enough for the factorization:
$Q^2 \gg W^2, \ \Lambda^2_{\text{QCD}}$ in $\gamma^* \gamma \rightarrow h\bar h$;
$Q^2 \gg |t|, \ \Lambda^2_{\text{QCD}}$ in $\gamma^* h \rightarrow \gamma h$.
By the $s$-$t$ crossing, the final hadron $\bar h$
with the momentum $p'$ becomes the initial hadron $h$ with $p$, which
indicates the momentum changes from $p$ and $p'$ in the GDAs
to $p'$ and $-p$ in the GPDs. It means that both variables are 
related by the relations \cite{gpds-gdas,diehl-2000}.
\begin{align}
z' \leftrightarrow \frac{1-x/\xi}{2}, \ \ \ \ 
\zeta \leftrightarrow \frac{1-1/\xi}{2}, \ \ \ \ 
W^2 \leftrightarrow t ,
\label{eqn:variables-relation}
\end{align}
and then the GDAs are GPDs are related to each other by
\begin{align}
& \! \!  
\Phi_q^{h\bar h} (z',\zeta,W^2) 
\nonumber \\
& \! \! 
\longleftrightarrow
H_q^h \left ( x=\frac{1-2z'}{1-2\zeta},
            \, \xi=\frac{1}{1-2\zeta}, \, t=W^2 \right ) .
\label{eqn:gda-gpd-relation}
\end{align}

The physical regions of the kinematical variables are
\begin{align}
& 0 \le z \le 1   , \ \ \  |1-2\zeta| \le 1 , \ \ \ W^2 \ge 0,
\nonumber \\
& |x| \le 1   , \ \ \ \ \ \ \,  |\xi| \le 1 , \ \ \ \ \ \ \ \ \ \ t \le 0.
\label{eqn:physical-regions}
\end{align}
However, the relation of Eq.\,(\ref{eqn:gda-gpd-relation})
indicates that the physical GDAs do not necessary correspond
to the physical regions in Eq.\,(\ref{eqn:physical-regions})
of the GPDs: 
\begin{align}
0 \le |x| < \infty, \ \ 
0 \le |\xi| < \infty, \ \ 
|x| \le |\xi| , \ \ 
t \ge 0 .
\label{eqn:gda-gpd-kinematics}
\end{align}
Namely, the GDAs could lead to the unphysical kinematical regions, 
$|x| > 1$, $|\xi| > 1$, and $t >0$,
of the GPDs. Equation\,(\ref{eqn:gda-gpd-relation}) also indicates 
the relation $|\xi| \ge |x|$, which is called as 
the Efremov-Radyushkin-Brodsky-Lepage (ERBL) region.
The ERBL region of the GPDs can be investigated, for example,
by the hadronic reaction $N + N \to N + \pi + B$ \cite{kss-gpd}.
However, GDA studies will provide another information on 
the ERBL GPDs although it is in the unphysical region of $t>0$.

\subsection{Radon transforms for GPDs and GDAs 
            by using double distributions}
\label{radon}

We explained definitions and basic properties of the GPDs and GDAs.
They are related with each other by the $s$-$t$ crossing.
The studies of the GDAs should be valuable for the GPD studies
and vice versa. In fact, both GPDs and GDAs are expressed by
the common double distributions (DDs) by different Radon transforms.
The Radon transform is defined in $n$ dimensions 
for an arbitrary function $f (x)$ by
\cite{radon-book} 
\begin{align}
\hat f (p, \xi) 
   = \int d^n x \, f(x) \, \delta (p - \xi \cdot x) ,
\label{eqn:readon-define}
\end{align}
where $x$ is the $n$-dimensional space coordinate
[$x=(x_1,\,x_2,\,\cdots,\,x_n)$]
and $\xi$ is the unit vector in $n$ dimensions
[$\xi=(\xi_1,\,\xi_2,\,\cdots,\,\xi_n)$].
Because of the $\delta$ function,
the integral is over the $n-1$-dimensional plane constrained by 
$p= \xi \cdot x$.

\begin{figure}[t!]
\vspace{+0.00cm}
\begin{center}
\hspace{+0.20cm}
   \includegraphics[width=7.0cm]{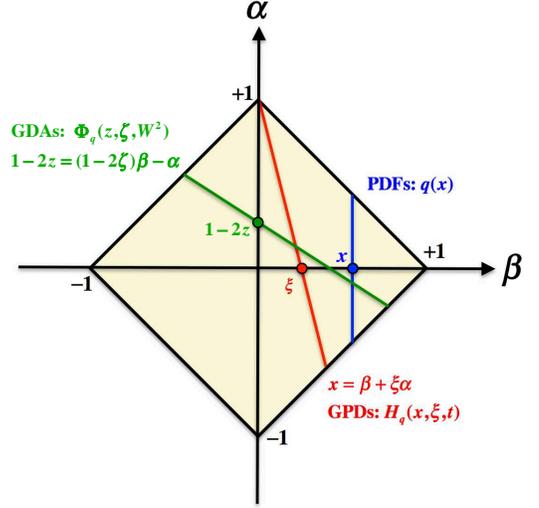}
\end{center}
\vspace{-0.6cm}
\caption{Kinematical support region of the double distributions
and integral paths for obtaining the GPDs, GDAs, and PDFs.}
\label{fig:dds-support}
\vspace{-0.3cm}
\end{figure}

Using this Radon transform, we can express the GPDs and GDAs 
in terms of double distributions (DDs),
$F_q (\beta,\alpha,t)$ and $G_q (\beta,\alpha,t)$,
defined by the matrix element
\cite{gpds-gdas,teryaev-2001}
\begin{align}
& \langle \, h (p ') \, | \, \bar q (-y/2) \, \slashed{y} \, q(y/2) \,
                  | \, h (p) \, \rangle _{y^2=0}
\nonumber \\                  
& = 2 P \cdot y
   \int d\beta \, d\alpha \, e^{-i\beta P \cdot y + i\alpha \Delta \cdot y/2}
        \, F_q (\beta,\alpha,t)
\nonumber \\                  
& \ \ 
- \Delta \cdot y
    \int d\beta \, d\alpha \, e^{-i\beta P \cdot y + i\alpha \Delta \cdot y/2}
        \, G_q (\beta,\alpha,t) ,
\label{eqn:double-def-pion}
\end{align}
for the scalar hadron $h$ like the pion.
The kinematical support region is given by
$|\beta|+|\alpha|\le 1$ for the DDs.
Using the Radon transform, we can express the GPDs
in terms of these DDs as
\begin{align}
H_q (x, \xi, t) = & \int d\beta \, d\alpha \, \delta (x-\beta- \xi \alpha) 
\nonumber \\
& \times \left [ \, F_q (\beta,\alpha,t) + \xi G_q (\beta,\alpha,t) \, \right ] .
\label{eqn:dd-gpds-pion}
\end{align}
Namely, the GPDs are obtained by integrating the DDs over the slight line
$x=\beta+ \xi \alpha$ as shown in Fig.\,\ref{fig:dds-support}.

The parton distribution functions (PDFs) are obtained as a special case
of this integral over the vertical line in Fig.\,\ref{fig:dds-support}
with the constraint of the forward limit ($t=0$),
and they are expressed as
\begin{align}
q (x) =  \int_{-1+x}^{1-x} d\alpha \,  F_q (\beta,\alpha,t=0) .
\label{eqn:dd-pdfs}
\end{align}
There are similar relations of the gluon DD to the gluon GPDs and PDF
\cite{gpds-gdas}.

As just an example, we introduce a simple parametrization 
for the DDs $F_q (\beta,\alpha)$, which 
are expressed by the corresponding PDF $q(\beta)$ 
multiplied by a profile function $h_q (\beta,\alpha)$ as 
\cite{radyushkin-dds}
\begin{align}
F_q (\beta,\alpha) = h_q (\beta,\alpha) \, q(\beta) .
\label{eqn:dds-paramet1}
\end{align}
The profile function may be expressed as
\begin{align}
h_q (\beta,\alpha) = \frac{\Gamma (2b+2)}{2^{2b+1} [\Gamma (b+1)]^2}
   \,  \frac{[(1-|\beta|)^2-\alpha^2 ]^b}{(1-|\beta|)^{2b+1}} ,
\label{eqn:profile-paramet1}
\end{align}
if the GPDs become the $\xi$-independent ones
$H_q (x,\xi) =\theta (x) \, q(x) - \theta (-x) \, \bar q(-x)$
in the limit $b \to \infty$.

The matrix element associated with the GDAs is also expressed
in the same way by the DDs as
\cite{gpds-gdas,teryaev-2001}
\begin{align}
& \langle \, h (p) \, \bar h (p ') \, 
  | \, \bar q (-y/2) \, \slashed{y} \, q(y/2) \,
                  | \, 0 \, \rangle _{y^2=0}
\nonumber \\                  
& 
= (p-p') \cdot y 
    \! \int  \! d\beta \, d\alpha \, e^{-i\beta (p-p') \cdot y/2 
    + i\alpha (p+p') \cdot y/2}
\nonumber \\[-0.20cm]
& \ \ \ \ \ \ \ \ \ \ \ \ \ \ \ \ \ \ \ \ \ \ \ \ \ \ \ \times
         F_q (\beta,\alpha,W^2)
\nonumber \\                  
& - (p+p') \cdot y 
  \! \int d\alpha \, e^{i\alpha (p+p') \cdot y/2}
        \, D_q (\alpha,W^2) .
\label{eqn:double-def-gdas}
\end{align}
Then, the GDAs can be expressed by the DDs as
\begin{align}
\Phi_q^{h \bar h} (z,\zeta,W^2)
& = -2 (1-2\zeta)  \int d\beta \, d\alpha 
\nonumber \\
& \! \! \! \! \! \! \! \! \! \! \! \! \! \! \! \!
  \! \! \! \! \! \! \! \! \!
\times \delta \big( 1-2z-(1-2\zeta)\beta+\alpha \big) \,
F_q (1-2z,\alpha,W^2)
\nonumber \\
& \ \ 
     -2 D_q (x/\xi,W^2) ,
\label{eqn:dd-gdas}
\end{align}
which indicates that the GDAs are obtained by 
the Radon transform along the different line
$1-2z-(1-2\zeta)\beta+\alpha =0$
as shown in Fig.\,\ref{fig:dds-support}.

We found that both GPDs and GDAs can be expressed by the DDs.
Therefore, experimental measurements of the GDAs should be
valuable also for the GPD studies through the determination of
the DDs and vice versa. In particular, the GDAs correspond to
specific kinematical regions of the GPDs as explained 
in Sec.\,\ref{gpd-gda}. These investigations from the direction 
of the GDAs could be supplementary to the direct GPD studies.
Furthermore, it is the advantage of the GDAs that exotic hadron
GDAs can be measured in future, whereas their GPDs cannot be studied
experimentally because there is no stable exotic-hadron target.
Considering these merits, we believe that our GDA project 
should be important for future developments on hadron 
tomography not only for ordinary hadrons such as the nucleons
and pions but also for exotic-hadron candidates.

\subsection{Timelike form factors of energy-momentum tensor 
and gravitational-interaction radii}
\label{gravitation-radius}

The GPDs and GDAs are measured in the DVCS and two-photon processes
which are, of course, electromagnetic interaction processes. However, 
their studies could also probe an aspect of gravitational interactions 
with quarks and gluons. In order to understand this fact, we explain it 
by taking the quark GPD and GDA definitions.
As given in Eqs.\,(\ref{eqn:gpd-n}) and (\ref{eqn:gda-def}),
the GPDs and GDAs are defined by the same non-local vector operator. 
For the GDAs, its moments multiplied by the momentum factor $2 (P^+ /2)^n$ 
are expressed by the derivatives as
\cite{gpds-gdas}
\begin{align}
&
\! \! \! \! \! \! \! \! 
2 \, (P^+ /2)^{n} \! \! \int_0^1 dz \, (2z-1)^{n-1} 
  \! \!  \int\frac{d y^-}{2\pi}e^{i (2z-1) P^+ y^- /2}
\nonumber \\
 & 
\! \!  \! \! \! \! \! \! \! \! 
\times
\overline{q}(-y/2) \gamma^+ q(y/2) \Big |_{y^+ = \vec y_\perp =0}
\! \! \!
 = \overline q (0) \gamma^+ \!
 \left ( i \overleftrightarrow \partial^+  \right )^{n-1} 
\! \! \!
 q(0) .
\label{eqn:tensor-int}
\end{align} 
where the derivative $\overleftrightarrow \partial$ is defined by
$f_1 \overleftrightarrow \partial f_2 
 = [ f_1 (\partial f_2)  - (\partial f_1) f_2 ]/2$.
For $n=2$, this operator is the energy-momentum tensor
of a quark, and it is a source of gravity, whereas
it is the vector-type electromagnetic current for $n=1$.

\begin{figure}[b]
\begin{center}
  \includegraphics[width=8.6cm]{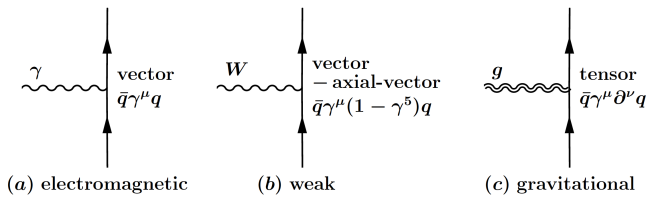}
\end{center}
\vspace{-0.50cm}
\caption{Electromagnetic, weak, and gravitational interactions 
         with a quark. The gravitational interactions $(c)$ are probed
         by the GPDs and GDAs. }
\label{fig:gravitational} 
\end{figure}

As shown in Fig.\,\ref{fig:gravitational},
$(a)$ the electromagnetic interaction is described by 
the vector current $\bar q \gamma^\mu q$,
$(b)$ the weak interaction is characterized by
the vector minus axial-vector current $\gamma^\mu (1-\gamma_5)$,
and $(c)$ the gravitational one is by the tensor interaction
given by $\bar q \gamma^\mu \partial^\nu q$ for a quark.
In Eq.\,(\ref{eqn:tensor-int}), the GPDs and GDAs
contain this factor as the energy-momentum tensor of a quark.
The charge radius of the proton is measured by elastic 
electron scattering in the form of the electric form factor
through the photon exchange process $(a)$.
In the similar way, the gravitational radius should be
measured by the graviton exchange process $(c)$
in principle. However, it is impossible to do an actual
scattering experiment directly at accelerator facilities 
for the gravitational interaction due to the ultra-weak interaction 
nature. On the other hand, it is possible to access such physics
through the GPDs and GDAs. Therefore, the gravitational
radii of hadrons are measurable quantities,
although it may be somewhat surprising that
a different physics aspect can be investigated 
through the electromagnetic processes.

In this way, the following integral of the quark GDAs over the variable $z$ 
is related to a matrix element of the quark energy-momentum
tensors $T_{q}^{\mu\nu}$
\cite{Polyakov-1999,diehl-2000,gda-form-factor-W2}:
\begin{align}
& \int_0^1 dz (2z -1) \, 
\Phi_q^{\pi^0 \pi^0} (z,\,\zeta,\,W^2) 
\nonumber \\[-0.20cm]
 & \ \ \ \ \ \ \  
 = \frac{2}{(P^+)^2} \langle \, \pi^0 (p) \, \pi^0 (p') \, | \, T_q^{++} (0) \,
       | \, 0 \, \rangle ,
\label{eqn:integral-over-z}
\end{align}
and there is a similar equation on the gluon matrix element.
The quark energy-momentum tensor is given by
\begin{align}
T_q^{\,\mu\nu} (x) & = \overline q (x) \, \gamma^{\,(\,\mu} 
   i \overleftrightarrow D^{\nu)} \, q (x), 
\label{eqn:energy-momentum-t}
\end{align}
where $D^\mu$ is the covariant derivative 
$D^\mu = \partial^{\,\mu} -ig \lambda^a A^{a,\mu}/2$ defined by 
the QCD coupling constant $g$ 
and the SU(3) Gell-Mann matrix $\lambda^a$.
The notation $X^{(\mu\nu)}$ is given by the symmetric combination
$X^{(\mu\nu)} \equiv (X^{\mu\nu}+X^{\nu\mu})/2$.
Here, $T_q^{++}$ indicates the lightcone $++$ components as expressed
in Eq.\,(\ref{eqn:tensor-int}), so that
it is specifically given by $T_q^{++} = (T_q^{00}+T_q^{03}+T_q^{30}+T_q^{33})/2$.
These equations indicate that the GDAs probe the energy-momentum tensors 
of quarks and gluons, in the same way as the GPDs, in the timelike process.

In an isolated system, the energy-momentum tensor is conserved
$\partial_\mu T^{\,\mu\nu}=0$. However, if there is an external force
and gravity, it satisfies \cite{weinberg-book}
\begin{align}
\frac{1}{\sqrt{-g}} \, \partial_\mu \left( \sqrt{-g} \, T^{\,\mu\nu} \right) 
= G^\nu - \Gamma^\nu_{\mu\lambda} \, T^{\mu\lambda} ,
\label{eqn:energy-momentum-cons}
\end{align}
where $G^\nu$ is the energy-force density, $g$ is defined by
the metric tensor $g_{\mu\nu}$ as $g=\det(g_{\mu\nu})$,
and $\Gamma^\nu_{\mu\lambda}$ is the affine connection tensor.
The second term on the right-hand side is the gravitational-force density.
As the electromagnetic interaction and weak interaction are characterized 
theoretically by the vector and vector minus axial-vector 
operators, $\bar q \gamma^\mu q$ and $\bar q \gamma^\mu (1-\gamma_5) q$,
respectively for quarks, the gravitational interaction is characterized
by the energy-momentum tensor $T^{\mu\nu}$.
Namely, the energy-momentum tensors of quarks and gluons 
are sources of gravitational interactions.
Now, the 3D structure-function studies are getting popular
in hadron physics, and this tensor appears in the 3D structure functions, 
in particularly in the GPDs and GDAs as illustrated
in Fig.\,\ref{fig:gravitational}. For example, the GDAs probe 
the 3D structure of a hadron in the form of the timelike form factors.
The GDAs are related to the energy-momentum tensor in 
Eq.\,(\ref{eqn:integral-over-z}), so that they probe
the gravitational interaction, for example, 
as the form factors of energy momentum tensor.
These form factors are explicitly defined later
in Eqs.\,(\ref{eqn:emt-ffs-timelike-0}) and (\ref{eqn:emt-ffs-timelike}).

The GPDs and GDAs contain information on spacelike and timelike
form factors, respectively. For example, the simple parametrization
of the GPDs is given in Eq.\,(\ref{eqn:gpd-paramet1})
expressed as the longitudinal PDF multiplied by
the two-dimensional transverse form factor.
In general, the two-dimensional transverse charge density $\rho_T ^h (r_\perp)$
is given by the Fourier transform of the spacelike electric form factor
of a hadron $h$ as
\begin{align}
\rho_T ^h (r_\perp) & = \int \frac{d^2 q_\perp}{(2\pi)^2} 
          \, e^{-i\vec q_\perp \cdot \vec r_\perp} \, F_T ^h (q_\perp)
\nonumber \\
& = \int_0^\infty \frac{d q_\perp}{2\pi} q_\perp 
              J_0 (q_\perp r_\perp) F_T ^h (q_\perp) ,
\label{eqn:2D-rho}
\end{align}
where $J_0$ is the Bessel function. The two-dimensional transverse
root-mean-square (rms) radius is then given by
\begin{align}
\! 
\langle \,  r_\perp^{\,2} \, \rangle _h
\equiv \int d^2 r_\perp \, r_\perp^{\,2} \,\rho_T ^h (r_\perp)
= - \left. 4 \, \frac{d F_T ^h (q_\perp)}{dq_\perp^{\,2}} 
    \right|_{q_\perp \to 0} .
\label{eqn:2D-radius}
\end{align}
The transverse form factors of the energy-momentum tensor are calculated 
by using a simple parametrization for the GPDs of the proton,
and the results indicate that they could be different from charge form factor
\cite{teryaev-emt-form}.

In the three-dimensional case, the charge density and the form factor
are related with each other by
\begin{align}
\rho^h (r) & = \int \frac{d^3 q}{(2\pi)^3} 
          \, e^{-i\vec q \cdot \vec r} \, F^h (q)
\nonumber \\
& = \int_0^\infty \frac{d q}{2\pi^2} |\vec q \, |^{\,2} 
              j_0 (q r) F^h (q) ,
\label{eqn:3D-rho}
\end{align}
where $j_0$ is the spherical Bessel function. 
The rms radius is obtained by
\begin{align}
\langle \,  r^{\,2} \, \rangle _h
\equiv \int d^3 r \, |\vec r \,|^{\,2} \,\rho ^h (r)
= - \left. 6 \, \frac{d F ^h (q)}{d|\vec q\,|^{\,2}} 
    \right|_{|\vec q \,| \to 0} .
\label{eqn:3D-radius-1}
\end{align}

For timelike form factors probed by the $e^+ e^-$ or $\gamma^* \gamma$
reactions, we can relate them to the spacial distributions by
using the dispersion relation. Considering that singularities of the form
factor $F^h (t)$ is in the positive real $t$ axis from $4 m_h^2$,
we can express the $t$-channel form factor by the dispersion integral 
over the real positive $t$ ($\equiv s$) as
\cite{form-factor-dispersion,form-factor-h}:
\begin{align}
F^h (t) & = \int_{4 m_h^2}^\infty \frac{ds}{\pi} 
            \frac{{\rm Im}\, F^h (s)}{s-t-i\varepsilon} .
\label{eqn:dispersion-form-1}
\end{align}
Namely, the $t$-channel form factor $F^h (t)$ can be
calculated from the $s$-channel one $F^h (s)$.
Then, using Eqs.\,(\ref{eqn:2D-rho}) and (\ref{eqn:dispersion-form-1}) 
together with consideration on the constituent-counting rule 
in the asymptotic region \cite{kk2014}, we have \cite{form-factor-h}
\begin{align}
\rho_T ^h (r_\perp) & = \int_{4 m_h^2}^\infty \frac{ds}{2\pi^2} 
              K_0 (\sqrt{s} \, r_\perp) \, {\rm Im} \,
              F_T ^h (s) ,
\label{eqn:rho-timelike}
\end{align}
where $K_0$ is the modified Bessel function of the second kind.
However, the imaginary part of the form factor, namely its phase,
is not available from the measurement of 
$\gamma^* \to h \bar h$
because its cross section is proportional to $| F^h (t) |^2$
and a theoretically model-dependent input is needed for estimating
the spacial charge distribution from the measurement on
the timelike form factor.
In Ref.\,\cite{form-factor-h}, the Gounaris-Sakurai amplitude
\cite{gs-1968} is used for ${\rm Im} F^\pi (t)$ to obtain 
the transverse charge radius 
$\sqrt{\langle \,  r_\perp^{\,2} \, \rangle ^\pi_{\text{ch}} }=0.53$ fm,
which corresponds to the three-dimensional one 
$ \langle \,  r^{\,2} \, \rangle ^\pi_{\text{ch}}
 = 1.5 \langle \,  r_\perp^{\,2} \, \rangle ^\pi_{\text{ch}}=0.42$ fm$^2$.
Here, ``ch" indicates the electric charge.
This value is comparable to the $\pi e$ scattering measurement value
$\langle \,  r^{\,2} \, \rangle ^\pi_{\text{ch}}
 = 0.439 \pm 0.008$ fm$^2$
\cite{pi-e-scattering}
for the charged pion.
These results are for electric charge radii probed by electromagnetic
interactions, whereas we investigate gravitational
radii for hadrons, particularly the pion in this work, by 
using the GDAs in the two-photon process $\gamma^* \gamma \to h\bar h$. 

The three-dimensional density is calculated 
by using Eqs.\,(\ref{eqn:3D-rho}) and (\ref{eqn:dispersion-form-1}) as 
\begin{align}
\rho ^h (r)  = \int_{4 m_h^2}^\infty \frac{ds}{4\pi^2} 
              \frac{e^{-\sqrt{s} r}}{r} \, {\rm Im} \,
              F^h (s) .
\label{eqn:rho3-timelike}
\end{align}
The three-dimensional rms radius is also obtained
by using Eqs.\,(\ref{eqn:3D-radius-1}) and (\ref{eqn:dispersion-form-1}) as
\begin{align}
\langle \,  r^2 \, \rangle _h
&  =  \left. \frac{6}{F^h (t=0)} 
\, \frac{d F ^h (t)}{dt} \right|_{|t| \to 0} 
\nonumber \\
& = \frac{6}{F^h (t=0)} 
\int_{4 m_\pi^2}^\infty \frac{ds}{\pi} \frac{{\rm Im} \, F^h(s)}{s^2} ,
\nonumber \\
F^h (t=0) & = \int_{4 m_\pi^2}^\infty \frac{ds}{\pi} 
\frac{{\rm Im} \, F^h(s)}{s} ,
\label{eqn:3D-radius-2}
\end{align}
where the normalization of the spacelike form factor is taken into 
account explicitly by the replacement $F^h (t) \to F^h (t)/ F^h (t=0)$.
The spacelike gravitational form factors $\Theta_1 (t)$ and $\Theta_2 (t)$ 
are defined by the energy-momentum tensor $T^{\mu\nu}$
\cite{gda-form-factor-W2,emt-pion,emt-pion1}.
In the GPD and GDA studies \cite{gda-form-factor-W2}, other notations $A(t)$ 
and $B(t)$ are often used. Here, $A$ and $B$ are used for expressing other 
quantities, so that we use the notations $\Theta_1 (t)$ and $\Theta_2 (t)$ 
for the gravitational form factors. In the spacelike process, they are
defined by
\begin{align}
& \langle \, \pi^a (p') \, | \, T_q^{\mu\nu} (0) \, | \, \pi^b (p) \, \rangle 
\nonumber \\
& \ \ 
= \frac{\delta^{ab}}{2} 
  \left [ \, \left ( t \, g^{\mu\nu} -q^\mu q^\nu \right ) \, \Theta_{1, q} (t)
                + P^\mu P^\nu \,  \Theta_{2, q} (t) \,
  \right ] ,
\label{eqn:emt-ffs-spacelike}
\end{align}
for a quark $q$. Here, the momenta are defined by $P=p+p'$ and $q=p'-p$.
We defined the form factors and the energy-momentum tensor
for one quark type (namely, flavor-$q$ quark and antiquark) 
in order to avoid confusions. 
In Ref.\,\cite{gda-form-factor-W2}, the form factors are
expressed by $A$ and $B$, and they are related to 
$\Theta_1 (t)$ and $\Theta_2 (t)$ by
\begin{align}
A_q (t) = \Theta_{2,q} (t), \ \ \ B_q (t) = - \frac{1}{4} \Theta_{1,q} (t) .
\label{eqn:ffs-theta-ab}
\end{align}

As discussed above Eq.\,(\ref{eqn:variables-relation}),
the variables $(p,p')$ (GPD) in the $t$ channel is changed 
for $(-p',p)$ (GDA) in the $s$ channel by the $s$-$t$ crossing.
Then, using the momentum notations $P=p+p'$ and $\Delta=p'-p$, 
we obtain the definition of the timelike form factors 
from Eq.\,(\ref{eqn:emt-ffs-spacelike}) as
\begin{align}
& \langle \, \pi^a (p) \, \pi^b (p') \, | \, T_q^{\mu\nu} (0) \, | \, 0 \, \rangle 
\nonumber \\
& 
= \frac{\delta^{ab}}{2} 
  \left [ \, \left ( s \, g^{\mu\nu} -P^\mu P^\nu \right ) \, \Theta_{1, q} (s)
                + \Delta^\mu \Delta^\nu \,  \Theta_{2, q} (s) \,
  \right ] .
\label{eqn:emt-ffs-timelike-0}
\end{align}
From Eq.\,(\ref{eqn:integral-over-z}) and this definition,
we can evaluate the gravitational form factors for the pion.

\section{Theoretical formalism}
\label{formalism}

We explain the cross section for the two-photon process 
$\gamma^* \gamma \rightarrow \pi^0 \pi^0 $ to express it 
in terms of the GDAs in Sec.\,\ref{cross}.
First, the situation of the pion distribution amplitude (DA), instead
of the GDAs, is explained in Sec.\,\ref{pion-da}, and $Q^2$ evolution 
of the DA and the GDAs are discussed in Sec.\,\ref{Q2-evolution}.
The $\zeta$ dependence of the GDAs is introduced in 
Sec.\,\ref{zeta-dependence}.
Then, the parametrization of the GDAs is introduced 
in Sec.\,\ref{gda-expression} to determine them from experimental data.
Contributions from $f_0$ and $f_2$ resonances are included
in the analysis, and coupling constants for the resonances
are explained in Sec.\,\ref{resonance-constants}, and
the $Q^2$ scale dependence of such resonance contributions 
is discussed in Sec.\,\ref{resonance-scale}.
In Sec.\,\ref{gravitational-ffs}, the relations between 
the gravitational form factors and the GDAs are derived.

\subsection{Cross section for the two-photon process 
\boldmath{$\gamma^* \gamma \rightarrow \pi^0 \pi^0 $}}
\label{cross}

\begin{figure}[b!]
\begin{center}
  \includegraphics[width=5.0cm]{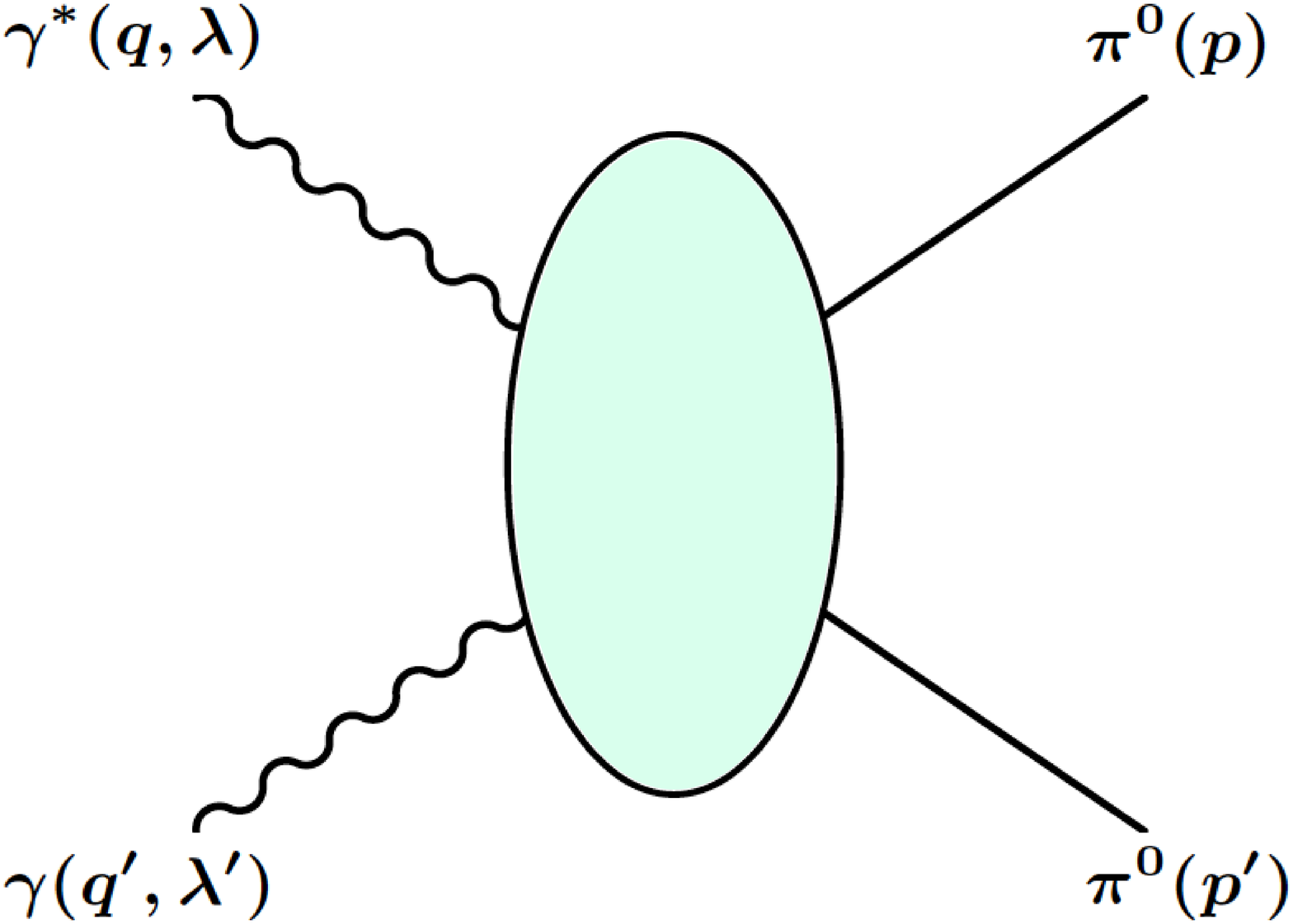}
\end{center}
\vspace{-0.50cm}
\caption{$\gamma^* \gamma \rightarrow \pi^0 \pi^0$ process.}
\label{fig:2gamma-fig} 
\end{figure}

The pion-pair production process $\gamma^* \gamma \rightarrow \pi^0 \pi^0$ 
is shown in Fig.\,\ref{fig:2gamma-fig}, and its cross section is
written by the matrix element ${\cal M}$ as \cite{Berger:1986ii} 
\begin{align}
d\sigma = \frac{1}{4 q\cdot q'}
\underset{\lambda, \lambda'}{\overline\sum}
& | {\cal M} (\gamma^* \gamma \to \pi^0 \pi^0 ) |^2 \,
\frac{d^3 p}{(2\pi)^3 \, 2E_p}  \frac{d^3 p'}{(2\pi)^3 \, 2E_{p'}} 
\nonumber \\
& \times
(2\pi)^4 \delta^4(q+q'-p-p') ,
\label{eqn:cross-section}
\end{align}
where one of the initial photons is taken on mass shell (${q'}^2=0$).
The matrix element ${\cal M} (\gamma^* \gamma \to \pi^0 \pi^0)$
is given by the hadron tensor ${\cal T}_{\mu\nu}$ as
\begin{align}
& \! \! \! \! \! 
i {\cal M} (\gamma^* \gamma \to \pi^0 \pi^0 ) 
  = \epsilon^\mu(\lambda) \, \epsilon^\nu(\lambda') \, 
  {\cal T}_{\mu\nu} ,
\nonumber \\
& \! \! \! \! \! 
{\cal T}_{\mu \nu } \!
= i \! \! \int \! d^4 y \, {e^{ - iq \cdot y}} \!
\left\langle \pi^0 (p) \pi^0 (p') \! \left| 
{TJ_\mu ^{em}(y)J_\nu ^{em}(0)} \right|0 \right\rangle ,
\label{eqn:matrix}
\end{align}
by the photon polarization vector $\epsilon^\mu(\lambda)$ 
and the electromagnetic current $J_\mu ^{em}(y)$.
In obtaining the total cross section, the cross section should be
divided by two due to two identical particles in the final state
to avoid the double counting. Alternatively, the cross section
is integrated over the half solid angle, in stead of the factor 1/2,
for calculating the total cross section. In any case, 
differential cross sections are discussed in this paper, so that
such a factor is not needed.

We define the helicity amplitudes $A_{i j}$ by
\begin{align}
A_{i j} & = \frac{1}{e^2} \,
\varepsilon _\mu ^{( i )}(q) \, \varepsilon _\nu ^{( j )}(q') \,
{{\cal T}^{\mu \nu }} , 
\nonumber \\
& \ \ \ \ \ \ \ 
i=-,\ 0, \ + \, , \ \ 
j=-,\ + \, .
\label{eqn:helicity-A}
\end{align}
If the kinematical condition 
$Q^2 \gg W^2, \Lambda^2$ is
satisfied, the two-photon process can be factorized into
the hard part ($\gamma^* \gamma \to q\bar q$) and
the soft part  ($q\bar q \to \pi^0 \pi^0$) as shown
in Fig.\,\ref{fig:gda-fig}.
In the Breit frame, $q$ is taken along the $z$ axis.
Introducing two timelike vectors $n=(1,0,0,1)/\sqrt{2}$ and 
$n^{\prime}=(1,0,0,-1)/\sqrt{2}$, we express the photon and 
quark momenta as
$q=(n-n') \sqrt{Q^2/2}$, 
$q'= n' (Q^2+W^2) / \sqrt{2 Q^2}$, 
$k=z n \sqrt{Q^2/2}$, and 
$k'=(1-z) n \sqrt{Q^2/2}$.
At large $Q^2$, the hadron tensor can be expressed by the factorized form as
\begin{align}
& {\cal T}_{\mu\nu} = i \int d^4y \, e^{-iq \cdot y}  
	\left \langle \pi^0 (p) \pi^0 (p') | T J_{\mu}^{em}(y)J_{\nu}^{em}(0)|0\right \rangle 
\nonumber \\
& \! \! \! \! \! \!
=\sum_q(-e^2e_q^2) \! \int \! \frac{d^4k}{(2\pi)^4} 
 \left[ \frac{\gamma^\mu (\slashed{k}-\slashed{q})\gamma^\nu}{(k-q)^2+i\epsilon}
 + \frac{\gamma^\nu (\slashed{q}-\slashed{k}')\gamma^\mu}{(q-k')^2+i\epsilon} 
 \right]_{ba} 
\nonumber \\
& \ \ \ \ \ \ \times
\int d^4y \, e^{-ik \cdot y}  \left \langle \pi^0 (p) \pi^0 (p') 
         | T\, \overline{q}_{b}(y)q_{a}(0)|0\right \rangle .
\label{eqn:tensor-1}
\end{align}
The first part describes the process $\gamma^* \gamma \to q\bar q$ 
of Fig.\,\ref{fig:gda-fig}, and the second one does the soft process
$q \bar q \to \pi^0 \pi^0$.
For the term $\overline{q}_{b}(y)q_{a}(0)$ in this equation, we use the Fierz identity 
\begin{align}
4 \, \overline{q}_{b} q_{a} & = \gamma_{ab}^{\lambda} \, 
\overline{q}\gamma_{\lambda} q
 + (\gamma^{\lambda}\gamma_5)_{ab} \, \overline{q}\gamma_{\lambda}\gamma_5 q 
\nonumber \\
& + I_{ab}^{\lambda} \, \overline{q} q + (\gamma_5)_{ab} \, 
\overline{q}\gamma_{5}q 
 + \sigma_{ab}^{\alpha\beta} \, \overline{q}\sigma_{\alpha \beta}q ,
\label{eqn:Fierz}
\end{align}
where the first two terms and the last one 
are the leading twist terms, 
while the third and fourth ones 
are twist-3 terms. 
Since the trace of an odd number of $\gamma_{\lambda}$ is zero, 
only the first two terms survive. 
However, the second term is the axial-vector current, 
which cannot exist for $\pi^0 \pi^0$ state 
due to the party invariance. After all, only the first term
contributes to the hadron tensor.

In the leading order of the running coupling constant $\alpha_s$,
the gluon GDA contribution is neglected and the hadron tensor
can be expressed by the quark GDAs 
by calculating the hard part of Eq.\,(\ref{eqn:tensor-1}) as
\cite{kk2014,Polyakov-1999,diehl-2000}
\begin{align}
\! \! \! \! 
{\cal T}^{\mu \nu } =  - g_{T}^{\, \mu \nu}{e^2} 
\sum\limits_q \frac{{e_q^2}}{2} \!
\int_0^1 \! \! {dz} \frac{{2z - 1}}{{z(1 - z)}}
\Phi_q^{\pi^0 \pi^0}(z,\zeta ,{W^2}) ,
\label{eqn:hadron-tensor-LT}
\end{align}
where $g_T^{\mu \nu }$ is defined by
\begin{alignat}{2}
g_{T}^{\, \mu \nu} & = -1  & \ \  & \text{for $\mu=\nu=1, \ 2$} ,
\nonumber \\
              & = 0  & \ \     & \text{for $\mu$, $\nu=$others} .
\label{eqn:hadron-tensor-LT}
\end{alignat}
The hadron tensor ${\cal T}^{\mu \nu }$ is generally written 
by the product of the two electromagnetic currents 
in Eq.\,(\ref{eqn:matrix}). In the leading twist, it is 
expressed by the matrix element of the vector current 
as given by the GDAs $\Phi_q^{\pi^0 \pi^0}$ in 
Eq.\,(\ref{eqn:pion-gdas-def1})
\cite{diehl-2000}. 
The situation is the same as the one in the hadron
tensor $W_{\mu \nu }$ in the charged-lepton deep inelastic 
scattering as expressed in the twist expansion 
\cite{jaffe-1985}.

Since only the non-vanishing terms are
$\varepsilon _\mu ^{( + )}(q)\varepsilon _\nu ^{( + )}(q')g_T^{\mu \nu } = 
 \varepsilon _\mu ^{( - )}(q)\varepsilon _\nu ^{( - )}(q')g_T^{\mu \nu } = - 1$,
the cross section is expressed by the helicity amplitude $A_{++}$ as 
\begin{align}
\frac{d\sigma}{d(\cos \theta)}
& = \frac{\pi \alpha^2}{4(Q^2+s)}
    \sqrt{1-\frac{4m_\pi^2}{s}} \, |A_{++}|^2  ,
\nonumber \\
A_{++}
& =\sum_q \frac{e_q^2}{2} \int^1_0 dz \frac{2z-1}{z(1-z)} 
   \Phi_q^{\pi^0 \pi^0} (z, \xi, W^2) ,
\label{eqn:cross2}
\end{align}
where the relation $A_{--} = A_{++}$ is used due to party conservation.
The gluon GDA contributes to the cross section through 
the amplitudes $A_{++}=A_{--}$ and $A_{+-}=A_{-+}$
in the next-to-leading order,
so that these terms are suppressed by the factor of $\alpha_s$.
There are also contributions from higher-twist amplitudes 
$A_{0+}$ and $A_{0-}$, which decrease as at least $1/Q$ because of
a helicity flip 
\cite{diehl-2000,nlo-2gamma}.

The $\gamma^* \gamma \to \pi^0 \pi^0$ cross section is expressed
by the GDAs in Eq.\,(\ref{eqn:cross2}). In order to determine the GDAs
from experimental data, we need to express the GDAs by a number of
parameters, which are then determined by a $\chi^2$ analysis
of the data on $d\sigma / d(\cos \theta)$.
There a number of studies on the pion distribution amplitudes; however,
it is the first attempt for the GDAs in comparison with actual 
experimental data.
Before discussing an appropriate functional form of the GDAs,
we explain the distribution amplitude (DAs), which are
related to the $z$-dependent part of the GDAs.
For example, the pion distribution amplitude $\Phi_\pi (z)$
is related to the GDAs by \cite{Polyakov-1999}
\begin{align}
\Phi_\pi (z) 
& = \Phi^{\pi\pi (-)} (z,\zeta=1, W^2=0) 
\nonumber \\
& = - \Phi^{\pi\pi (-)} (z,\zeta=0, W^2=0), 
\nonumber \\
& \! \! \! \! \! \! \! \! \! \! \! \! \! \! \! 
\Phi^{\pi\pi (+)} (z,\zeta=1, W^2=0) 
\nonumber \\
&  = \Phi^{\pi\pi (+)} (z,\zeta=0, W^2=0) =0 . 
\label{eqn:das-gdas}
\end{align}
In our analysis of $\gamma^* \gamma \to \pi^0 \pi^0$, we obtain
$\Phi^{\pi\pi (+)}$.

\subsection{Pion distribution amplitude}
\label{pion-da}

Before stepping into the details of the pion GDAs, we 
explain the pion distribution amplitudes (DAs).
The pion DAs are defined by the matrix element of a bilocal 
quark operator between the vacuum and the pion 
by taking the pion momentum along the positive $z$-axis as
\cite{diehl-2000,j-parc-gpd,mueller-1989}
\begin{align}
&
\langle \, \pi^a (p) \, | \, \overline \psi (y)_\alpha 
                  \, \psi (0)_\beta \,
                | \, 0 \, \rangle\Big |_{y^+ = \vec y_\perp =0}
\nonumber \\
&
= - \frac{i f_\pi}{4} \int_0^1 dz \, e^{i z p^+ y^-}
  ( \gamma_5 \, \pslash )_{\beta\alpha} \, \Phi_{\pi} (z,\mu) 
  + \cdots ,
\label{eqn:matrix-pi+}
\end{align}
where $a$ is the pion charge ($a$=$+$, $0$, $-$), 
$\overline \psi (y) \psi (0)$ indicates
$\bar u (y) d (0)$,
$[\bar u (y) u (0) -\bar d (y) d (0)] /\sqrt{2}$, or
$\bar d (y) u (0)$,
for $\pi^+$,  $\pi^0$, or  $\pi^-$, respectively,
and the ellipses indicate higher-twist terms.

The function $\Phi_\pi (z,\mu)$ is the leading-twist distribution 
expressed by the longitudinal momentum fraction $z$ of
a valence quark in the pion and the renormalization scale 
$\mu$ of the bilocal operator. 
The $\mu$ dependence is described by the 
ERBL evolution equations \cite{erbl}.
It is normalized as
\begin{align}
\int_0^1 dz \, \Phi_\pi (z, \mu)=1 ,
\label{norm-phi}
\end{align}
and $f_\pi$ is the pion decay
constant defined by
\begin{align}
\langle \, \pi^a (p) \, | \, \overline\psi (0) \gamma^\mu
\gamma_5 \, \psi (0) \, | \, 0 \, \rangle 
= - i f_\pi p^{\,\mu} .
\label{pion-decay}
\end{align}

In the asymptotic limit of $\mu \to \infty$, 
the pion distribution amplitude becomes
\begin{equation}
\Phi_\pi^{(\text{as})} (z) = 6 \, z \, (1-z) ,
\label{eqn:asymp}
\end{equation}
as it becomes obvious from the $Q^2$-evolution solution
of Eqs.\,(\ref{eqn:dga-solution}) and (\ref{eqn:GDAs-scaling-n=even}).
At finite $\mu$, it is generally expressed by using
the Gegenbauer polynomials as
\begin{equation}
\Phi_\pi (z,\mu) = 6 \, z \, (1-z) 
\sum_{n=0,2,4,\cdots}^\infty a_n (\mu) \, C_n^{3/2} (2z-1) ,
\label{eqn:Gegenbauer}
\end{equation}
where only the even terms contribute because 
the DA should satisfy the condition
$\Phi_\pi (1-z, \mu) = \Phi_\pi (z, \mu)$
under the exchange $z \leftrightarrow 1-z$.
It corresponds to the exchange of $q$ and $\bar q$ in the pion,
and the momentum distribution carried by a quark or antiquark
should be same under this exchange
because of positive C-parity of the axial current.
The Gegenbauer polynomials are 
$C_0^a (x)=1$, $C_1^a (x)=2ax$, $C_2^a (x)=-a+2a(1+a)x^2$, $\cdots$.
The current situation of the pion DA is explained in 
Ref.\,\cite{j-parc-gpd}.
Since the Gegenbauer polynomials are rapidly oscillating functions
at large $n$ and the coefficients $a_n(\mu)$ are small 
for large $\mu$, the $n\ge 4$ terms could be 
neglected at this stage.
As for the second coefficient $a_2$, there are theoretical
estimates by lattice QCD~\cite{Braun:2006dg} 
and QCD sum rules
~\cite{pionDA:CZ,Braun:1988qv,Bakulev,Ball:2006wn,Khodjamirian:1997tk,Braun:1999uj}
One of the well known functions was proposed by
Chernyak and Zhitnitsky (CZ) to take 
$a_2\,(\mu\simeq 0.5~{\rm GeV}) = 2/3$
as suggested by the QCD sum rule \cite{pionDA:CZ}:
\begin{align}
\! \! \! 
\Phi_\pi^{(\text{CZ})} (z, \mu) 
& = 6 \, z \, (1-z) \, \left[ 1 + \frac{2}{3}
     C_2^{3/2} (2z-1) \right] 
\nonumber \\
& \! \! \! \! \!
= 30 \, z \, (1-z) \, (2z-1)^2 
\ \ 
\text{at} \ \mu\simeq 0.5~{\rm GeV} ,
\label{eqn:CZ}
\end{align}
which is very different from the asymptotic form
because it has a minimum at $z=0.5$.
There are also recent theoretical suggestions 
on different $a_2$ values 
\cite{Braun:2006dg,Bakulev,Ball:2006wn,Khodjamirian:1997tk,Braun:1999uj,Kroll:2012sm}
and also $a_4$ and $a_6$ values \cite{pionDA:DSE}.
In principle, the different pion DAs can be tested by
experiments. The Belle measurements on the $\gamma \to \pi$ form factor
are close to the asymptotic DA form \cite{pion-da-exp}, whereas 
the BaBar data have a different tendency in the sense that
it is consistent with $a_2\,(\mu=2~{\rm GeV})=0.22$ \cite{Kroll:2012sm}.
Further measurements are needed to distinguish various theoretical DAs.

We comment on a slightly different convention from ours
in defining the distribution
amplitude because it may be sometimes confusing in using the decay constant
$f_\pi$ or $f_\pi /\sqrt{2}$.
In the Diehl's article of 2003 \cite{gpds-gdas}, the $\pi^0$ distribution
amplitude is defined for one quark flavor as
$ \langle \, \pi^0 (p) \, | \, \overline q (y)_\alpha 
   \, q (0)_\beta \, | \, 0 \, \rangle\Big |_{y^+ = \vec y_\perp =0}$
instead of the left-hand side of Eq.\,(\ref{eqn:matrix-pi+}).
Therefore, the decay constant definition of Eq.\,(\ref{pion-decay})
becomes 
$ \langle \, \pi^0 (p) \, | \, \overline u (0) \gamma^\mu
\gamma_5 \, u (0) \, | \, 0 \, \rangle 
= - \langle \, \pi^0 (p) \, | \, \overline d (0) \gamma^\mu
\gamma_5 \, d (0) \, | \, 0 \, \rangle 
= - i f_\pi p^{\,\mu}/\sqrt{2} $.
We should note that there is a factor of $\sqrt{2}$ in this expression.
However, this $\sqrt{2}$ is absorbed into the definition of 
distribution amplitude in his formalism
so that the decay constant $f_\pi$ stays the same as ours.

\subsection{Scale evolution of distribution amplitudes and 
generalized distribution amplitudes}
\label{Q2-evolution}

\begin{figure}[b!]
\begin{center}
  \includegraphics[width=4.1cm]{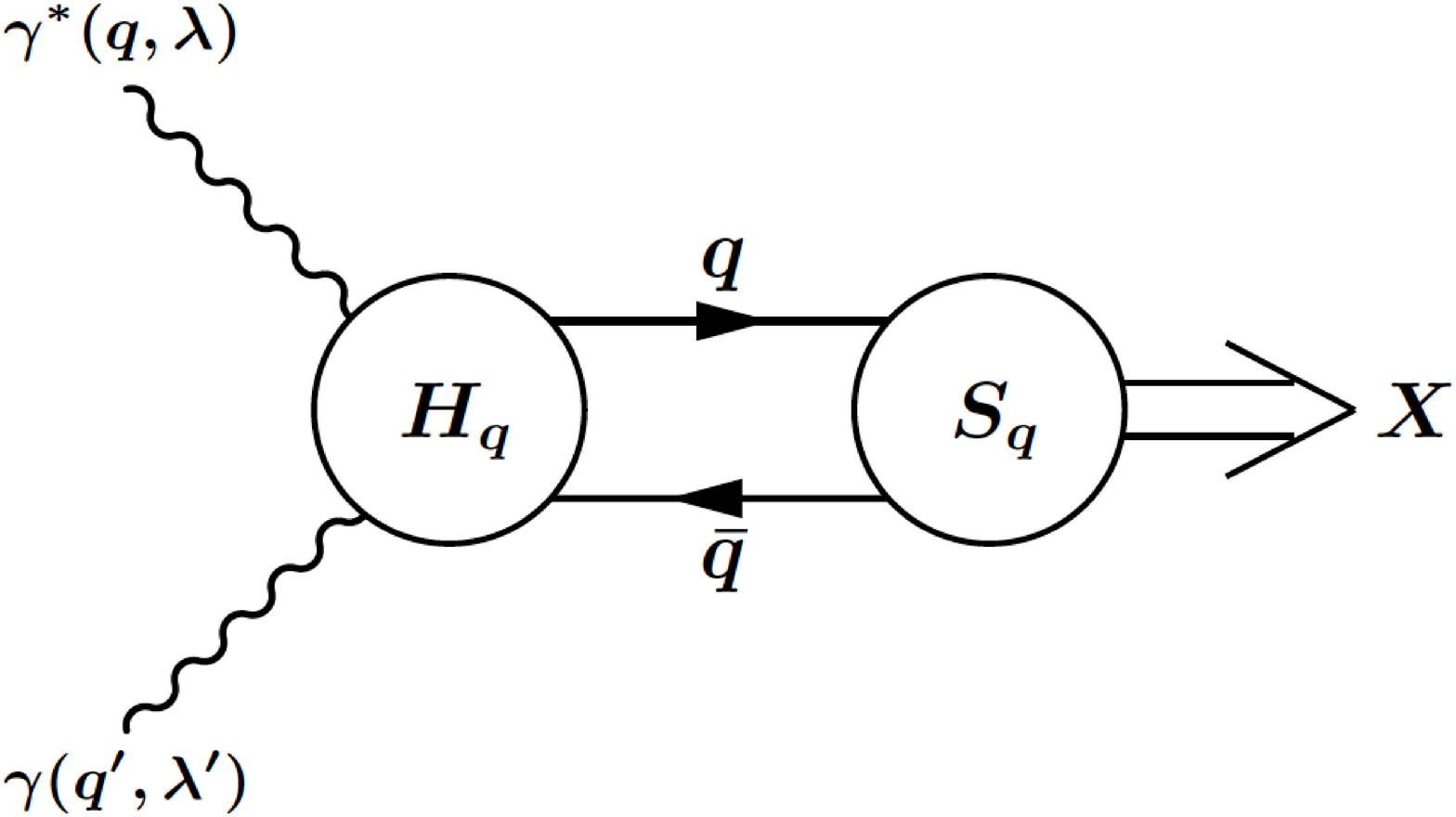}
  \vspace{1.0cm}
  \includegraphics[width=4.1cm]{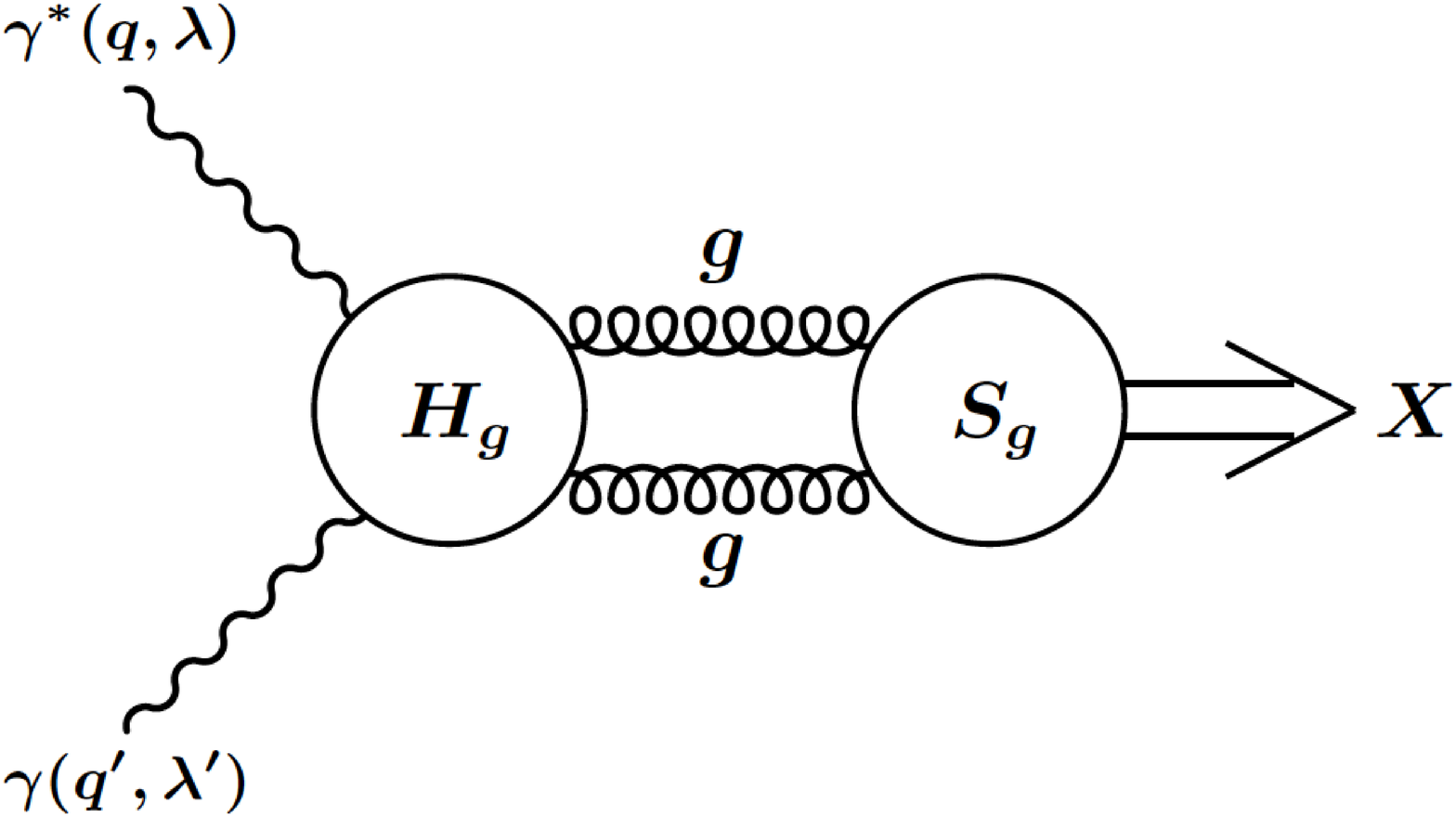}
\end{center}
\vspace{-1.50cm}
\caption{Factorization of $\gamma^* \gamma \rightarrow X$,
$X=\pi^0$ for DAs or $\pi^0 \pi^0$ for GDAs,
into the hard part $H_{q,g}$ and the soft one $S_{q,g}$.
For the isovector $\pi$, the right-hand-side process
with the two-gluon intermediate state does not exist
for the single $\pi^0$ production ($X=\pi^0$).}
\label{fig:2gamma-qg-x} 
\end{figure}

If the kinematical condition $Q^2 \gg W^2,\ \Lambda^2_{\text{QCD}}$ 
is satisfied, the process $\gamma^* \gamma \to \pi^0 \pi^0$ is 
factorized into the hard part $H_{q,g}$ and the soft one $S_{q,g}$
as shown in Fig.\,\ref{fig:2gamma-qg-x}.
Here, the final state $X$ is $\pi^0$ for the DAs or $\pi^0 \pi^0$ 
for the GDAs.
The hard part is calculated in perturbative QCD and the soft one
is expressed by the DAs or the GDAs. 
The $Q^2$ evolution equations of the DAs and GDAs are described 
by calculating the hard part in perturbative QCD. Since both reactions 
($\gamma^* \gamma \to \pi^0$ and $\gamma^* \gamma \to \pi^0\pi^0$)
have the same hard processes, the DAs and GDAs follow the same evolution 
equations, and their $z$ and scale-$\mu$ dependencies are represented 
by the functions $\Phi_q (z,\mu)$ and $\Phi_g (z,\mu)$ 
in the following discussions of this subsection
[$\Phi_q=\Phi_\pi$ for the DAs, 
 $\Phi_{q}=\Phi_{q}^{\pi\pi (+)}$ and
 $\Phi_{g}=\Phi_{g}^{\pi\pi}$ for the GDAs].

\begin{figure}[b!]
  \vspace{0.20cm}
\begin{center}
  \includegraphics[width=7.0cm]{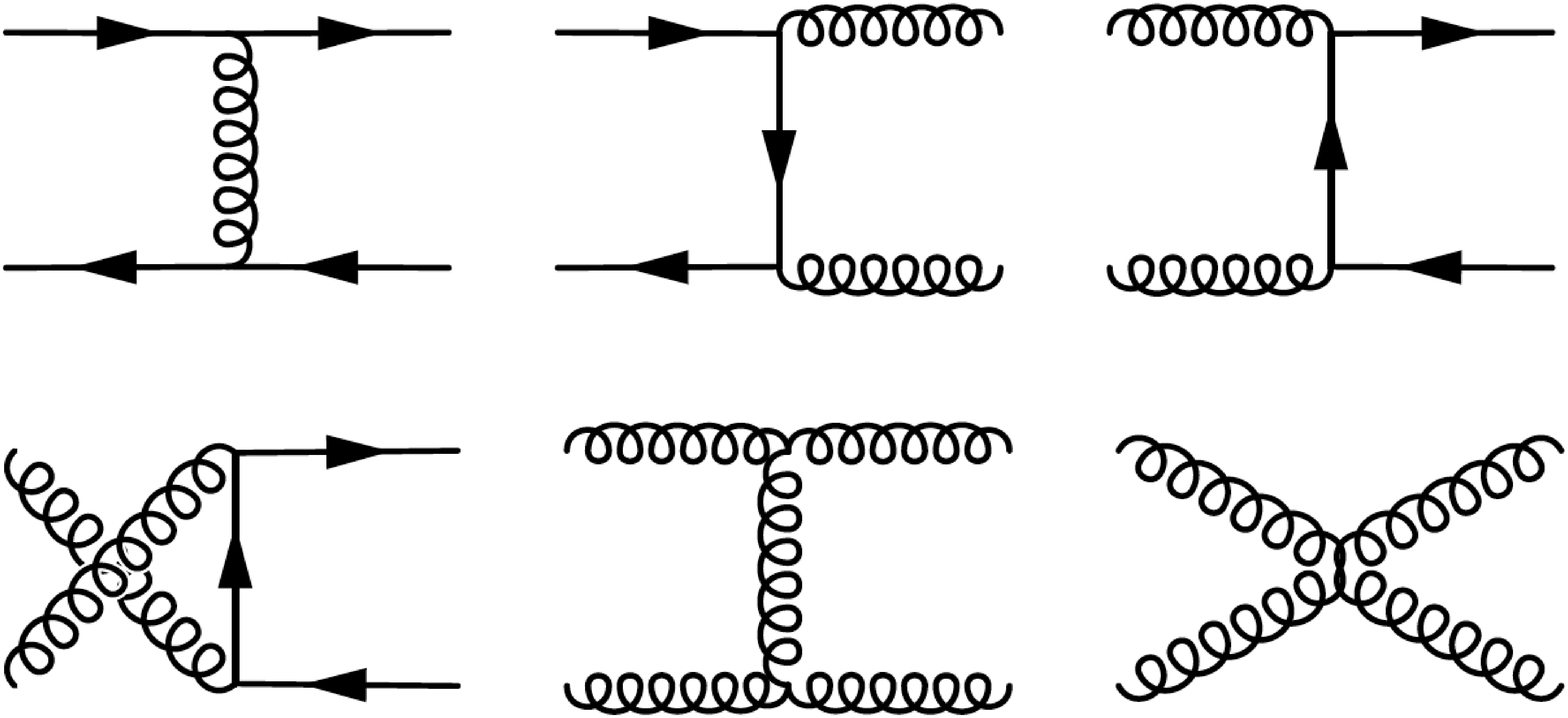}
\end{center}
\vspace{-0.40cm}
\caption{Leading contributions to the hard part $H_{q,g}$.}
\label{fig:2gamma-fact} 
\end{figure}

In order to describe the $Q^2$ evolution of the DAs and GDAs, 
we introduce the auxiliary quark and gluon functions $f_Q$ and $f_G$ 
defined by \cite{gpds-gdas,diehl-2000}
\begin{align}
z (1-z) f_Q (z,\mu) & = \sum_q \Phi_q (z,\mu), 
\nonumber \\
z^2 (1-z)^2 f_G (z, \mu) & = \Phi_g (z,\mu) .
\label{eqn:fqfg}
\end{align}
We introduce the variable $\tau$ defined by
\begin{align}
\tau = \frac{2}{\beta_0} 
        \ln \left[ \frac{\alpha_s (\mu_0^2)}{\alpha_s (\mu^2)} \right] ,
\label{eqn:tau-evolution}
\end{align}
for describing the evolution from $\mu_0^2$ to $\mu^2$ as usually
used in expressing the DGLAP evolution equations for the PDFs
\cite{dglap-solution}. Here, $\beta_0 = 11 -2 n_f/3$ and $\alpha_s$ 
is the running coupling constant. Then, the evolution equations are
expressed as
\begin{align}
\! \! \!
\frac{\partial}{\partial \tau} 
\left (   
\begin{matrix}
f_Q (z,\tau) \\
f_G (z,\tau)
\end{matrix}
\right ) \!
= \! \int_0^1 \! du \!
\left (   
\begin{matrix}
V_{QQ} (z,u) & \! V_{QG} (z,u) \\
V_{GQ} (z,u) & \! V_{GG} (z,u)
\end{matrix}
\right ) \!
\left (   
\begin{matrix}
f_Q (u,\tau) \\
f_G (u,\tau)
\end{matrix}
\right ) ,
\label{eqn:dga-evolution}
\end{align}
where the matrix $V$ is the kernel calculated in perturbative QCD.
The one-loop contributions to this kernel is shown in 
Fig.\,\ref{fig:2gamma-fact}.
The one-loop kernels have been obtained as \cite{gpds-gdas,diehl-2000}
\begin{align}
\! \! \! 
V_{QQ} & = C_F \left[ \, \theta (z-u) \, \frac{u}{z} 
                      \left( 1 + \frac{1}{(z-u)_+}  \right) 
                      +  \left\{ u,z \to \bar u,\bar z \right\}
                 \, \right] ,
\nonumber \\
\! \! \! 
V_{QG} & = 2 n_f T_F \left[ \, \theta (z-u) \, \frac{u}{z} \, (2z-u)
                      +  \left\{ u,z \to \bar u,\bar z \right\}
                 \, \right] ,
\nonumber \\
\! \! \! 
V_{GQ} & = \frac{C_F}{z\bar z} \left[ \, \theta (z-u) \, \frac{u}{z} \, 
                        (\bar z - 2 \bar u)
                      +  \left\{ u,z \leftrightarrow \bar u,\bar z \right\}
                 \, \right] ,
\nonumber \\
\! \! \! 
V_{GG} & = \frac{C_A}{z\bar z} \left[ \, \theta (z-u) \, 
                \bigg (  \frac{u \bar y}{(z-u)_+} -u \bar u
                    -\frac{u}{2z} \big\{ (2z-1)^2   \right.
\nonumber \\ 
       & \! \! \! \! \! \! \! \! \! \! \! \! \! 
         \left.            
                 + (2u-1)^2 \big\} \bigg)
                    \!  +  \left\{ u,z \leftrightarrow \bar u,\bar z \right\}
                 \, \right] 
                 -\frac{2}{3} n_f T_F \delta (u-z) ,
\label{eqn:evolution-kernel}
\end{align}
where $\bar u$ and $\bar z$ are defined by $\bar u = 1-u$ and $\bar z = 1- z$, 
and $C_F$, $T_F$, and $C_A$ are given by 
$C_F = (N_c^2-1)/(2N_c)$, $T_F=1/2$, and $C_A =N_c$ with
the number of colors $N_c=3$.
These equations are called ERBL evolution equations.

The integro-differential equations can be solved in the same way
with solving the DGLAP (Dokshitzer-Gribov-Lipatov-Altarelli-Parisi)
evolution equations \cite{dglap-solution} by using the anomalous
dimensions ($\gamma^{QQ}_n$, $\gamma^{QG}_n$, $\gamma^{GQ}_n$, $\gamma^{GG}_n$)
obtained from the kernel matrix $V$. From these anomalous dimensions, we define 
\begin{align}
\gamma_n ^\pm = \frac{1}{2}
\left [ \,
\gamma^{QQ}_n +\gamma^{GG}_n  
\pm \sqrt{ (\gamma^{QQ}_n -\gamma^{GG}_n)^2 + 4 \gamma^{QG}_n \gamma^{GQ}_n }
\, \right ] .
\label{eqn:anomalous-dimensions}
\end{align}
Then, the solution is written in terms of the Gegenbauer polynomials $C_n^{\, a}$
and the anomalous dimensions as
\begin{align}
& \! \! \! 
\sum_q^{n_f} \Phi_q^{+} (z,\mu) = z (1-z) 
         \! \! \!
         \sum_{\text{odd $n$}}
         \! \! \! A_n (\mu) \, C_n^{\, 3/2} (2z-1),
\nonumber \\
& \! \! \!
\Phi_g (z,\mu) = z^2 (1-z)^2
        \!  \! \! 
                 \sum_{\text{odd $n$}}
        \! \! \!
        A'_n (\mu) \, C_{n-1}^{\, 5/2} (2z-1),
\label{eqn:dga-solution}
\end{align}
where the coefficients are given by
\begin{align}
A_n (\mu) & = A_n^+ \left[ \frac{\alpha_s (\mu^2)}{\alpha_s (\mu^2_0)} \right]
                      ^{2\gamma_n^+ /\beta_0}
       \! \!   + A_n^- \left[ \frac{\alpha_s (\mu^2)}{\alpha_s (\mu^2_0)} \right]
                      ^{2\gamma_n^- /\beta_0} ,
\nonumber \\
A'_n (\mu) & = g_n^+ A_n^+ \left[ \frac{\alpha_s (\mu^2)}{\alpha_s (\mu^2_0)} \right]
                      ^{2\gamma_n^+ /\beta_0}
      \! \!   + g_n^- A_n^- \left[ \frac{\alpha_s (\mu^2)}{\alpha_s (\mu^2_0)} \right]
                      ^{2\gamma_n^- /\beta_0} ,
\label{eqn:coeff-an}
\end{align}
with the factor $g_n^\pm = (\gamma_n^\pm -\gamma_n^{QQ})/(3\gamma_n^{QG}/n)$.
The summations of Eq.\,(\ref{eqn:dga-solution}) are taken for odd $n$
($n=1,\, 3,\, \cdots$) in the $C=$even case.

In the $n=$odd summation, all the anomalous dimensions are positive except 
for $\gamma_1^- =0$, so that the only the $A_1^-$ terms survive 
in the scaling limit of 
$\mu^2 \to \infty$. Using the Gegenbauer polynomial $C_1^{\,a}(x)=2ax$,
we have the $C=$even (isoscalar) GPDs as
\begin{align}
\sum_q^{n_f} \Phi_q^+ (z,\mu\to\infty) & = 3 \, A_1^- \, z (1-z) (2z-1) ,
\nonumber \\
\Phi_g (z,\mu\to\infty) & =  g_1^- \, A_1^- \, z^2 (1-z)^2 .
\label{eqn:GDAs-scaling-n=odd}
\end{align}
Therefore, the $z$-dependent functional forms are uniquely given for the GDAs.
This fact should be taken into account for parametrizing the GDAs.

For the $C=\text{odd}$ (isovector) GDAs, the $n$ summation
of Eq.\,(\ref{eqn:dga-solution}) is for the even numbers
($n=0,\, 2,\, \cdots$).
In the scaling limit, only the $n=0$ term survives 
and the GDAs become
\begin{align}
\sum_q^{n_f} \Phi_q^- (z,\mu\to\infty) & = A_0 \, z (1-z) ,
\label{eqn:GDAs-scaling-n=even}
\end{align}
by using $C_0^{a}(x)=1$.
The above isovector GDAs have $z$ dependence $z (1-z)$ which 
is the same as the $\rho$-meson (pion \cite{Polyakov-1999}) isovector DA 
of Eq.\,(\ref{eqn:asymp}) in the scaling limit.

\subsection{{\boldmath$\zeta$} dependence of generalized distribution amplitudes}
\label{zeta-dependence}

The $Q^2$ evolution of the GDAs are calculated in perturbative QCD
as shown in the previous subsection, and the $z$ dependence is 
given by the Gegenbauer polynomials. The GDAs also depend on
other two variables $\zeta$ and $W^2$. Here, we discuss 
the $\zeta$ dependence. As shown in Fig.\,\ref{fig:gda-fig}
and Eq.\,(\ref{eqn:zeta-frac}), the variable $\zeta$ indicates
the momentum fraction for a produced pion in the final state
and it is expressed by the polar angle ($\theta$) of the pion.
Therefore, we may expand the coefficients $A_n$ and $A'_n$
in terms of orthogonal polynomials, which could be taken as
the Legendre polynomials $P_l$:
\begin{align}
A_n (\zeta, W^2) & = 6 \, n_f \sum_{l=\text{even}}^{n+1} 
                     B_{nl} (W^2) P_l (2\zeta-1) ,
\label{eqn:An-expand-Legendre}
\end{align}
where $n$ is odd ($l$ is even) for $C=+$, and
$n$ is even ($l$ is odd) for $C=-$.
Here, the factor 6 comes in the similar way to
the normalization of the pion DA
as shown in Eqs.\,(\ref{norm-phi}) and (\ref{eqn:asymp}),
the flavor number $n_f$ appears because of the flavor summation
in Eq.\,(\ref{eqn:dga-solution}), and
$\l$ is the angular momentum of the final pion pair.
In addition, the same equation exists for $A'_n (\zeta, W^2)$
in terms of $B'_{nl} (W^2)$.
The $C$ invariance relations of the GDAs are given in 
Eqs.\,(\ref{eqn:charge-conjugation}) and (\ref{eqn:charge-conjugation-g}),
so that the odd-$l$ terms do not contribute to the $C=+$ GDAs.

From the scale-dependence relations of Eq.\,(\ref{eqn:coeff-an}),
the coefficients $B_{nl}$ should follow the same relations:
\begin{align}
B_{n l} (W^2,\mu) & = B_{nl}^+ (W^2) 
             \left[ \frac{\alpha_s (\mu^2)}{\alpha_s (\mu^2_0)} \right]
                      ^{2\gamma_n^+ /\beta_0}
\nonumber \\
& \ \ 
              + B_{nl}^- (W^2)
             \left[ \frac{\alpha_s (\mu^2)}{\alpha_s (\mu^2_0)} \right]
                      ^{2\gamma_n^- /\beta_0} ,
\label{eqn:coeff-bnl}
\end{align}
and a similar equation for $B'_{n l} (W^2,\mu)$.
In the scaling limit $\mu\to\infty$, only the lowest terms survive
in $A_n (\zeta, W^2)$ and $A'_n (\zeta, W^2)$, and we obtain
\cite{gpds-gdas,diehl-2000}
\begin{align}
\sum_q^{n_f}
\Phi_q^{\pi\pi (+)} & (z, \zeta, W^2)  = 18 \, n_f \, z (1-z) (2z-1)
\nonumber \\[-0.35cm]
& \ \ 
\times
[B_{10}^-(W^2)+B_{12}^-(W^2) P_2(2\zeta-1)] , 
\nonumber \\[+0.10cm]
\Phi_g^{\pi\pi} & (z,  \zeta, W^2)  = 48 \, z^2 (1-z)^2 
\nonumber \\ 
& \ \ 
\times
[B_{10}^-(W^2)+B_{12}^-(W^2) P_2(2\zeta-1)] ,
\label{eqn:phi-paramet1}
\end{align}
where the Legendre polynomial $P_2 (x)$ is given by $P_2 (x) = (3x^2-1)/2$.
Since the Legendre polynomial term is given by
$P_2(2\zeta-1)=1-6\zeta (1-\zeta)$, the sum rule of Eq.\,(\ref{eqn:gda-sum-I=0})
is satisfied if the coefficients satisfy the relation
$B_{10} (W^2=0) = - B_{12} (W^2=0)$, 
which is considered in 
the parametrization in the next subsection.
This is the basic functional forms for $z$ and $\zeta$ dependencies
in the scaling limit. Next, we explain our actual parametrization
for the GDAs by following the essence of these basic functional forms.

\subsection{Expression of generalized distribution amplitudes}
\label{gda-expression}

With the basic knowledge of the pion DA and GDAs, we need to express 
the GDAs by a number of parameters. In particular, the $z$ dependence 
is given by Eqs.\,(\ref{eqn:GDAs-scaling-n=odd}) and 
(\ref{eqn:GDAs-scaling-n=even}) in the scaling limit.
Considering these functional forms, we express the GDAs with a number
of parameters. First, we neglect the higher-order $\alpha_s$ effects
and higher-twist effects, so that the gluon GDA does not appear.
Since $\pi^0\pi^0$-production data are analyzed in this work,
only the $C=\text{even}$ GDAs contribute to the cross section.
The $C=\text{even}$ function of Eq.\,(\ref{eqn:GDAs-scaling-n=odd}) is
$z(1-z)(2z-1)$. Since the $C=\text{even}$ isoscalar GDAs have $-$ sign 
under the change $z \to 1-z$ as given in Eq.\,(\ref{eqn:I=0-1-relations}),
the same parameter $\alpha$ is assigned for the powers of $z$ and $1-z$:
$\Phi_q^{\pi^0\pi^0}(z) \sim z^\alpha (1-z)^\alpha (2z-1)$.
The $2z-1$ factor comes from the lowest Gegenbauer polynomial 
$C_1^{\, 3/2} (2z-1)$, which survives in the scaling limit.
However, the detailed $z$ dependence is not determined from
the current data, so that we decide to take the lowest
Gegenbauer polynomial form of $2z-1$, 
supplementing by the phenomenological parameter $\alpha$ 
which will later appear to be close to the asymptotical value 1.

We use the following function for 
explaining the $\gamma^* \gamma \to \pi^0 \pi^0$ data
at a fixed $Q^2$ value:
\begin{align}
\Phi_q^{\pi\pi (+)}(z, \zeta, & W^2)  = 
N_\alpha z^\alpha(1-z)^\alpha (2z-1)
\nonumber \\
& \ \ 
\times
[B_{10}(W^2) +B_{12}(W^2) P_2(2\zeta-1)] , 
\label{eqn:phi-paramet1}
\end{align}
where $N_\alpha$ is the overall constant determined by the sum rule
(\ref{eqn:gda-sum-I=0}) as 
\begin{align}
N_\alpha = \frac{3 \, (2 \alpha +3 )}{5 \, B(\alpha+1,\alpha+1)} ,
\label{eqn:n-alpha}
\end{align}
with the beta function $B(a,b)$. The quark-momentum fraction factor 
$M_{2(q)}^\pi$
and the $W^2$-dependent form factor $F_q^\pi (W^2)$ 
are included in the coefficients $B_{nl} (W^2)$.
The $\zeta$ dependence can be re-expressed by the angle $\theta$ 
defined in Eq.\,(\ref{eqn:qqpp}) as
\begin{align}
 B_{10}(W^2) & +B_{12}(W^2)P_2(2\zeta-1)
 \nonumber \\
 & = \widetilde{B}_{10}(W^2)+\widetilde{B}_{12}(W^2)P_2(\cos\theta) ,
\label{eqn:b10-12-tilde}
\end{align}
where the invariant-mass dependent functions $\widetilde{B}_{nl}(W^2)$ 
and $B_{nl}(W^2)$ are related with each other by
\begin{align}
\widetilde{B}_{10}(W^2) & =B_{10}(W^2)-\frac{1-\beta^2}{2}B_{12}(W^2),  
\nonumber \\
\widetilde{B}_{12}(W^2) & =\beta^2B_{12}(W^2) . 
\label{eqn:bnl}
\end{align}
In the limit of $W^2\to 4m_\pi^2 \simeq 0$, they are given by 
\cite{Polyakov-1999,diehl-2000}
\begin{align}
B_{12}(0)= \frac{10}{9} M_{2(q)}^\pi ,
\label{eqn:b10b12-relation-W=0}
\end{align}
where $M_{2(q)}^\pi$ is the momentum fraction carried 
by the $q$-flavor quarks and antiquarks
in the pion ($\sum_q^{n_f} M_{2(q)}^\pi \simeq 0.5$).
This equation is obtained by considering the forward limit of the GPDs
and then the $s$-$t$ crossing to relate the GPDs and GDAs, so that
it should be a model-independent relation.
Then, the relation between $B_{10}(0)$ and $B_{12}(0)$ is studied
in a soft-pion theorem, and it was obtained as
\cite{Polyakov-1999,diehl-2000}
\begin{align}
B_{10}(0) = -B_{12}(0) .
\label{eqn:b10b12-relation-W=0-2}
\end{align}
Then, the $W^2$ dependence of $B_{10}(W^2)$ and $B_{20}(W^2)$ was 
studied at small $W^2$ as a possible constraint on the functional form 
of $W^2$ within a instanton model of QCD 
\cite{gda-form-factor-W2}.

The gluon GDA does not contribute to the cross section
because the higher-order and higher-twist terms are neglected
in our analysis. However, as discussed in Sec.\,\ref{Q2-evolution}, 
it affects the $Q^2$ evolution. It will be shown in 
Figs.\,\ref{fig:data-scaling-1} and \ref{fig:data-scaling-2}
that current Belle data are not accurate enough to probe
the scaling violation. 
The quark GDAs are provided at a fixed $Q^2$ scale which is
taken as the average $Q^2$ value (16.6 GeV$^2$) of the Belle data.
Then, the $Q^2$ evolution is not taken into account in our analysis
within the Belle-data range ($8.92 \le Q^2 \le 24.25$ GeV$^2$).
Therefore, the gluon GDA does not contribute in our analysis.

There are two terms, which correspond to the angular momenta,
$l=0$ and 2, of the pion pair. There are intermediate 
meson contributions to the cross section for $\gamma^* \gamma \to \pi^0\pi^0$,
so that the invariant-mass dependent factors $\widetilde{B}_{nl}$ have
imaginary parts expressed by the phase shifts $\delta_l (W)$:
\begin{align}
& \widetilde{B}_{nl}(W^2)=\bar{B}_{nl}(W^2) \, e^{i\delta_l (W)} . 
\label{eqn:b_phase}
\end{align}
Here, we use the $\pi\pi$ phase shifts by Bydzovsky, Kaminski, Nazari,
and Surovtsev \cite{pi-pi-code}. There is also another study 
on the phase shifts in Ref.\,\cite{pi-pi-warkentin}.
The relation of Eq.\,(\ref{eqn:b10b12-relation-W=0})
indicates that the $\bar{B}_{nl}(W^2=0)$ factors are given by
\begin{align}
\bar B_{10} (0) & = - \frac{3-\beta^2}{2} B_{12}(0)
= - \left( 1 +\frac{2m_\pi^2}{W^2} \right) B_{12}(0), 
\nonumber \\
\bar B_{12} (0) & = \beta^2 B_{12}(0)
= \left( 1 -\frac{4m_\pi^2}{W^2} \right) B_{12}(0) .
\label{eqn:b10b12bar-relation-W=0}
\end{align}
There are two types of contributions to $\widetilde{B}_{nl}(W)$.
One is the continuum and the other is from the intermediate resonances
expressed by 
\begin{align}
\! \! \! \! \! \! \!
\bar{B}_{nl}(W^2) & = 
\bar B_{nl}(0) \, F_q^\pi (W^2) 
\nonumber \\
& 
+ \! \sum_R \frac{c_{_R}}{\sqrt{(M^2_{R}-W^2)^2+\Gamma^2_{R} \, M^2_{R} }},
\label{eqn:b-cont-res}
\end{align}
where $M_R$ is the resonance mass, $\Gamma_R$ is its width, 
and $c_{_R}$ is a constant. 

The $W^2$ dependence of the continuum part of the pion GDAs
is given by the form factor, which could be parametrized as \cite{kk2014}
\begin{align}
F^{\,\pi}_q (W^2) 
= \frac{1}{\left[ 1 + (W^2-4 m_\pi^2)/\Lambda^2 \right]^{n-1}} .
\label{eqn:form-factorw2}
\end{align}
Here, $\Lambda$ is the cutoff parameter, 
which indicates the pion size, 
$n$ is the number of active constituents
according to the constituent-counting rule in perturbative QCD 
\cite{BC-1976}, and it is normalized as $F_{h(q)} (4 m_\pi^2)=1$.
It is the continuum part of 
the timelike forms factor of the energy-momentum tensor.
Here, the pion size means the gravitational-interaction size 
instead of the usual charge radius
in electromagnetic interactions as explained in
Sec.\,\ref{gravitation-radius}.
The high-energy behavior of the form factor is given by
the factor $n$, which is supposed to be $n=2$ for the pion
\cite{kk2014}. 

\subsection{Resonance terms and their coupling constants}
\label{resonance-constants}

\begin{figure}[b!]
     \includegraphics[width=5.0cm]{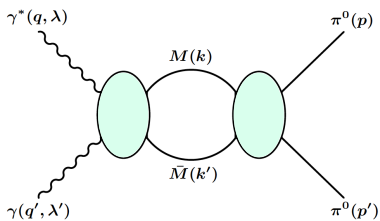}
\vspace{-0.20cm}
\caption{$\gamma^* \gamma \rightarrow \pi^0 \pi^0$ 
    through the intermediate states $M\bar{M}$.}
\label{fig:2gamma-int-meson}
\end{figure}

The resonance contributions are illustrated as the intermediate states
in Fig.\,\ref{fig:2gamma-int-meson}.
Above 1 GeV of the invariant mass $W$, the intermediate $K\bar K$ 
and $\eta\eta$ channels contribute to the process.
However, their contributions may not be as large as the pion ones,
and they are not explicitly considered in this work. 
As seen in Fig.\,\ref{fig:2gamma-int-meson}, we also need the kaon GDAs
in the formalism if the $K\bar K$ were introduced in the intermediate state.
The constant $c_{_R}$ in Eq.\,(\ref{eqn:b-cont-res}) is expressed 
by the $R\to\pi\pi$ coupling constant $g_{R\pi\pi}$,
the decay constant $f_R$, the mass $M_R$, and the width $\Gamma_R$.
As for the mesons with $I^G (J^{PC})=0^+ (0^{++}),\ 0^+ (2^{++})$,
we consider 
\begin{align}
0^+(0^{++}): &
\ f_0(500), \  f_0(980),
\nonumber \\
0^+(2^{++}): &
\ f_2(1270),
\label{eqn:resonance-included}
\end{align}
which could make conspicuous contributions to the cross section 
$\gamma^* \gamma \to \pi^0 \pi^0$ in the invariant mass region 
of $W \le 2.05$ GeV, in our analysis. In this energy region,
there are other possible resonances  
\begin{align}
0^+(0^{++}): &
\ f_0(1370), \  f_0(1500), \ f_0(1710) ,
\nonumber \\
0^+(2^{++}): &
\ f_2(1525), \ f'_2(1950), \ f_2(2010) , 
\label{eqn:resonance-neglected}
\end{align}
in principle. However, these meson effects have minor effects 
on the cross section, hence they are not included in our analysis.

The GDAs are defined by the matrix element of the nonlocal vector operator 
from the vacuum to the $\pi^0 \pi^0$ state, and it is expressed by three steps 
for describing the process with the intermediate $f_0$ meson
\cite{teryaev-2005-f2}. First, the $f_0$ meson is produced 
from the vacuum, it propagates, and then it decays into the pion pair:
\begin{align}
& \bra{\pi^0 (p) \pi^0 (p')}\overline{q}(-z/2)\gamma_{\mu} q(z/2)  \Ket{0}
= \bra{\pi^0 (p) \pi^0 (p')}  f_0 (P) \rangle 
\nonumber \\
& \times \frac{1}{m^2_{f_0}-P^2-i\Gamma M_{f_0}} 
\bra{f_0 (p)}\overline{q}(-z/2)\gamma_{\mu} q(z/2)  \Ket{0}+\cdots .
\label{eqn:b10a}
\end{align}
In Sec\,\ref{pion-da}, the pion distribution amplitude is defined by 
$\overline \psi \gamma_\mu \gamma_5 \psi$
instead of one quark flavor one 
$\overline q \gamma_\mu \gamma_5 q$. The above $f_0$ distribution amplitude
is related to the one defined by 
$\overline \psi \gamma_\mu \psi 
 = (\overline u \gamma_\mu u + \overline d \gamma_\mu d )/\sqrt{2}$ as
\begin{align}
\bra{f_0 (p)} &  \overline{q}(-z/2) \gamma_{\mu} q(z/2)  \Ket{0}
\nonumber \\
& = \frac{1}{\sqrt{2}} 
\bra{f_0 (p)}\overline{\psi}(-z/2)\gamma_{\mu} \psi(z/2)  \Ket{0} ,
\label{eqn:f0-da}
\end{align}
where $q=u$ or $d$.
The final $2\pi$-decay part is simply the coupling constant written as
\begin{align}
\bra{\pi^0 (p) \pi^0 (p')} f_0 (P) \rangle = -ig_{f_0 \pi^0 \pi^0}, 
\label{eqn:2pi-decay}
\end{align}
and the first $f_0$ production part is expressed by the distribution amplitude
for $f_0$ as discussed in the later part of this subsection.
Then, the $f_0$ contribution to $\widetilde{B}_{10}(W)$ is written as
\begin{align}
\widetilde{B}_{10}(W)=\frac{5g_{f_0\pi\pi} f_{f_0}}
                         {3 \sqrt{2} [m^2_{f_0}-W^2-i\Gamma M_{f_0}]},    
\label{eqn:b10b}
\end{align}
so that its absolute value is given by
\begin{align}
\bar{B}_{10}(W)=\frac{5g_{f_0\pi\pi} f_{f_0} }
   {3 \sqrt{2} \sqrt{[(M^2_{f_0}-W^2)^2+\Gamma^2_{f_0} M^2_{f_0} ]}},
\label{eqn:b10b2}
\end{align}
where the factor 5/3 comes from the convention difference in
defining the distribution amplitude, namely the overall factor
could be 30 or 18. This difference becomes $30/18=5/3$.
In the same way, the $f_2$ contribution is given by
\begin{align}
\bar{B}_{12}(W) =  \frac{10 \, g_{f_2\pi\pi} f_{f_2} M^2_{f_2} \beta^2 }
  {9 \sqrt{2} \sqrt{ (M^2_{f_2}-W^2)^2+\Gamma^2_{f_2} M^2_{f_2} }}, 
\label{eqn:b12ap1}
\end{align}
where the different factor $10 \, M^2_{f_2} /9$ comes from 
the tensor nature of $f_2$ in defining the coupling constant
to $2\pi$ and also the decay constant
\cite{teryaev-2005-f2,Anikin:2008bq}.
In Ref.\,\cite{teryaev-2005-f2},
the $\beta^2$ factor is included in Eq.\,(A26) of this paper.

As for the resonance terms, we use the $W$ dependence of 
$|\widetilde{B}_{nl}(W^2)|$ in Eqs.\,(\ref{eqn:b10b2}) and (\ref{eqn:b12ap1})
\cite{Anikin:2008bq}, although the resonance properties are also obtained 
by the Belle collaboration for the resonances $f_0(980)$ and $f_2(1270)$ 
\cite{Masuda:2015yoh}.
In Refs.\,\cite{Anikin:2008bq,braun-2016,cheng-2010-f2}, 
the constants are 
$f_{f_2}=0.101$ GeV at $Q^2=1$ GeV$^2$, 
$M_{f_2}=1.275$ GeV, and $\Gamma_{f_2}=0.185$ GeV
for $f_2(1270)$, and the decay constant $g_{f_2 \pi \pi}$ is defined by
$ g_{f_2 \pi \pi}=\sqrt{(2/3) 24 \pi \Gamma(f_2 \rightarrow \pi \pi) / M^3_{f_2}}$
with $\Gamma(f_2 \rightarrow \pi \pi)=0.85 \, \Gamma_{f_2}$.
Here, the factor of 2 in 2/3 comes from the identical particles 
of two $\pi^0$'s, and the factor 1/3 does from 
$\Gamma(f_2 \rightarrow \pi^0 \pi^0)=1/3\Gamma(f_2 \rightarrow \pi \pi)$.
As it will become clear in comparison with the actual measurements,
the Belle data indicate a clear peak of $f_2 (1270)$.

\begin{table}[t!]
\caption{Resonance constants in our analysis.
The decay constant $f_{f_2}$ is shown at $Q^2=1$ GeV$^2$ in this table.
In Table \ref{tab:analysis-results}, the decay constants are
listed at $Q^2=16.6$ GeV$^2$,
which is the average scale of the used Belle data.
The value $f_{f_2}=0.101$ at $Q^2=1$ GeV$^2$ corresponds to 
$f_{f_2}=0.0754$ at $Q^2=16.6$ GeV$^2$.}
\label{tab:resonaces}
\vspace{-0.50cm}
\begin{center}
\begin{tabular}{|c|cccc|} 
\hline
                 &              &              &                &             \\[-0.30cm]
\ Meson ($h$)  \ &  \ $M$ (GeV) & \ $\Gamma$ (GeV)  & $g_{_{h\pi\pi}}$   & $f_{h}$ (GeV) \ \\[+0.04cm] \hline
$f_0 \, (500)$   &  \  0.475    & 0.550        & 2.959 GeV \,    &  $-$  \        \\
$f_0 \, (980)$   &  \  0.990    & 0.055        & 1.524 GeV \,    &  $-$        \  \\
$f_2 (1270)$     &  \  1.275    & 0.185        & \ \ 0.157 \,GeV$^{-1}$ &  0.101  \      \\ \hline
\end{tabular}
\end{center}
\end{table}

For the S-wave resonances of $f_0 (500)$ and $f_0 (980)$, 
we have $M_{f_0 (500)}=0.475$ GeV and $ \Gamma_{f_0 (500)}=0.55$ GeV \cite{pdg-2016}, 
the decay constant $g_{f_0 \pi \pi}$ is defined by
$ g_{f_0 \pi \pi}=\sqrt{(2/3)  16\pi \Gamma(f_0 \rightarrow \pi \pi) M_{f_0}}$
with  $\Gamma(f_0 \rightarrow \pi \pi)= \Gamma_{f_0}$
for both $f_0 (500)$ and $f_0 (980)$.
The decay width of $f_0(980)$ is not well determined by experiments, 
and it is listed as 10$-$100 MeV. 
We use the middle values of 
the Particle Data Group \cite{pdg-2016}, namely
$\Gamma_\sigma=550$ MeV between 400 MeV and 700 MeV.
As for the decay constants $f_{f_0 (500)}$ and $f_{f_0 (980)}$, 
no experimental information is available. 
There are theoretical estimates on $f_{f_0 (980)}$ by the QCD sum-rule method.
However, they assume the $q\bar q$ configuration for $f_0 (980)$ and 
their decay-constant values seem to be inconsistent with 
the Belle data on the differential cross section
as shown in Sec.\,\ref{f0-980-contribution}.
There is no theoretical estimate on $f_{f_0 (500)}$ as far as we searched,
so it is simply terminated or it is considered as one of the parameters
in our analysis.
These numerical values are summarized in Table \ref{tab:resonaces}.

Next, we define decay constants and distribution amplitudes
for the resonances $f_0 (500)$, $f_0 (980)$, and $f_2 (1270)$.
In the reaction $\gamma^*\gamma\to \pi^0\pi^0$, the matrix elements
of a vector current between the vacuum and these meson states
are involved in its cross section.
First, the matrix element for the tensor meson $f_2 (1270)$ is 
expressed by the decay constant $f_{f_2}$ and the distribution amplitude 
$\Phi_{f_2} (z,\mu)$ as
\cite{teryaev-2005-f2,cheng-2010-f2}
\begin{align}
& \! \! \! 
\langle \, f_2 (p) \, | \, \overline \psi (y) \, \gamma_\mu
                  \, \psi (0) \,
                | \, 0 \, \rangle\Big |_{y^+ = \vec y_\perp =0}
= 
  \frac{\varepsilon_{\alpha\beta}^{(\lambda)*} 
        y^\alpha y^\beta}
       {(p \cdot y)^2}  
\nonumber \\
& \ \ \ \ \ \ \ 
\times 
f_{f_2} \, m_{f_2}^2 \, p_\mu
  \int_0^1 dz \, e^{i z p^+ y^-}
   \, \Phi_{f_2} (z,\mu) 
  + \cdots ,
\label{eqn:matrix-f2}
\end{align}
where $\varepsilon_{\alpha\beta}^{(\lambda)}$ is the polarization
vector of $f_2$ meson \cite{cheng-2010-f2}, 
and the higher-twist terms are not explicitly written.
The distribution amplitude for $f_2$ is given by the summation
of odd Gegenbauer polynomials due to the $C$-parity as explained
in Eq.\,(\ref{eqn:dga-solution}), and it is expressed as
\cite{teryaev-2005-f2,cheng-2010-f2}
\begin{align}
\Phi_{f_2} (z,\mu) & =   6 \, z \, (1-z) 
          \sum_{\text{odd }n=1}^\infty B_n (\mu) C_n^{3/2} (2z-1) 
\nonumber \\
& =  B_1 (\mu) \, 18 \, z \, (1-z) \, (2z-1) + \cdots ,
\label{eqn:da-f2}
\end{align}
whereas it is the even polynomials for the pion 
as shown in Eq.\,(\ref{eqn:Gegenbauer}).

In the same way, the matrix elements for the scalar mesons 
$f_0 (500)$ and $f_0 (980)$ 
are given by
\begin{align}
& \! \! \! 
\langle \, f_0 (p) \, | \, \overline \psi (y) \, \gamma_\mu
                  \, \psi (0) \,
                | \, 0 \, \rangle\Big |_{y^+ = \vec y_\perp =0}
\nonumber \\
& \ \ \ \ \ \ \ \  
= 
\, p_\mu
  \int_0^1 dz \, e^{i z p^+ y^-}
   \, \Phi_{f_0} (z,\mu)  ,
\label{eqn:matrix-f0}
\end{align}
where the distribution amplitude is defined
by including the decay constant $f_{f_0}$ for a practical purpose, 
because the combined quantity of $f_{f_0}$ and the amplitude
becomes finite even though $f_{f_0}$ itself vanishes.
We define the decay constants $f_{f_0}$ and $\bar f_{f_0}$
by the matrix elements for the vector and scalar operators as
\cite{cheng-2006-f0} 
\begin{align}
\langle \, f_0 (p) \, | \, \overline \psi (0) \, \gamma_\mu
                  \, \psi (0) \,
                | \, 0 \, \rangle 
& = f_{f_0} \, p_\mu ,
\nonumber \\
\langle \, f_0 (p) \, | \, \overline \psi (0) 
                  \, \psi (0) \,
                | \, 0 \, \rangle 
& = \bar f_{f_0} \, m_{f_0} .
\label{eqn:matrix-f0-y=0}
\end{align}
Writing the above vector current at the position $x$
as $J_\mu (x) =  \overline\psi (x) \, \gamma_\mu \, \psi (x) 
              = e^{i \hat p \cdot x} J_\mu (0) e^{-i \hat p \cdot x}$
and using the equation of motion, we relate the two decay constants as
\begin{align}
(m_{\bar q}-m_q) \, \bar f_{f_0} = m_{f_0} \, f_{f_0} ,
\label{eqn:decay-constants-relation}
\end{align}
where $m_q$ and $m_{\bar q}$ are quark and antiquark masses.
In the $f_0$-meson case, the masses are equal ($m_{\bar q}-m_q=0$).

Because of the conservation of the vector current or 
charge-conjugation invariance, the constant $f_{f_0}$
should vanish $f_{f_0}=0$.
However, the nonlocal matrix element of Eq.\,(\ref{eqn:matrix-f0})
does not vanish at finite $Q^2$, whereas 
it vanishes in the scaling limit $Q^2 \to\infty$
as we explain later in Sec.\,\ref{resonance-scale}.
Comparing Eqs.\,(\ref{eqn:matrix-f0}) and (\ref{eqn:matrix-f0-y=0}),
we obtain the relation
\begin{align}
\int_0^1 dz \, \Phi_{f_0} (z,\mu) = f_{f_0}=0 .
\label{eqn:da-f0-sum}
\end{align}
For the scalar mesons with $m_q \ne m_{\bar q}$,
the relation (\ref{eqn:decay-constants-relation})
can be used to relate the decay constants.
Therefore, according to Ref.\cite{cheng-2006-f0}, we may take that 
the $f_0$ distribution amplitude is expressed by $\bar f_{f_0}$ 
and the Gegenbauer polynomials as
\begin{align}
\Phi_{f_0} & (z,\mu)  =  \bar f_{f_0} \, 6 \, z \, (1-z) 
\nonumber \\
&
\times   \bigg [ \, B_0 (\mu) 
     + \sum_{\text{odd }n=1}^\infty B_n (\mu) C_n^{\, 3/2} (2z-1) \, \bigg ] .
\label{eqn:da-f0-1}
\end{align}
Then, the normalization of Eq.\,(\ref{eqn:da-f0-sum}) is satisfied
if $B_0$ is taken as $(m_{\bar q} - m_q )/m_{f_0} \equiv 1/\mu_{f_0}$.
The integral of the first term is $f_{f_0}$ and those of
the subsequent summation terms vanish identically.
The first term $\bar f_{f_0}/\mu_{f_0} = f_{f_0}$ vanishes
for the $f_0$ meson, so that it is given by
\begin{align}
\! \! \! \!
\Phi_{f_0} (z,\mu) & =  \bar f_{f_0}  6 \, z \, (1-z) 
          \sum_{\text{odd }n=1}^\infty B_n (\mu) C_n^{\, 3/2} (2z-1) 
\nonumber \\
& =  \bar f_{f_0} B_1 (\mu) \, 18 \, z \, (1-z) \, (2z-1) + \cdots ,
\label{eqn:da-f0-2}
\end{align}
where $C_1^{\, 3/2} (x)=3 x$ is used.

\subsection{Scale dependence of resonance contributions}
\label{resonance-scale}

There are finite contributions to the $\gamma^* \gamma \to \pi^0 \pi^0$
cross section from $f_2 (1270)$, $f_0 (500)$, and $f_0 (980)$ 
at small $Q^2$. However, as the $Q^2$ increases, they become smaller 
and smaller, and they eventually vanish in the scaling limit $Q^2 \to \infty$.
The scale dependence of the distribution amplitude is given 
by the anomalous dimensions $\gamma_n$ and the leading coefficient
$\beta_0 = (11 C_A - 4 T_R n_f)/3$ 
of the $\beta$ function with $C_A=N_c$ and $T_R=1/2$
as \cite{teryaev-2005-f2,cheng-2010-f2}
\begin{align}
& 
f_{f} (Q^2) B_n (Q^2) = f_{f} (Q_0^2) B_n (Q_0^2) \,
  \left [ \frac{\alpha_s (Q^2)}{\alpha_s (Q_0^2)} \right ]^{\gamma_n/\beta_0} ,
\nonumber \\
& \ \ \ \ \ \ 
\gamma_n = C_F \left [ \, 1 - \frac{2}{(n+1)(n+2)} 
            + 4 \sum_{j=2}^{n+1} \frac{1}{j} \, \right ] ,
\label{eqn:da-f02-evol}
\end{align}
where $C_F = (N_c^2-1)/(2 N_c)$ with the number of colors $N_c$.
Here, the meson $f$ indicates $f_0 (500)$, $f_0 (980)$, or $f_2 (1270)$,
and the decay constant $f_f$ is 
$\bar f_{f_0 (500)}$, $\bar f_{f_0 (980)}$, or $f_{f_2 (1270)}$.
One could express the scale evolution separately for 
the decay constant and the distribution amplitude as
\cite{cheng-2010-f2}
\begin{align}
f_{f} (Q^2) & = f_{f} (Q_0^2) \,
  \left [ \frac{\alpha_s (Q^2)}{\alpha_s (Q_0^2)} \right ]^{-4/\beta_0} ,
\nonumber \\
B_n (Q^2) & = B_n (Q_0^2) \,
  \left [ \frac{\alpha_s (Q^2)}{\alpha_s (Q_0^2)} \right ]^{(\gamma_n +4)/\beta_0} .
\label{eqn:da-f02-evol-2}
\end{align}
The leading Gegenbauer polynomial is taken in Eq.\,(\ref{eqn:da-f0-2}), 
and its anomalous dimension is given by $\gamma_1 = 2 \, C_F /3$.
This finite anomalous dimension indicates that the distribution 
amplitudes decrease with increasing $Q^2$ as shown in 
Eq.\,(\ref{eqn:da-f02-evol}). 

From Eq.\,(\ref{eqn:da-f02-evol-2}), it is possible to describe
the $Q^2$ evolution separately for the decay constant and 
the distribution amplitude. However, the overall scale dependence 
is given by Eq.\,(\ref{eqn:da-f02-evol}) in any case.
The scale dependence is often attributed only to the decay constant
\cite{teryaev-2005-f2,cheng-2010-f2}, namely 
\begin{align}
f_{f} (Q^2) = f_{f} (Q_0^2)  \,
  \left [ \frac{\alpha_s (Q^2)}{\alpha_s (Q_0^2)} 
      \right ]^{\gamma_n/\beta_0},
\label{eqn:ff-f02-evol-3}
\end{align}
and the distribution amplitude may be normalized 
in the scale-independent way as 
\begin{align}
\int_0^1 dz \, (2z-1) \, \Phi_f (z) = 1 ,
\label{eqn:da-f02-norm}
\end{align}
so that it becomes 
\begin{align}
\Phi_f (z) = 30 \, z \, (1-z) \, (2z-1) ,
\label{eqn:da-f02-normalized}
\end{align}
as the leading distribution. 
In Eqs.(\ref{eqn:da-f2}) and (\ref{eqn:da-f0-2}), the $f_2$ and $f_0$
distribution amplitudes are defined with the scale dependence.
However, the scale independent expression of Eq.\,(\ref{eqn:da-f02-normalized})
is used in this work, which means to take the $B_1$ factor as
$B_1 = 5/3$ \cite{cheng-2010-f2}.
This is a consistent description with Eq.\,(\ref{eqn:da-f02-evol-2}).
However, it may be somewhat confusing, so that
one should remember that the distribution amplitude vanishes
in the scaling limit $\Phi_f (z,\mu) =0$ at $\mu\to\infty$,
although the scale-independent expression (\ref{eqn:da-f02-norm}) 
is often used practically.

\subsection{Gravitational form factors for pion}
\label{gravitational-ffs}

As shown in Eq.\,(\ref{eqn:integral-over-z}), the GDAs probe
the $++$ component of the energy momentum tensor, and it is expressed 
by the form factors for $\pi^0$ as
\cite{emt-pion,emt-pion1}
\begin{align}
& \! \! \! \! 
\langle \, \pi^0 (p) \, \pi^0 (p') \, | \, T_q^{++} (0) \, | \, 0 \, \rangle 
\nonumber \\
& \! \! \! \! 
= \frac{1}{2} 
  \left [ \, \left ( s \, g^{++} -P^+ P^+ \right )  \Theta_{1, q} (s)
                + \Delta^+ \Delta^+   \Theta_{2, q} (s) \,
  \right ] .
\label{eqn:emt-ffs-timelike}
\end{align}
Calculating the $+$ components by using the momentum assignments
in Eq.\,(\ref{eqn:qqpp}) and using its relation to the GDAs in
Eq.\,(\ref{eqn:integral-over-z}), we obtain
\begin{align}
\! \! \! \!
&
\int_0^1 dz (2z -1) \, 
\Phi_q^{\pi^0 \pi^0} (z,\,\zeta, \,W^2) 
\nonumber \\
& 
= - \Theta_{1, q} (s) + \frac{\beta^2}{3} \, \Theta_{2, q} (s)
+ \frac{2 \beta^2}{3} \Theta_{2, q} (s)  P_2 (\cos\theta) .
\label{eqn:emt-ffs-expression}
\end{align}
On the other hand, from the GDA expression in terms of $\widetilde B_{10}$
and $\widetilde B_{20}$ in Eqs.\,(\ref{eqn:phi-paramet1}) and
(\ref{eqn:b10-12-tilde}) with the normalization of Eq.\,(\ref{eqn:n-alpha}),
the integral of the GDA is given by
\begin{align}
\int_0^1 dz & (2z -1) \, 
\Phi_q^{\pi^0 \pi^0} (z,\,\zeta, \,W^2) 
\nonumber \\
& 
= \frac{3}{5} \left [ \, \widetilde B_{10} (W^2)
+ \widetilde B_{12} (W^2) \, P_2 (\cos\theta) \right ] .
\label{eqn:gda-integral-1}
\end{align}
From Eqs.\,(\ref{eqn:emt-ffs-expression}) and (\ref{eqn:gda-integral-1}),
the gravitational form factor are expressed 
by the S- and D-wave components of the GDAs as
\begin{align}
\Theta_{1, q} (s) & 
= -\frac{3}{5} \widetilde B_{10} (W^2) + \frac{3}{10} \widetilde B_{20} (W^2) ,
\nonumber \\
\Theta_{2, q} (s) & 
= \frac{9}{10 \, \beta^2} \widetilde B_{20} (W^2) .
\label{eqn:emt-ffs-gdas}
\end{align}
Quark and antiquark contributions are added to obtain the timelike
gravitational form factors of the pion as
\begin{align}
\Theta_{n} (s) & = \sum_{i=q} \Theta_{n, i} (s),
\ \ \ n=1, \ 2 .
\label{eqn:quark-sum-theta12}
\end{align}
In this way, if the GDAs are determined from experimental measurements,
the gravitational form factors, consequently gravitational radii, 
are obtained for the pion.

Next, we discuss normalizations of the form factors. 
Using the Legendre polynomial expressed by $\zeta$ as
$P_2 (\cos\theta)= [ -12 \zeta (1-\zeta) + 3 - \beta^2]/(2 \beta^2)$,
we obtain the integral of Eq.\,(\ref{eqn:emt-ffs-expression}) as
\begin{align}
\! \! \! \!
&
\int_0^1 dz (2z -1) \, 
\Phi_q^{\pi^0 \pi^0} (z,\,\zeta, \,W^2) 
\nonumber \\
& 
= - \Theta_{1, q} (s) + \Theta_{2, q} (s)
  - 4 \zeta (1-\zeta) \Theta_{2, q} (s)  .
\label{eqn:emt-ffs-expression-2}
\end{align}
The right-hand-side of this equation should be equal to
the sum $- 4 M_{2(q)}^\pi$ given in Eq.\,(\ref{eqn:gda-sum-I=0})
at $W^2 = 4 m_\pi^2$, and it leads to the relations
\begin{align}
\Theta_{1, q} (s=4 m_\pi^2) 
= \Theta_{2, q} (s= 4 m_\pi^2)
= M_{2(q)}^\pi ,
\label{eqn:emt-ffs-norm-1}
\end{align}
in the scaling limit.
Quark and antiquark contributions are added to obtain the form factors:
$\Theta_{n} (s=4 m_\pi^2) = \sum_{i=q} \Theta_{n, i} (s=4 m_\pi^2)$.
Then, such sum of the right-hand side of Eq.\,(\ref{eqn:emt-ffs-norm-1}) is 
$\sum_{q} M_{2(q)}^\pi$.
Therefore, the normalizations of the form factors become
the momentum fraction carried by quarks and antiquarks in the pion:
\begin{align}
\Theta_{1} (s=4 m_\pi^2) = \Theta_{2} (s=4 m_\pi^2) 
= \sum_{q} M_{2(q)}^\pi ,
\label{eqn:emt-ffs-norm-2}
\end{align}
in the scaling limit.
The factor $\sum_{q} M_{2(q)}^\pi$ is written as $R_\pi$
in some articles. Here, the only the quark contributions are discussed,
so that the normalization becomes the quark (and antiquark) momentum fraction.
However, if the gluon contribution is added, the relation should be
$\Theta_{1} (s=4 m_\pi^2) = \Theta_{2} (s=4 m_\pi^2)=1$, which indicate
$A  (s=4 m_\pi^2) =1$ and $B  (s=4 m_\pi^2) =-1/4$
from Eq.\,(\ref{eqn:ffs-theta-ab}). Therefore, our timelike form factors
are consistent with the works in 
Refs.\,\cite{gda-form-factor-W2,emt-pion,emt-pion1}.
We should note that these normalizations are satisfied in
the scaling limit. However, the Belle measurements are 
at finite $Q^2$ with some resonance effects, so that
the actual values contain their effects. In fact, as we show later,
they are 
$\Theta_{1} (s=4 m_\pi^2) = \Theta_{2} (s=4 m_\pi^2) \sim 0.7$,
instead of $\sum_{q} M_{2(q)}^\pi = 0.5$,
in our GDA analysis.

\section{Results}
\label{results}

From these theoretical preparations, we proceed to the actual
analysis of experimental data. Here, the Belle data for 
$\gamma^* \gamma \to \pi^0 \pi^0$ \cite{Masuda:2015yoh} are used
for our study.
The invariant-mass dependent functions are parametrized with
the resonance contributions from 
$f_0(500)$, $f_0 (980)$ and $f_2(1270)$, and it is summarized as
\begin{align}
\Phi_q^{\pi\pi (+)} (z, \zeta, W^2) & = 
           N_\alpha z^\alpha(1-z)^\alpha (2z-1)
\nonumber \\
& \! \! \! \! \! \!
\times [\widetilde B_{10}(W^2) + \widetilde B_{12}(W^2) P_2(\cos \theta)] , 
\label{eqn:gda-parametrization}
\end{align}
where the normalization constant $N_\alpha$ is given 
in Eq.\,(\ref{eqn:n-alpha}).
The S and D wave terms are expressed by the contributions
from the continuum and the resonances as
\begin{align}
\! \! 
\widetilde{B}_{10}(W^2) 
&  = - \bigg[ \,  \left( 1+\frac{2\, m_\pi^2}{W^2} \right)
\frac{10}{9} M_{2(q)}^\pi   
F^{\,\pi}_q (W^2)
\nonumber \\[-0.05cm]
&  \! \! \!
+ \sum_{f_0}
\frac{5 \, g_{f_0\pi\pi} \, \bar f_{f_0}}
    {3 \sqrt{2} \sqrt{(M^2_{f_0}-W^2)^2+\Gamma^2_{f_0} M^2_{f_0} }} \, \bigg] \, 
     e^{i \delta_0 (W)}      ,
\nonumber \\
\! \! 
\widetilde{B}_{12}(W^2)
& =  \left( 1 - \frac{4\, m_\pi^2}{W^2} \right) \frac{10}{9}
 \bigg[  \,
 M_{2(q)}^\pi
F^{\,\pi}_q (W^2)
\nonumber \\[-0.05cm]
&  \! \! \!
+ \frac{g_{f_2\pi\pi} \, f_{f_2} M^2_{f_2} \beta^2 }
     {\sqrt{2} \sqrt{(M^2_{f_2}-W^2)^2+\Gamma^2_{f_2} M^2_{f_2} }} \, \bigg] \,
      e^{i\delta_2 (W)}  .
\label{eqn:B10B12-final}
\end{align}
The timelike form factor for the continuum
is given by the cut-off parameter $\Lambda$ and
the power of $n-1$ in Eq.\,(\ref{eqn:form-factorw2}).
The factor $n$ is suggested by the constituent
counting rule at high energies and it is $n=2$ for the pion.
Here, $f_0$ indicates $f_0 (500)$ and $f_0 (980)$.
However, the analyzed Belle data are not sensitive to $f_0 (980)$,
so that it is not included in our analysis.
The up and down quark GDAs are considered in our analysis,
and strange and charm quark contributions are neglected. 
We assigned a parameter $\alpha$ for the $z$-dependent functional form
of the quark GDAs.
This $z$-dependent function enters into the amplitude $A_{++}$
in Eq.\,(\ref{eqn:cross2}), and then the integral is given
in the form of $\int_0^1 dx (2z-1)^2 z^{\alpha-1} (1-z)^{\alpha-1}$.
This integral is expressed as the beta function as 
$B(\alpha,\alpha)/(2\alpha+1)$, and 
it plays a role of overall constant to explain 
the $\gamma^* \gamma \to \pi^0 \pi^0$ data. 

\begin{figure}[b]
     \includegraphics[width=7.0cm]{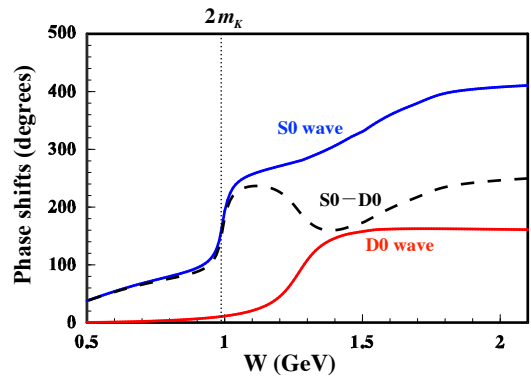}
\vspace{-0.30cm}
\caption{S-wave phase shift and D-wave phase shifts
by Bydzovsky, Kaminski, Nazari, and Surovtsev \cite{pi-pi-code}.}
\label{fig:phase}
\end{figure}

Next, we explain the S- and D-wave phase shifts used in our analysis.
The phase shifts $\delta_0$ and $\delta_2$ in Ref.\,\cite{diehl-2000} 
seem to work below $W=1$ GeV. The region of the center-of-mass energy 
is 0.525 GeV $\le \sqrt{s}=W \le 2.05$ GeV in the Belle data \cite{Masuda:2015yoh}.
In order to analyze the Belle data, we use the S-wave and D-wave 
$\pi\pi$ phase shifts obtained by Bydzovsky, Kaminski, Nazari,
and Surovtsev (BKNS) \cite{pi-pi-code}.
Their phase shifts are shown in Fig.\,\ref{fig:phase}.
They proposed a parametrization of the S- and D-wave phase shifts 
from analysis of the $\pi\pi$ scattering experimental data 
in the isospin$\,=0$ channel. 
Since only the difference of S- and D-wave phase shifts matters
for explaining the cross section
in our analysis of
Eqs.\,(\ref{eqn:gda-parametrization})
and (\ref{eqn:B10B12-final}), the difference is also shown
by the dashed curve.
Above the $K\bar K$ threshold at about 1 GeV, the phase difference
is, roughly speaking, a slowly varying function of $W$.
We note that the $K\bar K$ channel opens at the threshold energy
$2 m_{K^+}=0.987354$ GeV, so that only the $\pi\pi$ phase shifts
may not be sufficient. To be precise, the $K\bar K \to \pi\pi$ 
phase shifts should be introduced together with the kaon GDAs.
We may investigate such details step by step. 
In our GDA analysis, a simple case is considered by introducing 
a phase $\Delta \delta (W)$ for the S wave above the $K\bar K$ threshold,
$\delta_0 (W) =  \delta_0 (W)_{\text{BKNS}} + \Delta \delta (W)$,
$\delta_2 (W) =  \delta_2 (W)_{\text{BKNS}}$
with expectation that such effects are included in the modified part.
In our analysis, we introduce phase parameters in the S-wave as 
\begin{align} 
\delta_0 (W) = 
\delta_0 (W)_{\text{BKNS}} + a_\delta \, (W-2m_K)^{b_\delta} ,
\label{eqn:delta0-phase}
\end{align}
at $W > 2m_K$.
The parameters $a_\delta$ and $b_\delta$ are determined
by the $\chi^2$ analysis.

\subsection{\boldmath$f_0 (980)$ contribution}
\label{f0-980-contribution}

A possible complication or ambiguity is how to determine
the decay constants $\bar f_{f_0 (500)}$ and $\bar f_{f_0 (980)}$, 
whereas the constant $f_{f_2}$ is relatively 
well evaluated \cite{teryaev-2005-f2,cheng-2010-f2}.
It is because the internal configurations of $f_0 (500)$ and
$f_0 (980)$ are not well known. 
The evaluation of $\bar f_{f_0}$ is done for $f_0 (980)$
only by assuming that 
$f_0$ is a $q\bar q$-type meson, namely 
$(u\bar u+d\bar d)/\sqrt{2}$ ($\equiv n\bar n$),
$s\bar s$, or mixture of them \cite{cheng-2006-f0}. 
On the other hand, it is known that $f_0 (980)$ is likely 
to be a tetra-quark meson or $K\bar K$ molecule
so as to explain the experimental measurements on
$f_0 (980) \to \pi\pi$, $f_0 (980) \to\gamma\gamma$,
and $\phi \to f_0 (980) \gamma$ \cite{pdg-2016,f0-sk,f0-others}.
The theoretical decay constant is not evaluated unfortunately,
as far as we are aware, for the tetra-quark or $K\bar K$ configurations
for the $f_0 (980)$. Therefore, a realistic numerical estimate would not be
possible for $f_0 (980)$ in comparing with experimental data on
$\gamma^* \gamma \to \pi^0 \pi^0$.

Of course, the $f_0 (980)$ may be viewed as a $q\bar q$ state
at high energies, whereas it may be a $qq\bar q \bar q$ one
at low energies, because they could mix with each other.
In fact, there is an indication from the constituent-counting-rule 
studies on $\Lambda (1405)$ in comparison with the experimental
data on $\gamma + p \to \Lambda (1405) + K^+$ that 
$\Lambda (1405)$ looks penta-quark state ($qqqq\bar q$) 
at low energies, whereas it could be an ordinary three-quark one ($qqq$)
at high energies \cite{counting}. There is a possibility 
that the situation could be the same for $f_0 (980)$ 
on the energy-dependent composition.

\begin{figure}[t]
     \includegraphics[width=7.0cm]{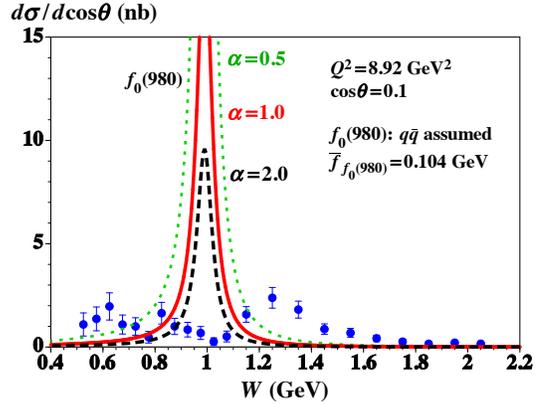}
\vspace{-0.30cm}
\caption{Expected $f_0 (980)$ contributions to the 
$\gamma^* \gamma \rightarrow \pi^0\pi^0$ cross section
by taking the decay constant 
$\bar f_{f_0 (980)}(Q^2=1 \ \text{GeV}^2)=0.104$,
which was obtained by assuming a $q\bar q$ configuration
for $f_0 (980)$ in the QCD sum rule. The cross sections are 
for the kinematics $Q^2=8.92$ GeV$^2$ and $\cos\theta=0.1$.
The parameter $\alpha$ is taken as $\alpha=0.5$, 1.0, and 2.0.
The $\bar f_{f_0 (980)}$ is evolved to $Q^2=8.92$ GeV$^2$.}
\label{fig:cross-f0-980}
\end{figure}

In any case, let us simply assume the decay constant $f_{f_0 (980)}$
by taking the $q\bar q$-type estimate in the QCD sum rule
in order to illustrate the situation and the issue.
As we will show later, the optimum value for the parameter $\alpha$
is roughly given by $\alpha\sim 1$. We find that it is difficult 
to accommodate the $f_0 (980)$ resonance with this parameter value.
Obtained cross sections are compared with the Belle data 
at $Q^2=8.92$ GeV$^2$ and $\cos\theta=0.1$ by taking
$\alpha =0.5, 1.0,$ and $2.0$ in Fig.\,\ref{fig:cross-f0-980}.
Here, the decay constant 
$\bar f_{f_0 (980)} (Q^2=1\ \text{GeV}^2)=0.104$ GeV 
was obtained in Ref.\,\cite{cheng-2006-f0} 
by considering the $u$ and $d$ quark contributions to the GDAs with 
$\bar f_n=0.35$ GeV at $Q^2=1$ GeV$^2$, the mixing angle 
$\theta_{f_0} = 32.5^\circ$, which is the middle of 
$25^\circ <  \theta_{f_0} < 40^\circ$, 
between $|\, n\bar n \, \rangle$ and $|\, s\bar s \, \rangle$,
$B_1 = -0.92$ by the QCD sum rule \cite{cheng-2006-f0},
and the conversion factor $18/30$ for the distribution amplitude
from Eq.\,(\ref{eqn:da-f0-2}) to Eq.\,(\ref{eqn:da-f02-normalized}).
The $Q^2$ evolution is also taken into account by using 
Eq.\,(\ref{eqn:ff-f02-evol-3}) from 
$Q^2=1$ GeV$^2$ to 8.92 GeV$^2$
in order to compare with the Belle data at $Q^2=8.92$ GeV$^2$.

From the comparison with the Belle data in Fig.\,\ref{fig:cross-f0-980},
we find that the $f_0 (980)$ peak structure is not obvious from the data
and that they are not consistent with the theoretical predictions
as long as $\alpha < 2$. Although the figure is one of the kinematical 
point of the Belle measurements, the comparisons with other data also
indicate a similar tendency. Here, we should note that the theoretical 
curves are shown by assuming the $q\bar q$ configuration of $f_0 (980)$
with the QCD sum rule estimate for the decay constant.
These results should suggest that $f_0 (980)$ could not be understood
mainly by the $q\bar q$ configuration. It is possibly a tetra-quark
(or $K\bar K$ molecule) state as widely known in the hadron-physics community. 
The decay constant could be very small if it is a tetra-quark ($qq\bar q\bar q$) 
type because the decay width is proportional to the matrix element of 
a bilocal operator.
Since the data do not show an obvious $f_0 (980)$ peak structure
and a theoretical estimate is not available for the decay width
by the tetra-quark picture, 
we do not include $f_0 (980)$ in our numerical analysis
in the analysis of Sec.\,\ref{analysis-results}.
If the measurements become more accurate in future, one may consider 
to include this contribution.

The $f_0 (980)$ effect is not conspicuous in the Belle data on
the differential cross section, for example, 
in Fig.\,\ref{fig:cross-f0-980}. However, it appears in the total 
cross section \cite{Masuda:2015yoh} and the $\gamma \to f_0 (980)$ 
transition form factor is investigated by using the Belle data 
\cite{kroll-2017}. Such studies indicate that $f_0 (980)$
is consistent with the $q\bar q$ configuration, which is 
different from our finding in Fig.\,\ref{fig:cross-f0-980}.
Because of these conflicting results, the $f_0(980)$ contribution 
and its internal configuration are not well understood.

\subsection{Analysis results}
\label{analysis-results}

\begin{table}[b]
\caption{Belle experimental data used in our analysis}
\label{tab:belle-data}
\begin{center}
\begin{tabular}{|c|c|c|} 
\hline
\ $Q^2$ (GeV$^2$)\ & \ $\cos \theta$\  & \ No. of data\  \\ 
\hline
8.92           &  \ 0.1, 0.3, 0.5, 0.7, 0.9 \     &  22 $\times$ 5           \\ 
10.93          &  \ 0.1, 0.3, 0.5, 0.7, 0.9 \     &  22 $\times$ 5 \\
13.37          &  \ 0.1, 0.3, 0.5, 0.7, 0.9 \     &  22 $\times$ 5 \\ 
17.23          & \ 0.1, 0.3, 0.5, 0.7, 0.9 \     &  22 $\times$ 5 \\
24.25          & \ 0.1, 0.3, 0.5, 0.7, 0.9 \     &  22 $\times$ 5 \\
\hline
total          &                       & 550       \\ 
\hline
\end{tabular}
\end{center}
\end{table}

The GDAs are expressed by a number of parameters, which are obtained
by a $\chi^2$ analysis of Belle experimental measurements on
$\gamma^* \gamma \to \pi^0 \pi^0$. The resonance part is fixed
as much as possible by other experimental and theoretical studies, 
and the used values are listed in Table \ref{tab:resonaces}.
The large uncertainty comes from the values of the decay constants,
$\bar f_{f_0 (500)}$, $\bar f_{f_0 (980)}$, and $f_{f_2 (1270)}$,
especially for the $f_0$ mesons. There are a number of reliable
theoretical studies on $f_{f_2 (1270)}$.
In Sec.\,\ref{f0-980-contribution}, we explained that the current
QCD sum rule estimate for $\bar f_{f_0 (980)}$ is much different
from the Belle measurements {\it if it is assumed as a $q\bar q$ state}.
There is no available estimate, as far as we are aware, for the decay
constant in the tetra-quark picture for $f_0 (980)$.
In any case, the data do not show a clear signature of $f_0 (980)$
in the $W \sim 1$ GeV region, so that $f_0 (980)$ is not included
in the following analysis. Furthermore, there is no theoretical
estimate on the decay constant for $\bar f_{f_0 (500)}$. 
We may simply assume that it is same as the $f_0 (980)$ value;
however, the results are inconsistent with the Belle data
in the same way with the $f_0 (980)$ case. 
Therefore, we consider two options in our studies:
\vspace{-0.15cm}
\begin{enumerate}
\item[(set 1)] Analysis without $f_0 (500)$ and $f_0 (980)$: \\
The GDAs are expressed by the parameters for only the continuum 
and $f_2 (1270)$, and they are determined by the $\chi^2$ analysis.
\vspace{-0.15cm}
\item[(set 2)] Analysis with $f_0 (500)$ and without $f_0 (980)$: \\
The decay constant $f_{f_0 (500)}$ is considered as an additional
parameter to be determined from the experimental data
in addition to the parameters in the set 1.
\end{enumerate}
\vspace{-0.15cm}
For the decay constants, the $Q^2$ evolution is taken into account 
by using Eq.\,(\ref{eqn:ff-f02-evol-3}) and taking the average scale 
of the Belle experiment as $\langle Q^2 \rangle =16.6$ GeV$^2$, 
which is a simple average of the minimum and maximum values, 
8.92 and 24.25 GeV$^2$, in the analyzed data in this work.

By considering the factorization condition of Eq.\,(\ref{eqn:hard-gpa}),
only the large $Q^2$ data with $Q^2 \ge 8.92$ GeV$^2$ are used 
in our analysis. Furthermore, the higher-order and higher-twist terms $A_{+-}$ and
$A_{0+}$ do not contribute significantly at large $Q^2$.
The $Q^2$ values to satisfy this condition are
$Q^2=$8.92, 10.93, 13.37, 17.23, and 24.25 GeV$^2$ in the Belle measurements.
In each $Q^2$, the pion angles are $\cos\theta=$0.1, 0.3, 0.5, 0.7, and 0.9
as listed in Table \ref{tab:belle-data}.
In each bin of $Q^2$ and $\cos\theta$, there are 22 data points, so that
the total number of data is 550.

\begin{figure}[b!]
     \includegraphics[width=6.0cm]{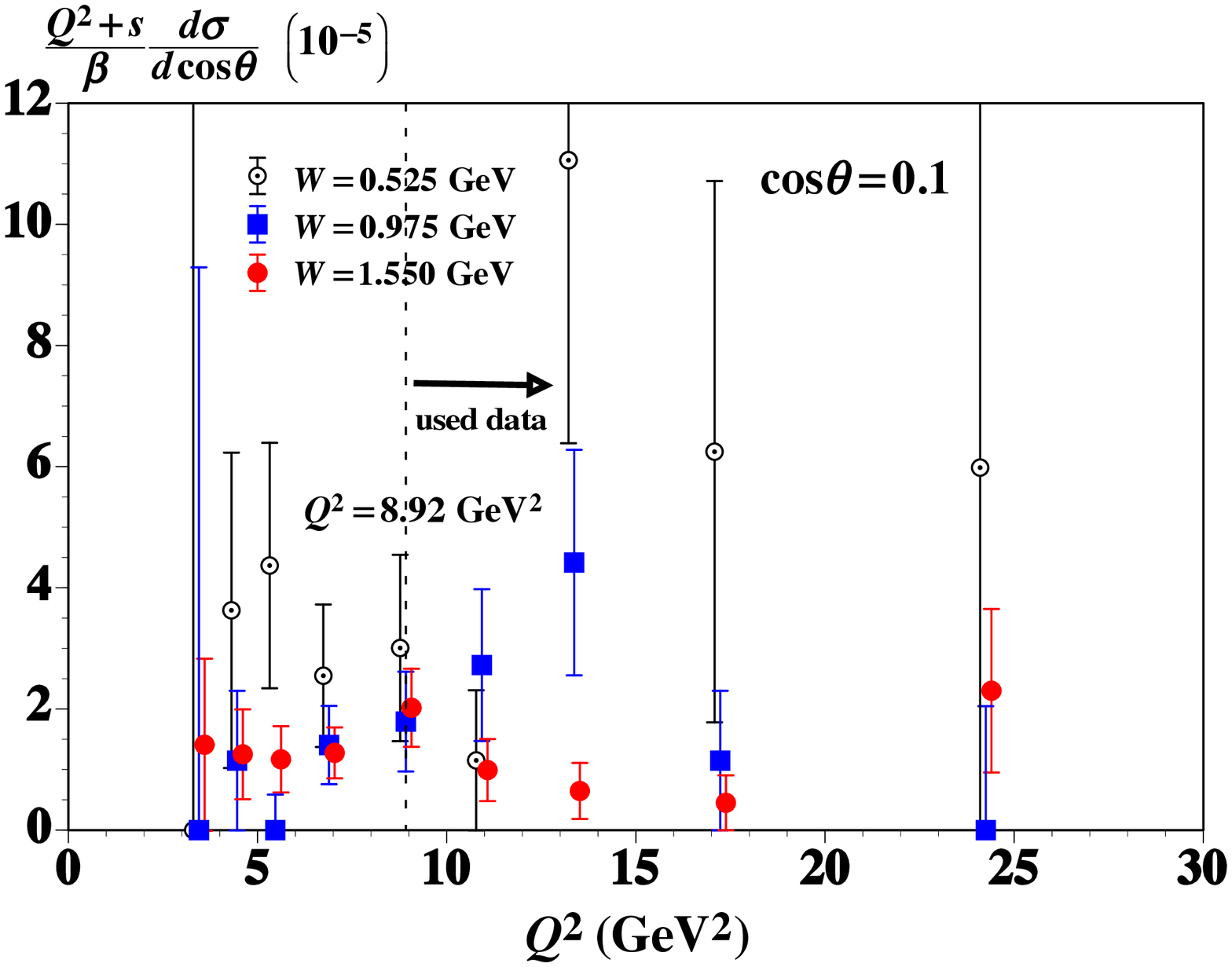}
\vspace{-0.30cm}
\caption{$Q^2$-scale dependence of the Belle data at $\cos\theta=0.1$.
The ordinate corresponds to the term with the GDAs integrated over $z$
in Eq.\,(\ref{eqn:cross2}).}
\label{fig:data-scaling-1}
\vspace{0.30cm}
     \includegraphics[width=6.0cm]{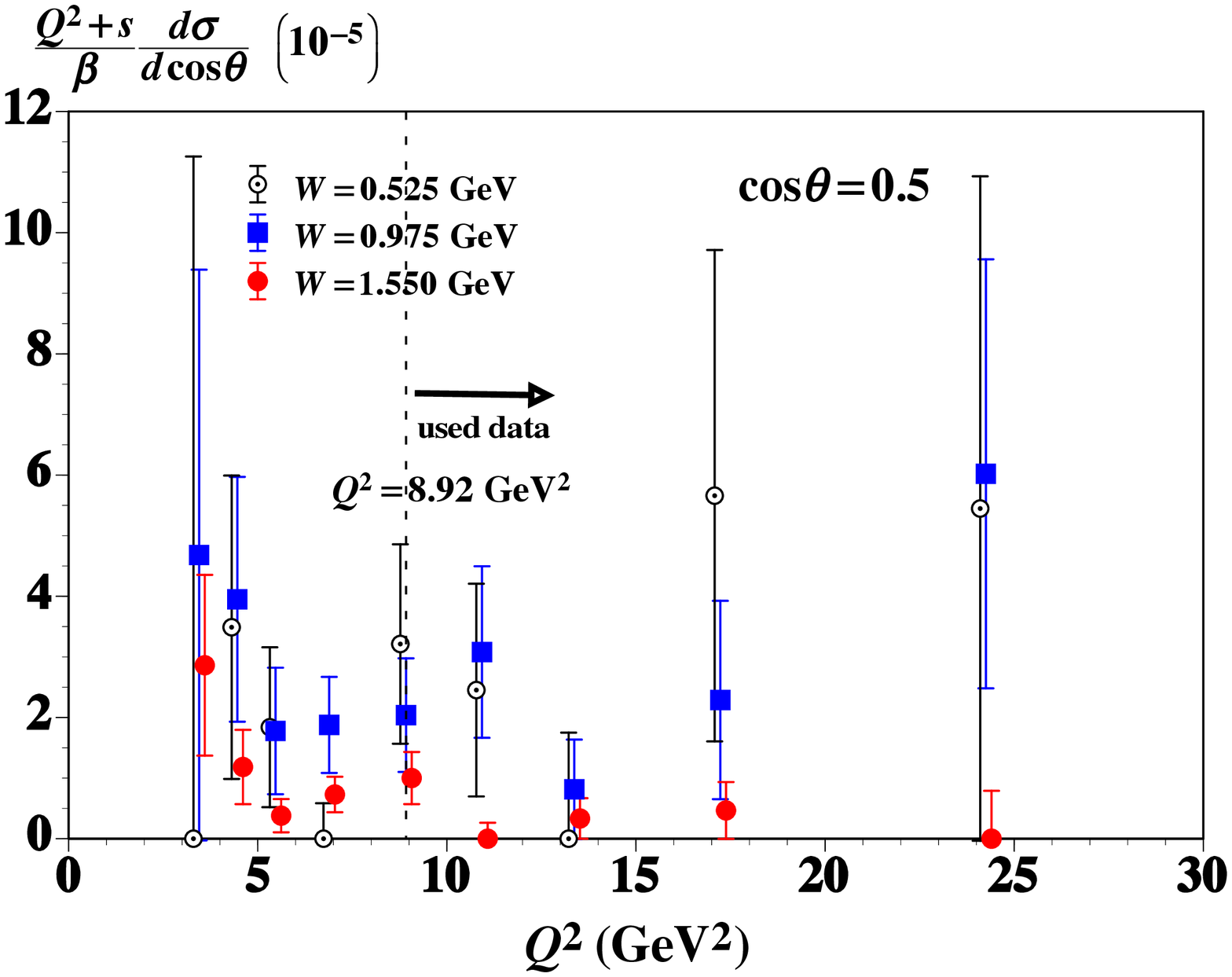}
\vspace{-0.30cm}
\caption{$Q^2$-scale dependence of the Belle data at $\cos\theta=0.5$.}
\label{fig:data-scaling-2}
\end{figure}

The GDAs are expressed by the three kinematical variables, $z$, $\zeta$, and $W^2$
without the scale $Q^2$ by considering the scaling region. 
Actual experiments are done at finite $Q^2$, so that
the GDAs extracted from the measurements may depend on $Q^2$.
In order to check the $Q^2$ dependence of the Belle data, 
we show the quantity $(Q^2 + s) d\sigma/\beta d (\cos\theta)$ 
in Figs.\,\ref{fig:data-scaling-1} and \ref{fig:data-scaling-2}
for $\cos\theta=0.1$ and $\cos\theta=0.5$, respectively
by choosing $W=0.525$, 0.975, and 1.550 GeV.
According to Eq.\,(\ref{eqn:cross2}), there is no scale dependence
for this quantity in the scaling limit.
As shown in these figures, the Belle data are not very accurate
at this stage to discuss whether $Q^2$ dependence exists.
However, there are tendencies for the scaling within the errors.
The $Q^2$ variations may be seen at $Q^2 < 6$ GeV$^2$ at $W=1.55$ GeV
and $\cos\theta=0.5$ in Fig.\, \ref{fig:data-scaling-2}; 
however, such data are irrelevant
in our analysis because only the data with $Q^2 \ge 8.92$ GeV$^2$
are used.

\begin{table}[b]
\caption{Constants and parameters determined by the $\chi^2$ analysis.
Here, the $\bar f_{f_0(500)}$ and $f_{f_2(1270)}$ values are provided 
at $Q^2=16.6$ GeV$^2$, and they correspond to 
$\bar f_{f_0(500)}=0.0246 \pm 0.0045$ and
$f_{f_2(1270)}=0.101$ at $Q^2=1$ GeV$^2$.}
\label{tab:analysis-results}
\begin{center}
\begin{tabular}{|c|c|c|} 
\hline
\  Parameter \                 & \ set 1 \                      
                               & \ set 2 \                          \\
\hline
$\alpha$                       & \,     0.801 $\pm$ 0.042 \,  
                               & \, \,  1.157 \, $\pm$ 0.132 \,  \ \ \  \\
$\Lambda$ (GeV)                & \,     1.602 $\pm$ 0.109 \,       
                               & \, \,  1.928 \, $\pm$ 0.213 \,  \ \ \  \\
\, $\bar f_{f_0(500)}$ (GeV) \,     & \,     0 \ \ \ \ \ \ \ (fixed) \, 
                               & \,     0.0184 $\pm$ 0.0034 \,          \\
\, $f_{f_2(1270)} \!$ (GeV) \, & \,     0.0754 (fixed)    \,     
                               & \,     0.0754 (fixed)  \,  \ \ \       \\
$a_\delta$                     & \,     3.878 $\pm$ 0.165 \,       
                               & \, \,  3.800 \, $\pm$ 0.170 \, \ \ \   \\
$b_\delta$                     & \,     0.382 $\pm$ 0.040 \, 
                               & \, \,  0.407 \, $\pm$ 0.041 \, \ \ \   \\
\hline
$\chi^2$/d.o.f.                &  1.22  
                               &  1.09                                 \\
\hline
\end{tabular}
\end{center}
\end{table}

In the analysis 1, there are four parameters,
$\alpha$, $\Lambda$, $a_\delta$, and $b_\delta$,
and the others are fixed. For example, $n=2$ is taken by the constituent-counting 
rule, and $\sum_q M_{2(q)}^\pi=0.5$ from pion-structure function studies.
The $f_0 (500)$ contribution is terminated by taking $\bar f_{f_0(500)}=0$.
The obtained parameter values are listed in Table \ref{tab:analysis-results}.
A reasonable fit is obtained in this analysis with $\chi^2/\text{d.o.f.}=1.22$.
Assigning the decay constant $\bar f_{f_0(500)}$ as an additional
parameter in the analysis 2, we obtained a better agreement with the data
with $\chi^2/\text{d.o.f.}=1.09$. In both cases, the parameter $\alpha$
is close to the asymptotic value $\alpha=1$.
For the pion distribution amplitude, a more concave functional form
is suggested at finite $Q^2$ \cite{Braun:2006dg,pionDA:DSE,convave-da}.
However, the pion distribution amplitude is related to the $C$-odd GDAs
as shown in Eq.\,(\ref{eqn:das-gdas}), and our current analysis is
for the $C$-even GDAs, so that there is no direct connection.

\begin{figure}[b!]
     \includegraphics[width=8.5cm]{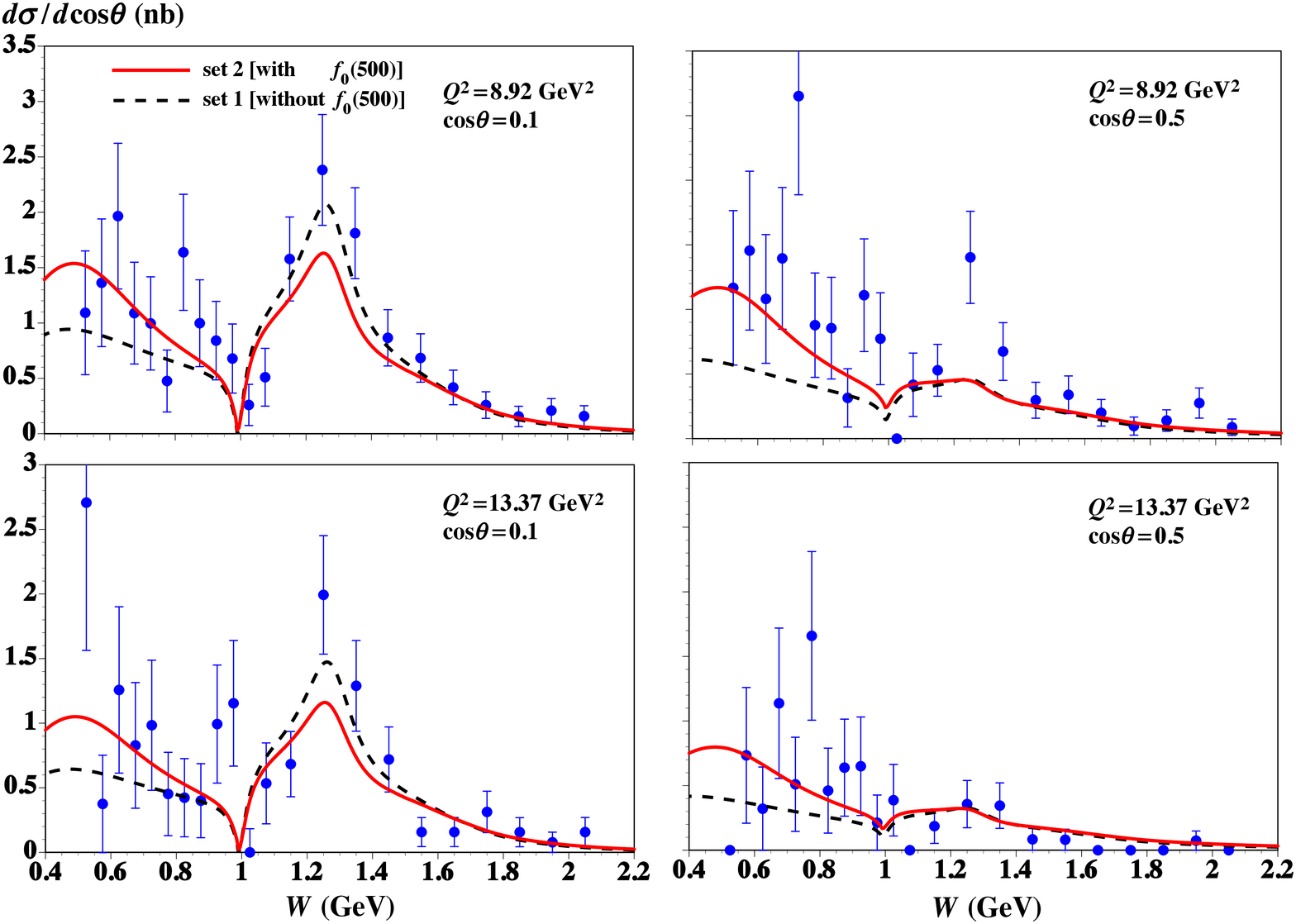}
\vspace{-0.30cm}
\caption{Comparison with the Belle cross sections measurements 
at $Q^2=8.92$ and 13.37 GeV$^2$ with $\cos\theta=0.1$ and 0.5.
The dashed and solid curves indicate our analysis results
for set 1 (without $f_0 (500)$) and set 2 (with $f_0 (500)$),
respectively.}
\label{fig:cross-1}
\vspace{0.30cm}
     \includegraphics[width=8.5cm]{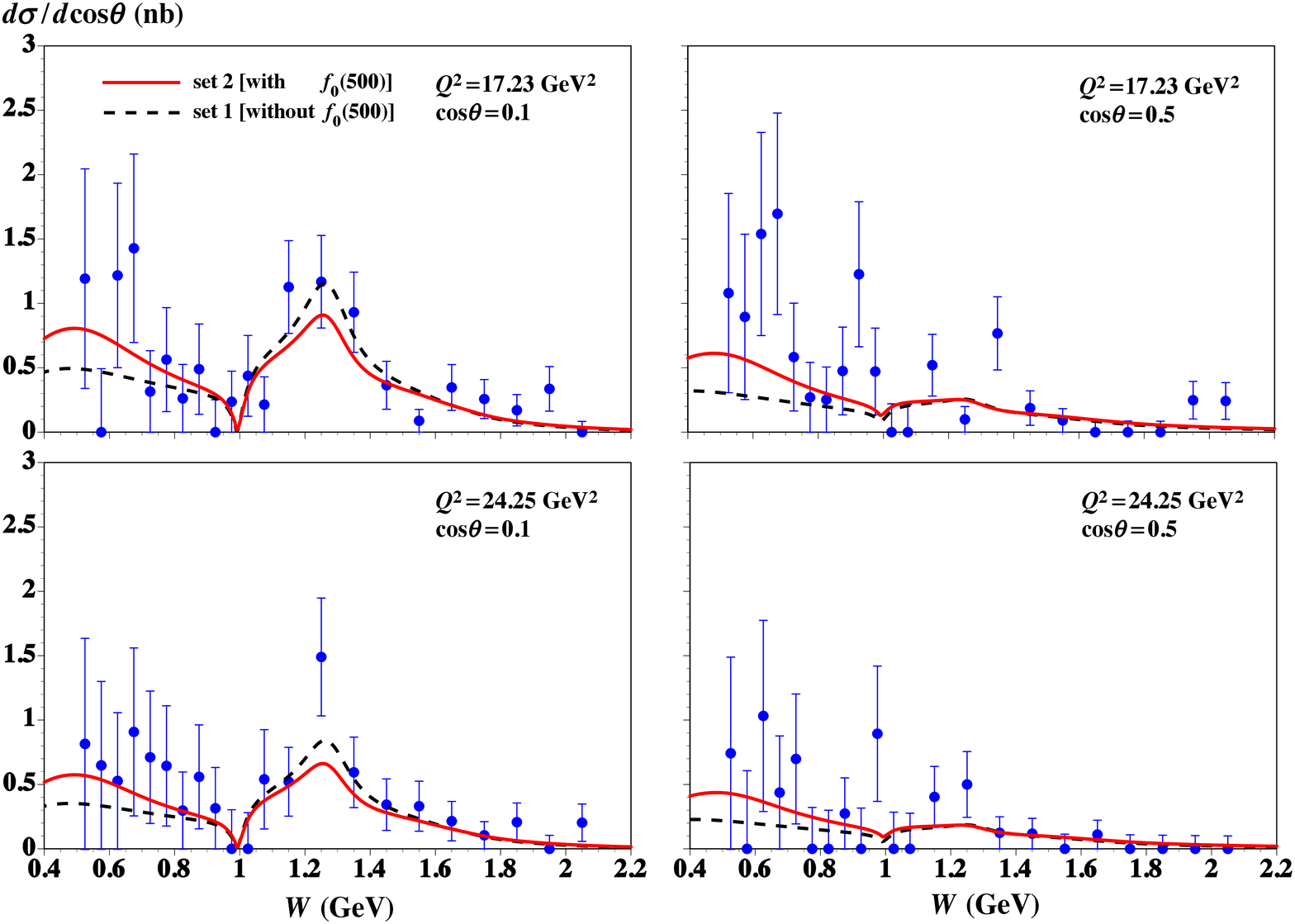}
\vspace{-0.30cm}
\caption{Comparison with the Belle cross sections measurements 
at $Q^2=17.23$ and 24.25 GeV$^2$ with $\cos\theta=0.1$ and 0.5.
The dashed and solid curves indicate our analysis results
for set 1 (without $f_0 (500)$) and set 2 (with $f_0 (500)$),
respectively.}
\label{fig:cross-2}
\end{figure}

The cutoff parameter is in the range $1.6 < \Lambda < 2.0$ GeV, which
is larger than the cutoff of the nucleon's electromagnetic
form factors. The set 2 provides a better description of the Belle data,
as indicated by the $\chi^2/\text{d.o.f.}$ value,
especially at small $W (< 0.8$ GeV).
In both analyses, the values of $a_\delta$ and $b_\delta$ stay
at almost same values, $a_\delta \simeq 3.8$ and $b_\delta \simeq 0.4$.
In order to explain the Belle data, the decay constant of $f_0(500)$
is $\bar f_{f_0(500)}=0.0183$ GeV at $Q^2=16.6$ GeV$^2$.
It becomes $\bar f_{f_0(500)}=0.0246$ GeV at $Q^2=1$ GeV$^2$,
and this value is much smaller than the one
for $\bar f_{f_0(980)}$ with the $q\bar q$ picture 
($\bar f_{f_0(980)}=0.104$ GeV at $Q^2=1$ GeV$^2$)
\cite{cheng-2006-f0}.

The actual comparisons with the Belle data sets are shown
in Figs.\,\ref{fig:cross-1} and \ref{fig:cross-2}
for $Q^2=$8.92, 13.37, 17.23, and 24.25 GeV$^2$ and
$\cos\theta=$0.1 and 0.5.
The dashed and solid curves are our theoretical results
for set 1 (without $f_0 (500)$) and set 2 (with $f_0 (500)$).
There is a dip around $W=1$ GeV, which is caused by cancellations
between the S- and D-wave terms.
The $f_2 (1270)$ contribution is obvious at $\cos\theta =0.1$
but it is relatively suppressed at larger $\cos\theta$ ($=0.5$).
As mentioned before, the $f_0 (980)$ effects do not appear in
the data. However, since the $\chi^2$ value is slightly smaller
in the analysis set 2 in comparison with the set-1 value,
the $f_0 (500)$ could be needed for interpreting the data
in the small $W$ range ($W<0.8$ GeV).

The whole cross section decreases with increasing $Q^2$
as shown in Figs.\ref{fig:cross-1} and \ref{fig:cross-2}.
Especially, at reasonably large $\cos\theta$, the $f_2$ 
resonance effects becomes small. Due to the scale dependence
of the decay constants $\bar f_{f_0}$ and $f_{f_2}$, the resonance
contributions should become small in comparison with the continuum
as $Q^2$ becomes large.
At high-energy $e^+ e^-$ colliders such as the international
linear collider (ILC), large $Q^2$ measurements should be done
and such experiments are suitable for probing the continuum
part of the GDAs. They are valuable for the studies of 
the GDAs as one of three dimensional structure functions
and their relations to the GPDs.

\begin{figure}[t!]
     \includegraphics[width=6.0cm]{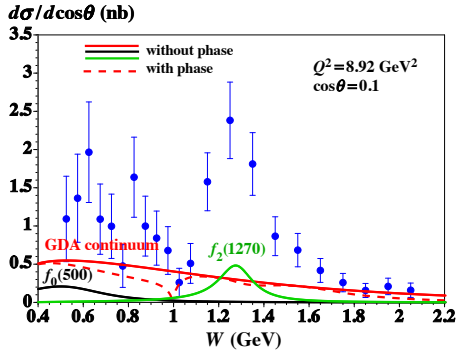}
\vspace{-0.30cm}
\caption{Each contribution to the 
$\gamma^* \gamma \rightarrow \pi^0\pi^0$ cross section
is shown for the 
the kinematics $Q^2=8.92$ GeV$^2$ and $\cos\theta=0.1$.
The solid curves indicate cross sections by terminating
other contributions and phase shifts. 
Three curves are for only $f_0 (980)$, continuum, or $f_2 (1270)$.
The dashed curve shows the continuum cross section 
by turning on the phase shifts.
The parameter values of the set-2 results are used here.}
\label{fig:each-cross}
\end{figure}

In order to see each term contribution to
the $\gamma^* \gamma \rightarrow \pi^0\pi^0$ cross section,
we show the cross section solely coming from
$f_0 (500)$, GDA continuum, or $f_2 (1270)$ in Fig.\,\ref{fig:each-cross} 
by terminating other terms and the phase shifts 
in Eq.\,(\ref{eqn:gda-parametrization})
for the kinematics of $Q^2=8.92$ GeV$^2$ and $\cos\theta=0.1$.
In the solid curves, the phase shifts
are also terminated, whereas the dashed curve indicates
the GDA continuum with the phase shifts.
For example, the solid GDA continuum curve is obtained
by setting $\bar f_{f_0 (500)}=f_{f_2 (1270)}=0$ and $\delta_0 = \delta_2 =0$.
Here, the parameters of the set 2 are used for drawing these curves. 
In comparison with the solid curve in Fig.\,\ref{fig:cross-1} 
for $Q^2=8.92$ GeV$^2$ and $\cos\theta=0.1$, these distributions seem 
to be very small. However, the continuum and $f_2$ contribute 
to the cross section constructively with almost the same magnitude, 
so that each contribution is about 1/4 of the cross section 
of Fig.\,\ref{fig:cross-1} if other terms are terminated. 
As expected, $f_0 (500)$ contributes only in the low-energy region 
of $W<0.8$ GeV, and it is much smaller than the continuum
according to the set-2 analysis. 
However, it depends on the $f_0 (500)$ decay constant,
which is taken as one of the parameters in our analysis
because of the lack of theoretical information.
The $f_2 (1270)$ contributes especially in the $W=1.27$ GeV region,
and its magnitude is comparable to the continuum.
The GDA continuum is a slowly varying function of $W$ 
and it is distributed in the wide $W$ range. 

\subsection{Gravitational form factors and radii for pion}
\label{emt-form-factor}

\begin{figure}[t!]
     \includegraphics[width=6.0cm]{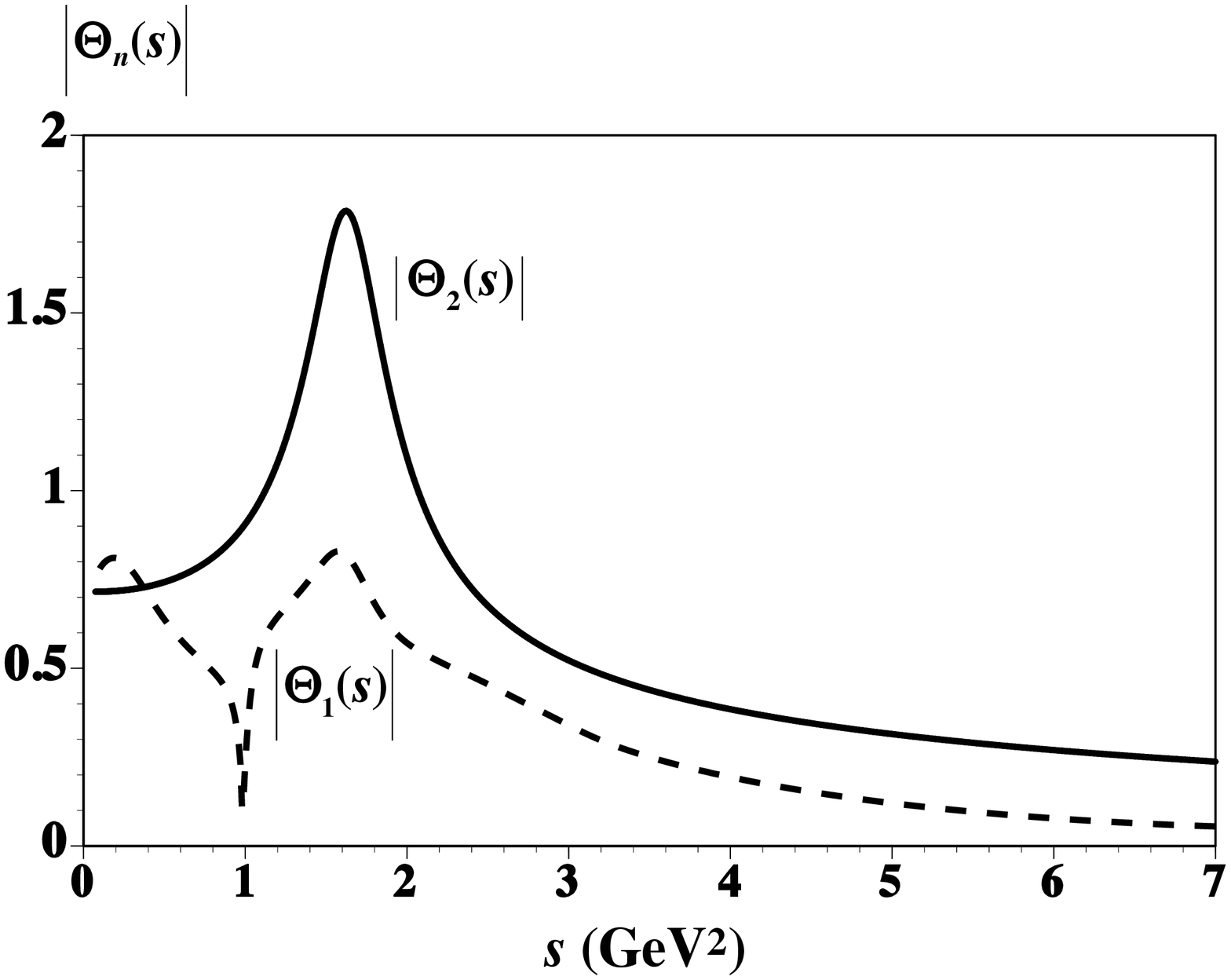}
\vspace{-0.30cm}
\caption{Absolute values of the timelike gravitational form
factors $\Theta_1(s)$ and $\Theta_2(s)$ of the pion.}
\label{fig:abs-timelike-ffs}
\vspace{0.30cm}
     \includegraphics[width=6.0cm]{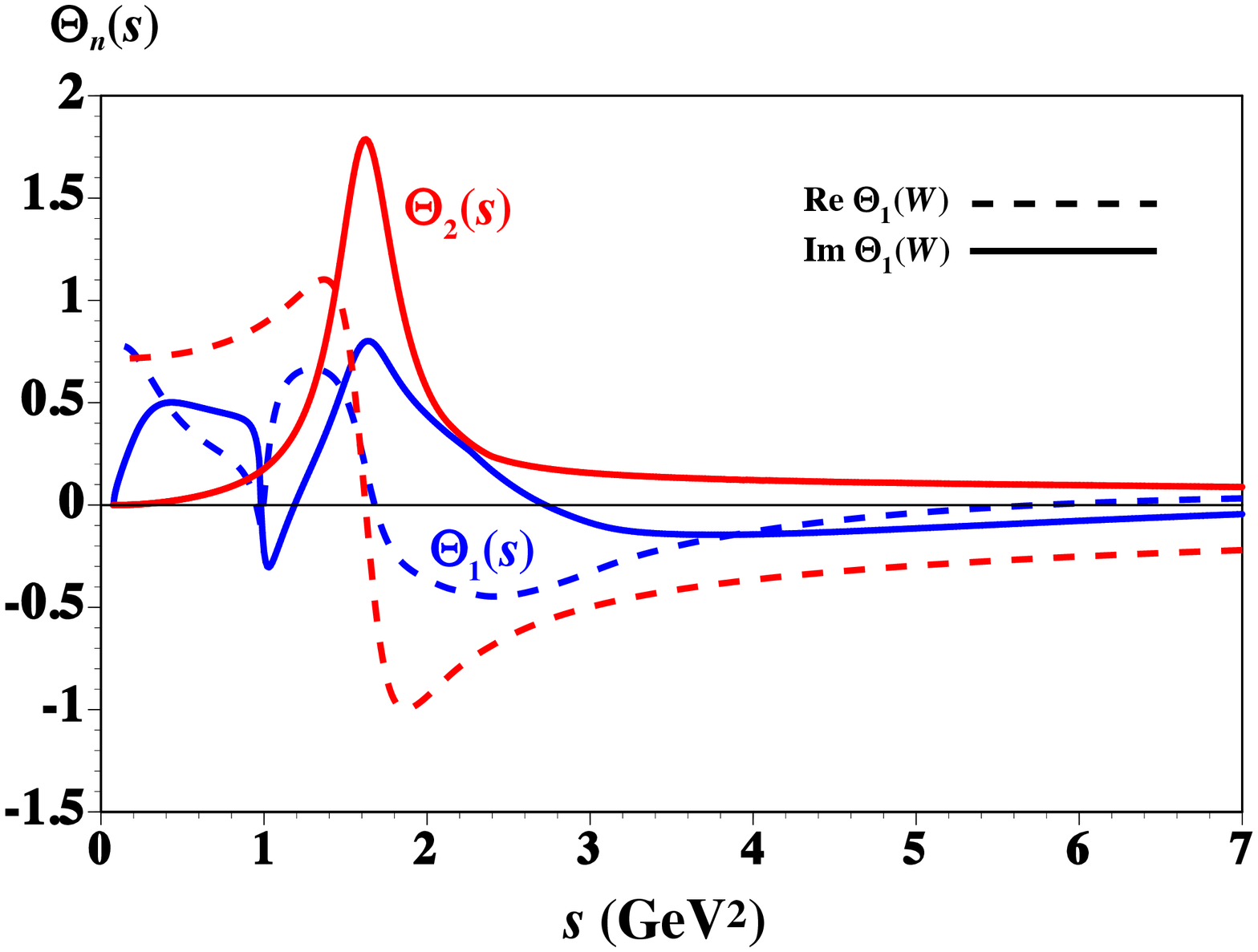}
\vspace{-0.30cm}
\caption{Real and imaginary parts of the timelike gravitational form
factors $\Theta_1(s)$ and $\Theta_2(s)$ of the pion.}
\label{fig:re-im-timelike-ffs}
\end{figure}

Since the optimum GDAs are determined from the Belle data,
the timelike gravitational form factors are calculated
by Eqs.\,(\ref{eqn:emt-ffs-gdas}) and (\ref{eqn:quark-sum-theta12}).
Their absolute values are shown in Fig.\,\ref{fig:abs-timelike-ffs},
and individual real and imaginary parts are in 
Fig.\,\ref{fig:re-im-timelike-ffs}.
The form factor $\Theta_2$ comes from the D-wave contribution
and it is peaked at the $f_2 (1270)$ position, whereas
the function $\Theta_1$ has a dip due to 
the interference between the S- and D-wave terms.
The imaginary part of $\Theta_2$ is peaked at the $f_2 (1270)$
resonance and its real part changes the sign.
The real and imaginary parts of $\Theta_1$
have both features on the interference and the $f_2 (1270)$ resonance.
As for the electric form factor of the pion in the timelike region,
there are recent theoretical studies by the holographic QCD
and lattice QCD \cite{timelike-pion}.

\begin{figure}[b]
     \includegraphics[width=6.0cm]{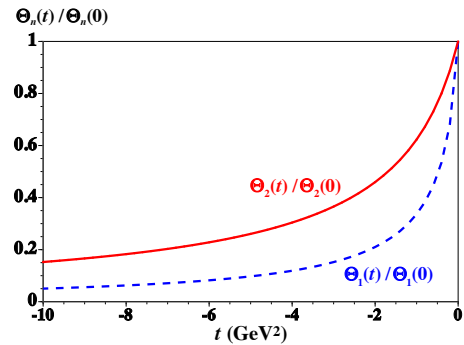}
\vspace{-0.30cm}
\caption{Spacelike gravitational form factors normalized to 
their values at $t=0$.}
\label{fig:spacelike-theta12}
\end{figure}

In order to find the space distributions and gravitational radii,
the timelike form factors should be transformed to the spacelike one
by using the dispersion relation of Eq.\,(\ref{eqn:dispersion-form-1}).
The obtained spacelike form factors are shown in 
Fig.\,\ref{fig:spacelike-theta12}. They are slowly decreasing function
of $-t$ and the slope is steeper for $\Theta_1$ than 
the one for $\Theta_2$ due to the additional S-wave term
in Eq.\,(\ref{eqn:emt-ffs-gdas}).

The substantial difference between the form factors certainly 
contradicts to the soft-pion theorem \cite{emt-pion1,Shifman:1988zk} 
which guarantees that Goldstone bosons in gravitational field 
are insensitive to the scalar curvature \cite{Voloshin:1982eb}.
As the gravity is coupled to the conserved energy-momentum tensor 
including the gluon contributions, it may be the signal
that gluon GDA, whose contribution to the considered 
two-photon process is suppressed, is essential.    

\begin{figure}[t!]
     \includegraphics[width=6.0cm]{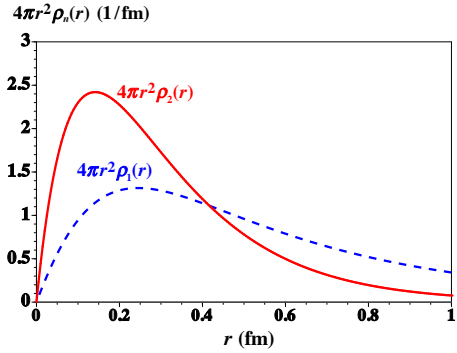}
\vspace{-0.30cm}
\caption{Gravitational 
densities $\rho_1 (r)$ and $\rho_2 (r)$.}
\label{fig:mass-pressure-density12}
\end{figure}

Then, the gravitational densities and their radii
are calculated by Eqs.\,(\ref{eqn:rho3-timelike})
and (\ref{eqn:3D-radius-2}), respectively.
The gravitational densities $\rho_1 (r)$ and $\rho_2 (r)$,
which are obtained from $\Theta_1 (t)$ and $\Theta_2 (t)$,
respectively, are shown for the pion
in Fig.\,\ref{fig:mass-pressure-density12}.
It is known that the spacelike electric form factor of the proton
is known as the dipole form 
$F_p (q) = 1/(1+\vec q^{\, 2}/\Lambda^2)^2$, which leads
to the exponential charge density 
$\rho_p (r) = (\Lambda^3/(8 \pi)) e^{-\Lambda r}$
by the Fourier transform. Typical functional forms 
of charge densities and form factors are given 
in Table \ref{tab:density-formfactor}
for hadrons and nuclei.
The charge form factors and densities of light nuclei 
are typically given by the Gaussian functional forms,
whereas the densities become flat ones for large nuclei.
The pion form factor is roughly given by the monopole form 
$F_\pi (q) = 1/(1+\vec q^{\, 2}/\Lambda^2)$ as suggested
by the constituent counting rule, and
its space distribution is given by the Yukawa form
$\rho_\pi (r) = (\Lambda^2 /(4\pi r)) e^{-\Lambda r}$.
It is a divergent function as $r \to 0$, so that 
it is more appropriate to show the density by $4 \pi r^2 \rho (r)$
rather than $\rho (r)$ itself as usually done
for the nucleons and nuclei.

\begin{table}[t!]
\caption{Typical densities and form factors.}
\label{tab:density-formfactor}
\begin{center}
\begin{tabular}{|c|c|c|} 
\hline
\ Hadrons\ & \ $\rho (r)$\  & \ $F(q)$\  \\ 
\hline
Quark           &  \ $\frac{1}{4\pi r^2} \delta (r)$ \   &  $1$    \\ 
Pion            &  $ \frac{\Lambda^2}{4\pi r} e^{-\Lambda r}$                               
                &  $\frac{1}{1+\vec q^{\, 2}/\Lambda^2}$       \\
Proton          &  \ $\frac{\Lambda^3}{8\pi} \, e^{-\Lambda r}$ \     
                &  $\frac{1}{(1+\vec q^{\, 2}/\Lambda^2 )^2}$      \\ 
Light nuclei    & \ $\left ( \frac{\Lambda^2}{\pi} \right ) ^{3/2} e^{-\Lambda^2 r^2}$\  
                &  $e^{- \vec q^{\, 2}/(4 \Lambda^2)}$ \\
\ Heavy nuclei\ \ & \ $\frac{3}{4\pi R^3} \, \theta (R-r)$ \   
                &  $\frac{3 j_1 (qR)}{qR}$ \\
\hline
\end{tabular}
\end{center}
\end{table}

To understand the physics meaning of the energy-momentum tensor and
the gravitational form factors, the static energy-momentum tensor is
defined in the Breit frame as \cite{energy-momentum}
\begin{align}
\! \! 
T^{\mu\nu}_q (\vec r \,) = \! 
\int \! \frac{d^3 q}{(2\pi)^3 \, 2E} e^{i \vec q \cdot \vec r} 
\left\langle \pi^0 (p') \! \left| 
\, T^{\mu\nu}_q (0) \, \right| \! \pi^0 (p) \right\rangle ,
\label{eqn:static-energy}
\end{align}
where $E=\sqrt{m_\pi^2 +\vec q^{\ 2}/4}$.
The $\mu\nu=00$ component satisfies the mass relation
\begin{align}
\int d^3 r \, T^{00}_q (\vec r ) = m_\pi \Theta_{2,q} (0) .
\label{eqn:T00}
\end{align}
Therefore, the $\Theta_2$ reflects the mass (energy)
distribution in the pion.
The $\mu\nu = ij$ ($i,\, j =1,\, 2,\, 3$) components are
expressed by the pressure $p(r)$ and shear force $s(r)$ as
\begin{align}
T^{\, ij}_q (\vec r \,) = p_q (r) \, \delta_{ij} 
    + s_q (r) \left ( \frac{r_i r_j}{r^2} - \frac{1}{3} \delta_{ij}  \right ) .
\label{eqn:Tij}
\end{align}
Using the definition of the energy-momentum-tensor form factors, 
we find that $p(r)$ and $s(r)$ are expressed by $\Theta_1$.
Namely, the $\Theta_1$ is the mechanical form factor
which contains information on the pressure and shear force.
The conservation of the energy-momentum tensor $\partial_\mu T^{\mu\nu} =0$
indicates the stability condition for the pressure $p(r)$ as
\cite{energy-momentum,stability}
\begin{align}
\int_0^\infty dr \, r^2 \, p(r) =0 .
\label{eqn:stability}
\end{align}
It is satisfied in our formalism due to the finite 
$\Theta_1 (t=0)$, as also noticed in Ref.\,\cite{emt-pion1},
because of the $\delta$ function in the $r$ integration.

According to the definition (\ref{eqn:static-energy}), the mass (energy) density 
is given mainly by the form factor $\Theta_2 (t)$; however, 
$\Theta_1 (t)$ also contributes at finite $t$.
On the other hand, pressure and shear-force densities are given
solely by the form factor $\Theta_1 (t)$. 
Therefore, we may use the terminologies ``mass" (or energy) and ``mechanical"
(pressure and shear force)
for $\Theta_2 (t)$ [$\rho_2 (r)$, $\langle r^2 \rangle _2$] and
$\Theta_1 (t)$ [$\rho_1 (r)$, $\langle r^2 \rangle _1$].

The gravitational densities $4 \pi r^2 \rho_1 (r)$
and $4 \pi r^2 \rho_2 (r)$ are peaked at 
$r=0.1 \sim 0.2$ fm region in Fig.\,\ref{fig:mass-pressure-density12}.
However, the mechanical density $\rho_1 (r)$ is distributed
in larger-$r$ region, which is our interesting finding
for studying the gravitational physics of the pion.
The mechanical density contains the shear force,
which could be dominant in the surface region,
so that the $\rho_1 (r)$ may be distributed 
in the relatively large-$r$ region.
From the densities $\rho_1 (r)$ and $\rho_2 (r)$
or the spacelike form factors $\Theta_1 (t)$ and $\Theta_2 (t)$,
the gravitational radii can be calculated.
We obtained the radii
$\sqrt {\langle r^2 \rangle _2} \equiv
 \sqrt {\langle r^2 \rangle _{\text{mass}}}$ and
$\sqrt {\langle r^2 \rangle _1} \equiv
 \sqrt {\langle r^2 \rangle _{\text{mech}}}$ as
\begin{align}
\! \! \! \! \! \!
\sqrt {\langle r^2 \rangle _{\text{mass}}} = 0.39 \, \text{fm}, \ 
\sqrt {\langle r^2 \rangle _{\text{mech}}} = 0.82 \, \text{fm}  \ 
\text{(set 2)} .
\label{eqn:g-radii-pion-2}
\end{align}
It is interesting that we found a mass radius which is much smaller 
than the charged one 
$\sqrt {\langle r^2 \rangle _{\text{charge}}} =0.672 \pm 0.008$ fm;
however, the mechanical radius is slightly larger as indicated in
the density $4 \pi r^2 \rho_1 (r)$ of
Fig.\,\ref{fig:mass-pressure-density12}. 
It is because that
there is also the S-wave term $\widetilde{B}_{10}$
in addition to the D-wave one $\widetilde{B}_{20}$.
In physics, the pressure and shear-force distributions
have different nature from the mass distribution.

We should note that there is uncertainty in our analysis in the sense
that only the relative phase $\delta_0 (W)-\delta_2 (W)$ 
affects the cross section; however, 
their absolute phases are not. It means that 
the phase $\Delta \delta (W)$ could be attributed 
to $\delta_2$ instead of $\delta_0$ in Eq.\,(\ref{eqn:delta0-phase}).
We repeated our $\chi^2$ analysis
with this extreme option and obtained the radius values as
$\sqrt {\langle r^2 \rangle _{\text{mass}}} = 0.32 \, \text{fm}$ and
$\sqrt {\langle r^2 \rangle _{\text{mech}}} = 0.88 \, \text{fm}$.
Therefore, it is fair to state at this stage 
that the evaluated gravitational radii are in the ranges:
\begin{align}
\sqrt {\langle r^2 \rangle _{\text{mass}}} 
  & = 0.32 \sim 0.39 \, \text{fm}, 
\nonumber \\
\sqrt {\langle r^2 \rangle _{\text{mech}}} 
  & = 0.82 \sim 0.88 \, \text{fm} .
\label{eqn:g-radii-pion-range}
\end{align}
It is encouraging that similar radii are obtained in totally 
different analyses. 
These corrections lead to interesting findings that
the mass radius is much smaller than the pion charge radius
$\sqrt {\langle r^2 \rangle _{\text{charge}}} =0.672 \pm 0.008$ fm
and that the mechanical radius, defined by the slope of 
the form factor $\Theta_1$ in Eq.\,(\ref{eqn:3D-radius-2}), 
is slightly larger the charge radius.
It is interesting to find that the mass radius is much smaller 
than the pion charge radius as suggested in Ref.\cite{teryaev-emt-form}.

Lattice QCD calculations on the energy-momentum tensor indicate
similar tendency that the mechanical radius is larger than the mass radius
\cite{lattice-g-forms}. Here, we should note the definition difference
from our form factor, namely
the factor of $-4$ in $B(t)$ and $\Theta_1 (t)$ as explained 
below Eq.\,(\ref{eqn:emt-ffs-norm-2}).
The actual radii are not shown in the lattice calculations 
\cite{lattice-g-forms}; however, spacelike form factors
and radii have similar tendencies with our results.
In addition, a theoretical estimate of D-term, which corresponds to
$\Theta_1$ in our studies, also shows a similar result \cite{d-term}.

Because the pion GDAs were obtained in this work, it is possible to
study their relation to the pion GPDs as explained 
in Secs.\,\ref{gpd-gda} and \ref{radon}.
In order to discuss the pion GPDs, we need to find appropriate 
double distributions from the GDAs and then to calculate the GPDs.
Since it is a significant work, we leave it as our future project.
In addition, the Belle collaboration has been investigating other 
hadron-pair production processes including $p\bar p$. Once such data become
available, it is possible to determine nucleonic GDAs in comparison
with the GPDs obtained in spacelike reactions.

This kind of studies has a bright prospect in the sense 
that the Belle collaboration has been analyzing other
meson productions $\gamma^* \gamma \to h\bar h$
from the two photon. The experimental errors of
Figs.\,\ref{fig:cross-1} and \ref{fig:cross-2}
are dominated by the statistical errors. The KEKB
was just upgraded to super-KEKB, so that the errors
should be much smaller in the near future.
Furthermore, if the ILC is realized, the two-photon cross section 
$\gamma^* \gamma \to h\bar h$
should be obtained in a very different kinematical region,
namely at large $Q^2$, and the ILC measurement should be
valuable for probing especially the continuum part
of the GDAs. 

For a long time, the GDAs had been considered as a purely 
theoretical subject. We showed in this work that it becomes possible
to investigate the GDAs experimentally with the appropriate
theoretical formalism. This study is merely a starting point.
Interesting prospects are waiting for us for investigating
gravitational physics for hadrons in the quark-gluon level.
For example, the equivalence principle indicates that
the anomalous gravitomagnetic moment should vanish
in the nucleon \cite{gravitomagnetic}. Therefore, 
the equivalence principle could be tested in the microscopic
particle physics by investigating the GPDs and GDAs for
the nucleon.

\section{Summary}\label{summary}

The GDAs are one of three-dimensional structure functions,
and they are related to the GPDs by the $s$-$t$ crossing
relation.
We analyzed the Belle data of the two-photon cross sections
$\gamma^* \gamma \to \pi^0 \pi^0$ for determining
the pion GDAs. This work is the first work to obtain the GDAs 
from the actual experimental data, and our results should be valuable 
for probing the three-dimensional structure of hadrons, especially
for future applications to unstable hadrons including exotic-hadron 
candidates which cannot be used in fixed-target experiments. 

Including the $f_0 (500)$ and $f_2 (1270)$ meson contributions
to the cross section, we expressed the pion GDAs by a number 
of parameters which were determined by analyzing the data.
The obtained $z$-dependence is close to the scaling one ($\alpha=1$). 
If we include $f_0 (980)$ contribution with constants
estimated by assuming it as a $q\bar q$ state, 
theoretical differential cross sections are much larger than 
the Belle measurements. 
The $f_0 (980)$ meson was not included in our actual analysis.
The GDAs contain the timelike gravitational form factors 
$\Theta_1 (s)$ and $\Theta_2 (s)$ of the energy-momentum tensor, 
and we calculated them from the obtained GDAs.
The function $\Theta_2 (s)$
is determined only by the D-wave part, whereas both S- and D-waves 
contribute to $\Theta_1 (s)$. Therefore, they have different
functional behaviors. This is the first time that 
the gravitational form factors are obtained from actual 
experimental measurements.

The timelike gravitational form factors are converted 
to the spacelike ones by the dispersion relation. 
Then, the gravitational mass and mechanical densities are
shown, and their radii are calculated.
We obtained
$ \sqrt {\langle r^2 \rangle _{\text{mass}}} 
    =  0.32 \sim 0.39 \, \text{fm}$
and  
$ \sqrt {\langle r^2 \rangle _{\text{mech}}} 
   = 0.82 \sim 0.88 \, \text{fm}$
from the form factors $\Theta_2$ and $\Theta_1$, respectively.
They indicate that the gravitational mass radius is much 
smaller than the charge radius
$\sqrt {\langle r^2 \rangle _{\text{charge}}} =0.672 \pm 0.008$ fm
and that the mechanical radius is slightly larger.
Future super-KEKB measurements should improve this situation.
We hope that this work will open a new field of gravitational
physics in the quark-gluon level.

\begin{acknowledgements}
\vspace{-0.20cm}

The authors thank Drs. M. Masuda and S. Uehara for explanations
and suggestions on Belle experimental measurements and two-photon processes;
M. Wakamatsu for discussions on the GPDs and energy-momentum tensor;
P. Bydzovsky, R. Kaminski, V. Nazari, and Y.~S.~Surovtsev
for supplying their code on the $\pi\pi$ phase shifts \cite{pi-pi-code},
which were used in this work. They also thanks V. Braun,
H.-Y. Cheng, H.-C. Kim, H.-D. Son, and K.-C. Yang
for their information on decay constants and form factors.
This work was supported by Japan Society for the Promotion of Science (JSPS)
Grants-in-Aid for Scientific Research (KAKENHI) Grant Number JP25105010.
Q.-T.S is supported by the MEXT Scholarship for foreign students 
through the Graduate University for Advanced Studies.
O.V.T thanks the Japan Society for the Promotion of Science
for its support on his visit to KEK, where this work was started.
SK and Q.-T.S thank the Institute for Nuclear Theory at 
the University of Washington for its hospitality and the
Department of Energy for partial support during the completion
of this work.
\end{acknowledgements}

\section*{Appendix: Errata in the first and second manuscripts}\label{appendix}
\vspace{-0.20cm}

In the first and second versions of the article arXiv:1711.08088 
\cite{kst-08088}, the factor of $1/\pi$ was missing 
in Eq.\,(54) for calculating the gravitational root-mean-square (rms) radius,
whereas the $1/\pi$ factor exists in Eq.\,(\ref{eqn:dispersion-form-1}). 
The correct rms radius should be calculated by Eq.\,(\ref{eqn:3D-radius-2}) 
of this article.
Although all the figures are right, the numerical values of gravitational
radii should be multiplied by $1/\sqrt{\pi}$.
In calculating the the rms radius by Eq.\,(\ref{eqn:3D-radius-2})
of the first and second versions \cite{kst-08088}, the normalization 
was properly taken into account 
by the replacement $F^h (t)  \to F^h (t)/F^h (t=0)$.
Because of the extra $1/\pi$ factor,
the rms radii of Eq.\,(131) should be multiplied by $1/\sqrt{\pi}$
as shown in Eqs.\,(\ref{eqn:g-radii-pion-2}) and 
(\ref{eqn:g-radii-pion-range}). The radius values are written 
in the abstract and summary section, and they were corrected accordingly.

These corrections lead to interesting findings that
the mass radius is much smaller than the pion charge radius
$0.672 \pm 0.008$ fm
and that the mechanical radius
is slightly larger the charge radius.
It is important to investigate physics origins of the differences
between the gravitational and charge radii.



\end{document}